\documentclass[aps,prb,reprint,longbibliography,amsmath,amssymb,superscriptaddress,floatfix]{revtex4-2}
\usepackage[T1]{fontenc}
\usepackage{graphicx}
\usepackage{array}
\usepackage{booktabs,multirow}
\usepackage{bm}
\usepackage{microtype}
\usepackage{braket}
\usepackage{mathrsfs}
\usepackage{hyperref}
\hypersetup{
  colorlinks=true,
  linkcolor=blue,
  citecolor=blue,
  urlcolor=blue,
  pdftitle={Universal eigenvector statistics of non-Hermitian random matrices}
}
\allowdisplaybreaks[2]
\DeclareMathOperator{\tr}{Tr}
\newcommand{\bigO}{\mathcal O}

\begin{document}
\title{Universal Eigenvector Statistics of Non-Hermitian Random Matrices}

\author{Ze Chen}
\thanks{These two authors contributed equally to this work.}
\affiliation{Department of Physics, Princeton University, Princeton, New Jersey 08544, USA}
\affiliation{Department of Electrical and Computer Engineering, Princeton University, Princeton, New Jersey 08544, USA}
\author{Zhenyu Xiao}
\thanks{These two authors contributed equally to this work.}
\affiliation{Princeton Quantum Initiative, Princeton University, Princeton, New Jersey 08544, USA}
\affiliation{Princeton Center for Theoretical Science, Princeton University, Princeton, New Jersey 08544, USA}
\author{Shinsei Ryu}
\affiliation{Department of Physics, Princeton University, Princeton, New Jersey 08544, USA}

\date{September 18, 2026}
\begin{abstract}
Eigenvector overlaps quantify nonorthogonality and govern the response
and dynamics of non-Hermitian systems.
We extend the universality of non-Hermitian random matrices to the
statistics of these overlaps.
We obtain analytical expressions for eigenvector overlaps in the
spectral bulk and near the origin, covering ten symmetry classes in
the limit of large matrix size.
Using fermionic replica nonlinear $\sigma$ models, we relate these
overlaps to Hermitian level statistics, symmetry class by symmetry class
and topological sector by topological sector.
Numerical calculations in various physical models support the
universality of the normalized overlaps in the regimes studied.
Our work establishes a duality between Hermitian level statistics and
non-Hermitian eigenvector overlaps.
\end{abstract}
\maketitle

\tableofcontents
\section{Introduction}

Random matrix theory (RMT) provides a statistical framework for complex
quantum and dynamical
systems~\cite{Wigner1951,Wigner1958,Dyson1962c,Mehta2004,Forrester2010,Huang2015,Stephanov1996,Osborn2004,AkemannWettig2004,May1972,Sommers1988,Sompolinsky1988}.
For Hermitian systems, the conjecture that quantum-chaotic spectra
exhibit random-matrix statistics has been extensively
tested~\cite{Bohigas1984,Haake2018,McDonaldKaufman1979,StoeckmannStein1990}.
This correspondence makes spectral correlations a benchmark for
identifying ergodic behavior and departures from it, underlying
applications to disordered
systems~\cite{Efetov1983,Altshuler1988,Shklovskii1993,Mirlin2000,Evers2008},
mesoscopic transport~\cite{Beenakker1997}, and interacting many-body
systems~\cite{OganesyanHuse2007,PalHuse2010,SerbynMoore2016,DAlessio2016,Abanin2019}.
Non-Hermitian random matrices extend this description to complex
spectra~\cite{Ginibre1965,Girko1985,Feinberg1997,Byun2024} and play an
analogous role in dissipative quantum
chaos~\cite{Grobe1988,Grobe1989,Denisov2019,Akemann2019,Hamazaki2019MBL,Hamazaki2020,Sa2020,Li2021,GarciaGarcia2022,Kulkarni2022,Shivam2023,GarciaGarcia2023},
providing spectral benchmarks for open quantum and optical
systems~\cite{Ashida2020,ElGanainy2018}.

While much of the literature has focused on the spectral correspondence
between chaotic physical systems and random matrices, it is natural to
ask whether this correspondence extends beyond eigenvalues.
Eigenvector statistics characterize localization in both
Hermitian~\cite{FyodorovMirlin1997,TikhonovMirlin2019,TikhonovMirlin2021}
and non-Hermitian systems~\cite{Ghosh2023}, providing information that
is not contained in the spectrum alone.
Generic non-Hermitian systems additionally exhibit eigenvector nonorthogonality:
their distinct left and right eigenvectors give rise to nontrivial
biorthogonal overlaps~\cite{chalker1998eigenvector,MehligChalker2000,Janik1999,Fyodorov2018,BourgadeDubach2020}.
For a nondegenerate eigenvalue, the diagonal overlap quantifies the
nonorthogonality of the corresponding mode, and is always unity in
Hermitian systems. It also measures spectral sensitivity to
perturbations~\cite{TrefethenEmbree2020,FyodorovSavin2012,Gros2014},
including enhanced sensitivity near exceptional
points~\cite{Wiersig2023,Wang2020Petermann}, and quantifies excess
spontaneous-emission noise and laser-linewidth
enhancement~\cite{Siegman1989,Schomerus2000Petermann}.
Off-diagonal overlaps describe correlations between distinct
nonorthogonal modes and, together with diagonal overlaps, govern
transient amplification and
relaxation~\cite{SavinSokolov1997,MehligChalker2000,ErdosKrugerRenfrew2018,TarnowskiNeriVivo2020,GudowskaNowak2020,Bao2025}.
These dynamical questions arise for decaying
states~\cite{Schomerus2017,Sommers1999}, Lindblad
evolution~\cite{Lindblad1976,Gorini1976,Prosen2008,Prosen2010,Can2019,Sa2020Liouvillian,Lieu2020},
and wave propagation~\cite{Prado2025,Akemann2025b}.
These correlations also connect to transport and nonequilibrium
response~\cite{DavyGenack2019,FyodorovOsman2022,Fyodorov2025Entropy}.

As in the study of eigenvalue statistics, generic complex random matrices
provide a natural starting point for non-Hermitian eigenvector statistics.
For the Ginibre ensemble, previous work obtained explicit
formulas for diagonal and off-diagonal
overlaps~\cite{chalker1998eigenvector,MehligChalker2000} and characterized
eigenvector nonorthogonality at given
eigenvalues~\cite{BourgadeDubach2020,Fyodorov2018,AkemannTribeTsareasZaboronski2020}.
Calculations have been extended to other
ensembles~\cite{Janik1999,NowakTarnowski2018}.
The universality of diagonal-overlap statistics has been established in several
settings~\cite{Dubach2021Spherical,Osman2026Overlaps}.
A natural next step is to understand how symmetry modifies these
statistics, as it does for eigenvalues.
Symmetry effects have been examined in real~\cite{Fyodorov2018,WurfelCrumptonFyodorov2024}
and quaternionic Ginibre
matrices~\cite{AkemannForsterKieburg2020,AkemannByunNoda2026}, and in the
diagonal overlaps of complex symmetric matrices~\cite{Akemann2025b}.
These results for specific ensembles motivate a common description of
how symmetry organizes eigenvector statistics across classes and
spectral regimes.

Non-Hermitian random matrices admit 38 symmetry classes~\cite{Bernard2002,Kawabata2019a},
enriching the tenfold Hermitian Altland--Zirnbauer (AZ) classification~\cite{Zirnbauer1996,Altland1997}.
Despite this richer classification, local level statistics in the generic
complex bulk, away from spectral boundaries, symmetry axes, and the
origin, are organized solely by transposition-based time-reversal
symmetry (TRS${}^{\dagger}$), giving a threefold symmetry classification.
Numerical level-spacing studies support this threefold
structure~\cite{Hamazaki2020}, and recent analytical work has
characterized the spectral correlations in these
classes~\cite{Kulkarni2025,Akemann2025a,ChenXiaoLiuRyu2026,XiaoChenLiuRyu2026}.
Additional symmetries become important in other spectral regions:
particle-hole symmetry (PHS) and sublattice symmetry (SLS) give rise to hard-edge statistics
near the origin that depend on the symmetry class and topological
sector~\cite{Akemann2002,Osborn2004,GarciaGarcia2022,Xiao2024,Kawabata2026,ChenXiaoLiuRyu2026},
while symmetries involving complex or Hermitian conjugation lead to
distinct statistics on and around the real axis~\cite{Xiao2022}.
For the bulk and the vicinity of the origin, a recently established
duality yields analytical results by relating non-Hermitian level
statistics to known Hermitian spectral correlations~\cite{ChenXiaoLiuRyu2026}.
This connection motivates us to ask whether the same duality also
captures how symmetry controls eigenvector nonorthogonality in these
spectral regions.

In this work, we obtain analytical expressions for eigenvector overlaps
across ten symmetry classes of non-Hermitian random matrices in the limit of large matrix size, extending this duality from eigenvalue to
eigenvector statistics.
Using replica nonlinear $\sigma$ models, we show that the same Hermitian
spectral correlations determine both non-Hermitian level statistics and
eigenvector overlaps.
In the bulk, the normalized off-diagonal overlap has three universal
forms, distinguished solely by TRS${}^{\dagger}$.
Near the origin, the diagonal overlap depends on all symmetries and the topological sectors.
We further determine the mean overlaps at fixed eigenvalues, revealing
how symmetry controls the enhancement of nonorthogonality and spectral
sensitivity as eigenvalues approach one another or the origin.
Numerical comparisons with Gaussian ensembles, non-Hermitian
Sachdev--Ye--Kitaev (SYK) and Anderson models, and a dissipative fermion model
on a bipartite lattice support the universality of the normalized
overlaps in the regimes studied.

The remainder of this work is organized as follows.
Section~\ref{sec:main-results} defines the symmetries and observables and
summarizes the results. We present the replica calculation in
Sec.~\ref{sec:derivations} and compare its predictions with numerical
results for Gaussian ensembles and physical models in
Secs.~\ref{sec:numerics} and~\ref{sec:physical-models}, respectively.
Section~\ref{sec:conclusion} concludes our work.
Appendices~\ref{app:class-a-self}--\ref{app:class-d-hard-edge} contain
the detailed derivations. {Appendix~\ref{app:schur} collects the
group contractions used, Appendix~\ref{app:unfolding-normalization}
describes the unfolding and normalization used in the numerical
comparisons, and Appendix~\ref{app:petermann_factor_density_correlation}
derives the diagonal overlap correlation.}

\section{Summary of Main Results}
\label{sec:main-results}
\label{sec:definitions-review}

\subsection{Symmetry Classification}
\label{subsec:symmetry-classification}

We first briefly review the tenfold AZ symmetry
classification of Hermitian random matrices~\cite{Dyson1962c,Altland1997}.
It is defined by two antiunitary symmetries, time-reversal symmetry
(TRS) and PHS, and one unitary symmetry,
chiral symmetry (CS):
\begin{equation}
\begin{aligned}
 \mathrm{TRS}:&\quad \mathcal T H^*\mathcal T^{-1}=H,
 &\quad&\mathcal T\mathcal T^*=\pm I,\\
 \mathrm{PHS}:&\quad \mathcal C H^*\mathcal C^{-1}=-H,
 &&\mathcal C\mathcal C^*=\pm I,\\
 \mathrm{CS}:&\quad \Gamma H\Gamma^{-1}=-H,
 &&\Gamma^2=I.
\end{aligned}
\label{eq:classification-hermitian}
\end{equation}
Here, $\mathcal T$, $\mathcal C$, and $\Gamma$ are unitary matrices.
The TRS constraint is preserved under a shift of the energy origin: if
$H$ obeys TRS, so does $H-E_0I$ with the same symmetry matrix
$\mathcal T$, for any real $E_0$. The absence of TRS, and its presence
with $\mathcal T\mathcal T^*=+I$ or $-I$, gives the three standard
classes A, AI, and AII, i.e., Dyson's threefold way. PHS and CS, by
contrast, pair $E$ with $-E$ and thereby single out $E=0$. They give the
seven additional classes, known as the nonstandard classes.

For a non-Hermitian matrix, $H^*\ne H^T$, so the Hermitian
antiunitary symmetry constraints admit two distinct extensions,
one based on complex conjugation and the other on transposition.
TRS${}^{\dagger}$ plays a
role analogous to that of TRS in the Hermitian case, and gives three bulk
universality classes~\cite{Hamazaki2020}. It is defined by
\begin{equation}
 \mathcal C_+H^T\mathcal C_+^{-1}=H,
 \quad \mathcal C_+\mathcal C_+^*=\pm I,
 \label{eq:classification-trs-dagger}
\end{equation}
where $\mathcal C_+$ is unitary. If $H$ obeys TRS${}^{\dagger}$, so does
$H-E_0I$ with the same $\mathcal C_+$, for any complex $E_0$.
The other symmetry constraints are generally not preserved under complex shifts of
the energy origin. They relate eigenvalues at distinct symmetry-related
positions and single out lines or points where the local statistics
change, such as the real axis~\cite{ForresterNagao2007,BorodinSinclair2009,KieburgVerbaarschotZafeiropoulos2013,AkemannByunKang2022,Xiao2022}.
The possible combinations of these symmetries give an enlarged
classification with 38 classes~\cite{Bernard2002,Magnea2008,Kawabata2019b,Kawabata2019a}.

This distinction explains why the real and quaternionic Ginibre
ensembles share the class-A bulk statistics away from the real
axis~\cite{Ginibre1965,Grobe1989,Byun2024,AkemannByunKang2022}:
their complex-conjugation symmetry relates eigenvalues in separate
local spectral regions.
For the real Ginibre ensemble, exact correlation functions were obtained~\cite{Sommers2007,ForresterNagao2007}, and their agreement with class A in the bulk and at the edge away from the real axis was proved rigorously~\cite{BorodinSinclair2009}.

In this paper, we focus on random-matrix ensembles with a statistical
rotation symmetry, i.e., $H$ and $e^{i\theta}H$ have the same
probability distribution for every real $\theta$. The compatible
symmetry constraints are TRS${}^{\dagger}$, together with
transposition-based PHS and SLS, defined by
\begin{align}
 \mathrm{PHS}:&\quad
 \mathcal C_-H^T\mathcal C_-^{-1}=-H,
 &\quad&\mathcal C_-\mathcal C_-^*=\pm I,
 \label{eq:classification-phs}\\
 \mathrm{SLS}:&\quad \mathcal SH\mathcal S^{-1}=-H,
 &&\mathcal S^2=I,
 \label{eq:classification-sls}
\end{align}
where $\mathcal C_-$ and $\mathcal S$ are unitary.
These define the ten-fold Altland–Zirnbauer$_0$ (AZ$_0$) symmetry
classes~\cite{Xiao2024}, listed in Table~\ref{tab:symmetry-classification}.

\begin{table*}[!t]
\caption{The ten AZ$_0$ classes. For TRS${}^{\dagger}$ and PHS,
$0$ denotes absence and $\pm1$ the sign in
Eqs.~\eqref{eq:classification-trs-dagger}
and~\eqref{eq:classification-phs}. SLS is absent ($0$) or present ($1$).
Replacing TRS${}^{\dagger}$ and SLS by TRS and CS gives the corresponding
Hermitian class, whose spectral correlations determine the eigenvector
statistics of the non-Hermitian class through the relations derived in
this work. Here $\beta$ labels bulk statistics, and $d$ is the generic
nonzero-eigenvalue multiplicity.}
\label{tab:symmetry-classification}
\centering
\small
\renewcommand{\arraystretch}{1.15}
\setlength{\tabcolsep}{10pt}
\begin{tabular}{@{}ccccccc@{}}
\toprule
Class & TRS${}^{\dagger}$ & PHS & SLS & Hermitian class & $\beta$ & $d$\\
\midrule
$\mathrm A$ & $0$ & $0$ & $0$ & $\mathrm A$ & $2$ & $1$\\
$\mathrm{AI}^{\dagger}$ & $+1$ & $0$ & $0$ & $\mathrm{AI}$ & $1$ & $1$\\
$\mathrm{AII}^{\dagger}$ & $-1$ & $0$ & $0$ & $\mathrm{AII}$ & $4$ & $2$\\
\midrule
$\mathrm{AIII}^{\dagger}$ & $0$ & $0$ & $1$ & $\mathrm{AIII}$ & $2$ & $1$\\
$\mathrm{BDI}_0$ & $+1$ & $+1$ & $1$ & $\mathrm{BDI}$ & $1$ & $1$\\
$\mathrm{CII}_0$ & $-1$ & $-1$ & $1$ & $\mathrm{CII}$ & $4$ & $2$\\
$\mathrm D$ & $0$ & $+1$ & $0$ & $\mathrm D$ & $2$ & $1$\\
$\mathrm C$ & $0$ & $-1$ & $0$ & $\mathrm C$ & $2$ & $1$\\
$\mathrm{CI}_0$ & $+1$ & $-1$ & $1$ & $\mathrm{CI}$ & $1$ & $1$\\
$\mathrm{DIII}_0$ & $-1$ & $+1$ & $1$ & $\mathrm{DIII}$ & $4$ & $2$\\
\bottomrule
\end{tabular}
\end{table*}

TRS${}^{\dagger}$ relates the left and right eigenvectors of the same
eigenvalue. It determines the statistics of eigenvalues away from $z=0$
and from the outer spectral edge, which we refer to as the bulk
statistics. Its absence and its signs $+1$ and $-1$ give classes A,
$\mathrm{AI}^{\dagger}$, and $\mathrm{AII}^{\dagger}$, with $\beta=2$,
$1$, and $4$~\cite{Hamazaki2020,ChenXiaoLiuRyu2026}, in analogy with the
Hermitian classes A, AI, and AII. For $\mathcal C_+\mathcal C_+^*=-I$,
Kramers degeneracy makes each generic eigenspace two-dimensional, so
that $d=2$. Otherwise, $d=1$. PHS and SLS pair $z$ with $-z$. They leave
the bulk statistics unchanged but give distinct statistics near $z=0$
in the other seven classes, known as the hard-edge
statistics~\cite{Xiao2024,ChenXiaoLiuRyu2026}. Hence, the bulk formulas
below depend only on $\beta$, whereas the hard-edge formulas depend on
the full symmetry class and its topological sector.

\subsection{Eigenvector Overlaps and Normalization}
\label{subsec:general-class-definitions}

We now set up the conventions and normalizations used throughout the
paper. For the ensembles we study, $N$ is the effective matrix size, i.e., the matrix size for classes without Kramers
degeneracy and the number of Kramers pairs otherwise. The matrix size is
thus $dN$, with $d=1$ and $d=2$ for classes without and with Kramers
degeneracy, respectively.
Unless stated otherwise, the Gaussian ensembles are normalized to the spectral radius
$\sqrt N$ and bulk density $R_1=1/\pi$.
For eigenvalue pairs, we write
$z=(z_1+z_2)/2$ and $\omega=z_1-z_2$.

Let \(H\) be diagonalizable, with \(N\) distinct eigenvalues
\(\lambda_a\) and spectral projectors \(\Pi_a\). Then,
\begin{equation}
  H=\sum_{a=1}^{N}\lambda_a\Pi_a,
  \quad
  \sum_{a=1}^{N}\Pi_a=I,
  \quad
  \Pi_a\Pi_b=\delta_{ab}\Pi_a.
\end{equation}
Each \(\Pi_a\) projects onto the full eigenspace of \(\lambda_a\). In
particular, a Kramers doublet is described by a single rank-two
projector.
For nondegenerate eigenvalues, $\Pi_a$ is given by left and right eigenvectors $\bra{L_a}$ and $\ket{R_a}$ as $\Pi_a = (\braket{L_a|R_a})^{-1} \ket{R_a}\bra{L_a}$, where $H\ket{R_a} = \lambda_a \ket{R_a}$ and $\bra{L_a} H = \lambda_a \bra{L_a}$.
We define the projector overlaps
\begin{equation}
  O_{ab}:= \frac{1}{d}\tr\,\left(\Pi_a\Pi_b^\dagger\right),
  \label{eq:general-class-rank-normalized-overlap}
\end{equation}
where we assume no degeneracy other than Kramers degeneracy, so that
each eigenspace has dimension $d$. The factor $1/d$ ensures that
$O_{aa} = 1$ if and only if $\Pi_a$ is a Hermitian projector.
For nondegenerate eigenvalues ($d=1$), this gives
\begin{equation}
  \begin{aligned}
    O_{aa}&=\frac{\braket{L_a|L_a}\braket{R_a|R_a}}
                     {|\braket{L_a|R_a}|^2},\\
    O_{ab}&=\frac{\braket{L_a|L_b}\braket{R_b|R_a}}
                     {\braket{L_a|R_a}\braket{R_b|L_b}}.
  \end{aligned}
  \label{eq:overlaps-left-right}
\end{equation}

Following Ref.~\cite{chalker1998eigenvector}, we define the density-weighted
diagonal and off-diagonal overlaps as
\begin{equation}
  \label{eq:cm_o1}
    \mathcal O_1(z)
    :=\frac1N\left\langle
      \sum_{a=1}^{N}O_{aa}\,
      \delta^{(2)}(z-\lambda_a)\right\rangle,
\end{equation}
\begin{equation}
    \mathcal O_2(z_1,z_2)
    \!:=\!\frac1N\!\Biggl\langle\!
      \sum_{\substack{a,b=1\\a\ne b}}^{N}\!O_{ab}
      \delta^{(2)}(z_1\! -\!\lambda_a)
      \delta^{(2)}(z_2\! -\!\lambda_b)\!\Biggr\rangle.
  \label{eq:cm_o2}
\end{equation}
Here, \(a\ne b\) means
that \(a\) and \(b\) label distinct eigenspaces, and
\(\delta^{(2)}(z)\) is the Dirac-$\delta$ function in the complex plane.

The quantities $\mathcal O_1$ and $\mathcal O_2$ are the central objects
of this work. They obey two exact properties that follow directly from
the definitions. First, since $\tr\,\Pi_a=d$, the individual diagonal
overlaps obey
\begin{equation}
  O_{aa} \ge 1.
  \label{eq:projector-overlap-lower-bound}
\end{equation}
Second, the overlaps obey the sum rule
\begin{equation}
  \sum_{b} O_{ab}
  = \frac{1}{d}\tr\,(\Pi_a)
  = 1.
  \label{eq:projector-overlap-sum-rule}
\end{equation}
The lower bound concerns an individual overlap, not its density-weighted
average.

We define the eigenvalue densities, with each Kramers pair counted
once, as
\begin{align}
  R_1(z)
  &:=\left\langle\sum_a
    \delta^{(2)}(z-\lambda_a)\right\rangle,
  \label{eq:conditional-r1-definition}\\
  R_2(z_1,z_2)
  &:=\left\langle\sum_{a,b}
    \delta^{(2)}(z_1-\lambda_a)
    \delta^{(2)}(z_2-\lambda_b)\right\rangle.
  \label{eq:conditional-r2-definition}
\end{align}
The inequality~\eqref{eq:projector-overlap-lower-bound} gives $\mathcal O_1(z)\ge R_1(z)/N$.
Integrating Eq.~\eqref{eq:cm_o2} over the second eigenvalue and using
Eq.~\eqref{eq:projector-overlap-sum-rule} gives
\begin{align}
    \int_{\mathbb C}\!\mathrm d^2z_2\,\mathcal O_2(z_1,z_2)
    &=\frac1N\left\langle
      \sum_a(1-O_{aa})
      \delta^{(2)}(z_1-\lambda_a)\right\rangle \notag\\*
    &=\frac1N R_1(z_1)-\mathcal O_1(z_1).
  \label{eq:definitions-overlap-sum-rule}
\end{align}

\subsection{Bulk Overlaps}
\label{sec:other-symmetry-classes}

In the rest of this section, we summarize our main results.
For the Gaussian ensembles in each symmetry class, the leading bulk
diagonal overlap is
\begin{equation}
  \pi\mathcal O_1(z)
  =\frac{\beta}{2}
    \left(1-\frac{|z|^2}{N}\right).
  \label{eq:general-class-bulk-o1}
\end{equation}
Here, $\beta$ is fixed by TRS${}^{\dagger}$
(Table~\ref{tab:symmetry-classification}), so the amplitude of
$\mathcal O_1$ differs among classes A, $\mathrm{AI}^{\dagger}$, and $\mathrm{AII}^{\dagger}$. The $\beta=1$
and $\beta=2$ results for classes $\mathrm{AI}^{\dagger}$ and A
reproduce the bulk mean overlaps obtained in
Refs.~\cite{Akemann2025b} and~\cite{chalker1998eigenvector}, respectively.
We derive the $\beta=4$ result for class $\mathrm{AII}^{\dagger}$ in
Sec.~\ref{sec:brief-class-a-o1}.
This amplitude difference is not a multiplicity factor from Kramers
degeneracy: the overlaps in
Eq.~\eqref{eq:general-class-rank-normalized-overlap} already include
the normalization $1/d$, and each Kramers pair is counted once.
Like the eigenvalue
density $R_1$, this amplitude is not universal, and in physical models
it must be measured or determined independently. The universal
predictions are the off-diagonal overlap normalized by $\mathcal O_1$
and the hard-edge profile relative to the bulk value.

We next define the Poisson transform of the level correlation of the
Hermitian partner class~\cite{ChenXiaoLiuRyu2026},
\begin{equation}
  \label{eq:def_p2}
  \mathcal{P}_2(|\omega|^2)
  = \frac{1}{\pi} \int_{-\infty}^\infty
  \frac{|\omega|^2/2\pi}{x^2 + (|\omega|^2/2\pi)^2}
  R_2^{\mathrm H}(x) \, \mathrm d x,
\end{equation}
where $R_2^{\mathrm H}$ is the unfolded eigenvalue pair-correlation
function of the Hermitian partner class.
The non-Hermitian eigenvalue pair-correlation function $R_2$ follows from
$\mathcal{P}_2$ as~\cite{ChenXiaoLiuRyu2026}
\begin{equation}
  \label{eq:r_2_P}
  \pi^2 R_2(\omega) = \frac{1}{2}
  \frac{\mathrm d^2}{\mathrm d(|\omega|^2)^2} [|\omega|^4 \mathcal{P}_2(|\omega|^2)].
\end{equation}
Here, $R_2$ includes the self-correlation $\delta$ function, which is
discussed in Appendix~\ref{subsubsec:recover_self_overlap}.
\par
Our first main result is that $\mathcal{O}_2$ is also determined by
$\mathcal{P}_2$. For $\omega\ne0$ with $|\omega| \ll \sqrt{N}$, we find
\begin{equation}
  \frac{2\pi\mathcal O_2(z_1,z_2)}
       {\mathcal O_1(z)}
  =-1+\frac{\mathrm d}{\mathrm d|\omega|^2}
    \left[|\omega|^2\mathcal P_2(|\omega|^2)\right].
  \label{eq:general-class-bulk-o2-ratio}
\end{equation}
We derive this relation for class A in Sec.~\ref{sec:brief-class-a-o2}
and extend it to classes $\mathrm{AI}^{\dagger}$ and
$\mathrm{AII}^{\dagger}$.

\begin{table*}[!t]
\caption{Bulk diagonal overlaps and universal normalized off-diagonal
overlaps, $\omega=z_1-z_2\ne0$. Diagonal amplitudes use Gaussian
ensembles with spectral radius $\sqrt N$.}
\label{tab:replica-bulk-summary}
\vspace{2pt}
\centering
\footnotesize
\renewcommand{\arraystretch}{1.30}
\setlength{\tabcolsep}{2.5pt}
\begin{tabular}{@{}clll@{}}
\toprule
\bfseries Class
      & \multicolumn{1}{c}{
          {\bfseries Diagonal overlap}
          }
      & \multicolumn{1}{c@{}}{
          {\bfseries Off-diagonal overlap}} \\
            & \multicolumn{1}{c}{\(\displaystyle\pi\mathcal O_1(z)\)} & \multicolumn{1}{c}{\(\displaystyle
            \pi\mathcal O_2(z_1,z_2)/\mathcal O_1(z)\)} \\
      \midrule
    \(\mathrm A\)
      & \(\displaystyle 1-\frac{|z|^2}{N}\)
      & \(\displaystyle -\frac{1-(1+|\omega|^2)e^{-|\omega|^2}}{|\omega|^4}\) \\[2pt]
    \(\mathrm{AI}^{\dagger}\)
      & \(\displaystyle \frac12\left(1-\frac{|z|^2}{N}\right)\)
      & \(\displaystyle -2\frac{\mathrm d}{\mathrm d|\omega|^2}\!\left[
          \frac{\sinh^2(|\omega|^2/2)}{|\omega|^2}\frac{\mathrm d}{\mathrm d|\omega|^2}\!\left(
            \frac{|\omega|^2E_1(|\omega|^2/2)}{\sinh(|\omega|^2/2)}
          \right)\right]\) \\[2pt]
    \(\mathrm{AII}^{\dagger}\)
      & \(\displaystyle 2\left(1-\frac{|z|^2}{N}\right)\)
      & \(\displaystyle \frac12\frac{\mathrm d}{\mathrm d|\omega|^2}\!\left[
          \frac{e^{-2|\omega|^2}}{|\omega|^2}\frac{\mathrm d}{\mathrm d|\omega|^2}\!\left(
            |\omega|^2e^{|\omega|^2}\operatorname{Shi}(|\omega|^2)
          \right)\right]\) \\[2pt]
\bottomrule
\end{tabular}
\end{table*}

\begin{table*}[tb]
\caption{Hard-edge diagonal overlaps $\pi\mathcal O_1(z)$ for $|z|>0$, from
Eq.~\eqref{eq:general-class-hard-edge-o1} and the Poisson transforms in
Ref.~\cite{ChenXiaoLiuRyu2026}; see Sec.~\ref{sec:class-d-hard-edge-resolvent}
for the even-dimensional class-D derivation.
Here $I_m$ and $K_m$ are modified Bessel functions,
$\mathcal K_m(s)=\int_s^\infty K_m(v)\,\mathrm dv$ and
$\mathcal I_m(s)=\int_0^s I_m(v)\,\mathrm dv$.
}
\label{tab:replica-results-summary}
\vspace{2pt}
\centering
\footnotesize
\renewcommand{\arraystretch}{1.30}
\setlength{\tabcolsep}{2.5pt}
\begin{tabular}{@{}ccll@{}}
\toprule
\bfseries Class & \bfseries Sector & \multicolumn{2}{c@{}}{\bfseries Hard-edge\quad $\pi\mathcal O_1(z)$}\\
\midrule
    \(\mathrm{AIII}^{\dagger}\) & \(\nu\in\mathbb Z\)
      & \multicolumn{2}{l@{}}{
        \(\displaystyle \frac12\{1+|z|^2[I_{|\nu|}(|z|^2)K_{|\nu|}(|z|^2)
        +I_{|\nu|+1}(|z|^2)K_{|\nu|-1}(|z|^2)]\}+\frac{|\nu|}{2|z|^2}\)} \\[2pt]
    \(\mathrm{BDI}_0\) & \(\nu\in\mathbb Z\)
      & \multicolumn{2}{l@{}}{
        \(\displaystyle \frac14\{1+|z|^2[I_{|\nu|}(|z|^2)K_{|\nu|}(|z|^2)
        +I_{|\nu|+1}(|z|^2)K_{|\nu|-1}(|z|^2)]
        +I_{|\nu|}(|z|^2)\mathcal K_{|\nu|}(|z|^2)\}+\frac{|\nu|}{4|z|^2}\)} \\[2pt]
    \(\mathrm{CII}_0\) & \(\nu\in\mathbb Z\)
      & \multicolumn{2}{l@{}}{
        \(\displaystyle 1+2|z|^2[I_{2|\nu|}(2|z|^2)K_{2|\nu|}(2|z|^2)
        +I_{2|\nu|+1}(2|z|^2)K_{2|\nu|-1}(2|z|^2)]
        -K_{2|\nu|}(2|z|^2)\mathcal I_{2|\nu|}(2|z|^2)+\frac{|\nu|}{|z|^2}\)} \\[2pt]
    \(\mathrm D\) & \(\nu\in\mathbb Z_2\)
      & \multicolumn{2}{l@{}}{
        \(\displaystyle 1+(-1)^\nu\frac{1-e^{-2|z|^2}}{4|z|^2}
        +\frac{\nu}{2|z|^2}\)} \\[2pt]
    \(\mathrm C\) & ---
      & \multicolumn{2}{l@{}}{
        \(\displaystyle 1-\frac{1-e^{-2|z|^2}}{4|z|^2}\)} \\[2pt]
    \(\mathrm{CI}_0\) & ---
      & \multicolumn{2}{l@{}}{
        \(\displaystyle \frac14\left\{1+\frac {|z|^2}2[(I_0(|z|^2)+I_2(|z|^2))K_0(|z|^2)
        +2I_1(|z|^2)K_1(|z|^2)]\right\}\)} \\[2pt]
    \(\mathrm{DIII}_0\) & \(\nu\in\mathbb Z_2\)
      & \multicolumn{2}{l@{}}{
        \(\displaystyle 1+|z|^2[(I_0(2|z|^2)+I_2(2|z|^2))K_0(2|z|^2)
        +2I_1(2|z|^2)K_1(2|z|^2)]
        +(-1)^\nu\left[\frac1{2|z|^2}-K_1(2|z|^2)\right]+\frac{\nu}{|z|^2}\)} \\
\bottomrule
\end{tabular}
\end{table*}
\par
The unfolded pair-correlation functions of the Hermitian classes
A~\cite{Mehta1960}, AI, and AII~\cite{Dyson1962b} are well known. With
$S(x)=\sin(\pi x)/(\pi x)$ and the self-correlation included, they read
\begingroup
\medmuskip=2mu
\thickmuskip=4mu
\begin{gather}
 R_2^{\mathrm{A},\mathrm H}(x)
 =\delta(x)+1-S(x)^2,
 \label{eq:summary-hermitian-a-r2}\\
 R_2^{\mathrm{AI},\mathrm H}(x)
 =\delta(x)+1-S(x)^2-S'(|x|)\!\int_{|x|}^{\infty}\!S(v)\,\mathrm dv,
 \label{eq:summary-hermitian-ai-r2}\\
 R_2^{\mathrm{AII},\mathrm H}(x)
 =\delta(x)+1-S(2x)^2+S'(2x)\!\int_0^{2x}\!S(v)\,\mathrm dv.
 \label{eq:summary-hermitian-aii-r2}
\end{gather}
\endgroup
In class AII, each Kramers pair is counted once. Substituting these into
Eq.~\eqref{eq:def_p2} gives~\cite{ChenXiaoLiuRyu2026}
\begin{align}
 \mathcal P_2^{\mathrm A}(|\omega|^2)
 &=1+\frac{2(1-e^{-|\omega|^2})}{|\omega|^4},
 \label{eq:summary-a-p2}\\
 \mathcal P_2^{\mathrm{AI}^{\dagger}}(|\omega|^2)
 &=1-\frac{4\sinh^2(|\omega|^2/2)}{|\omega|^4}\notag\\
 &\quad\times\frac{\mathrm d}{\mathrm d|\omega|^2}
 \left[\frac{|\omega|^2E_1(|\omega|^2/2)}{\sinh(|\omega|^2/2)}\right],
 \label{eq:summary-ai-p2}\\
 \mathcal P_2^{\mathrm{AII}^{\dagger}}(|\omega|^2)
 &=1+\frac{e^{-2|\omega|^2}}{|\omega|^4}
 \frac{\mathrm d}{\mathrm d|\omega|^2}
 \left[|\omega|^2e^{|\omega|^2}\operatorname{Shi}(|\omega|^2)\right],
 \label{eq:summary-aii-p2}
\end{align}
where
\begin{equation}
 E_1(s)=\int_s^\infty\frac{e^{-v}}v\,\mathrm dv,
 \quad \operatorname{Shi}(s)=\int_0^s\frac{\sinh v}v\,\mathrm dv.
\end{equation}
Applying Eq.~\eqref{eq:general-class-bulk-o2-ratio} to these three
transforms yields the normalized off-diagonal overlaps in
Table~\ref{tab:replica-bulk-summary}. In particular, for class A it reproduces
the result of Ref.~\cite{chalker1998eigenvector},
\begin{equation}
 \label{eq:summary-class-a-off-diagonal-overlap}
 \frac{\pi\mathcal O_2}{\mathcal O_1}
 =-\frac{1-(1+|\omega|^2)e^{-|\omega|^2}}{|\omega|^4}.
\end{equation}
At fixed bulk midpoint and $1\ll |\omega|^2\ll N$, all three families
have
\begin{equation}
 \label{eq:summary-bulk-overlap-tail}
 \mathcal O_2\sim-\frac{2\mathcal O_1(z)}{\beta\pi |\omega|^4}.
\end{equation}
The
$|\omega|^{-4}$ decay is thus common to the three families, with a
$\beta$-dependent coefficient. The three families are further
distinguished by the behavior of
$\mathcal{O}_2(z_1,z_2)/\mathcal{O}_1(z)$ at small $|\omega|$ and by its
subleading corrections at large $|\omega|$, both discussed in
Sec.~\ref{subsec:conditional-overlaps}.

The bulk relations also give a direct connection between the
non-Hermitian eigenvalue statistics and the off-diagonal overlap.
Combining Eqs.~\eqref{eq:r_2_P} and~\eqref{eq:general-class-bulk-o2-ratio},
we obtain, for $|\omega|>0$ at fixed bulk midpoint,
\begin{equation}
  \label{eq:bulk-overlap-from-eigenvalue-correlation}
  \frac{\mathcal O_2(z_1,z_2)}{\mathcal O_1(z)}
  =\frac{2}{\pi|\omega|^4}
    \int_0^{|\omega|}r^3[\pi^2R_2(r)-1]\,\mathrm dr,
\end{equation}
where $R_2(r)$ is the eigenvalue pair-correlation function $R_2(\omega)$ evaluated at $|\omega|=r$, in the normalization $\pi R_1=1$.

\subsection{Hard-Edge Overlaps}
\label{subsec:hard-edge-results}

For the seven hard-edge classes in
Table~\ref{tab:symmetry-classification}, i.e., classes
$\mathrm{AIII}^{\dagger}$, $\mathrm{BDI}_0$, $\mathrm{CII}_0$,
$\mathrm D$, $\mathrm C$, $\mathrm{CI}_0$, and $\mathrm{DIII}_0$,
the $\mathrm U(1)$ invariance of the ensembles implies that the
hard-edge statistics depend only on $|z|^2$. We define
the Poisson transform~\cite{ChenXiaoLiuRyu2026}
\begin{equation}
  \label{eq:def_p1}
  \mathcal{P}_1(|z|^2)
  = \frac{1}{\pi} \int_{-\infty}^\infty
  \frac{|z|^2/\pi}{x^2 + (|z|^2/\pi)^2}
  R_1^{\mathrm H}(x) \, \mathrm d x,
\end{equation}
where $R_1^{\mathrm H}$ is the hard-edge spectral density of the
Hermitian partner class, including the $\delta$ functions of protected
zero modes when present.
{The number of protected zero modes is set by the topological index
$\nu$ of the ensemble. For the chiral classes $\mathrm{AIII}^{\dagger}$,
$\mathrm{BDI}_0$, and $\mathrm{CII}_0$, $\nu\in\mathbb Z$ is the
difference between the dimensions of the $\mathcal S=+1$ and
$\mathcal S=-1$ eigenspaces, i.e., $\nu=\tr\,\mathcal S$ (divided by
two for $\mathrm{CII}_0$, where both eigenspaces are Kramers
degenerate), and $|\nu|$ zero modes are protected. For classes $\mathrm D$ and $\mathrm{DIII}_0$,
$\nu\in\mathbb Z_2$ is the parity of the matrix dimension (of half the
matrix dimension for $\mathrm{DIII}_0$), and a single zero mode (a single
Kramers pair for $\mathrm{DIII}_0$) is protected when $\nu=1$. Classes
$\mathrm C$ and $\mathrm{CI}_0$ have no protected zero modes. The
hard-edge results below depend only on $|\nu|$.}
The transform $\mathcal P_1$ determines the hard-edge eigenvalue density
{at nonzero eigenvalues, $|z|>0$,}
through~\cite{ChenXiaoLiuRyu2026}
\begin{equation}
  \label{eq:r1_by_p1}
 \pi R_1(z)=\frac{\mathrm d}{\mathrm d|z|^2}[|z|^2\mathcal P_1(|z|^2)]{,\qquad |z|>0}.
\end{equation}
\par
Our second main result is that $\mathcal{P}_1$ also determines
$\mathcal{O}_1$ {at nonzero eigenvalues},
\begin{equation}
  \pi\mathcal O_1(z)
  =
  \frac{\beta}{4}
    \left[1+\mathcal P_1(|z|^2)\right]{,\qquad |z|>0}.
  \label{eq:general-class-hard-edge-o1}
\end{equation}
For even-dimensional class D, the explicit result is
\begin{equation}
 \pi\mathcal O_1(z)=\frac12[1+\mathcal P_1(|z|^2)]
 =1+\frac{1-e^{-2|z|^2}}{4|z|^2}.
 \label{eq:draft-d-hard-edge-result}
\end{equation}
Thus, $\pi\mathcal O_1$ approaches $3/2$ at the origin and tends to
its central bulk value $1$ at large $|z|$.
The derivation for class D is given in
Sec.~\ref{sec:class-d-hard-edge-resolvent} and can be generalized to the
other symmetry classes. We test these results numerically in
Sec.~\ref{sec:numerics}.

Protected zero modes contribute a $\delta$ function to
$R_1^{\mathrm{H}}$, which generates a $1/|z|^2$ term in $\mathcal P_1$. This
term does not affect $R_1$ [Eq.~\eqref{eq:r1_by_p1}], since
$\mathrm d(|z|^2\cdot |z|^{-2})/\mathrm d|z|^2=0$. It does, however, enter the
overlap density at nonzero eigenvalues through
Eq.~\eqref{eq:general-class-hard-edge-o1}.

\par
Substituting the hard-edge densities of the Hermitian AZ
classes~\cite{Altland1997,Verbaarschot1993,Verbaarschot1994,Sener1998,Ivanov2002,Nishigaki2003}
into Eq.~\eqref{eq:def_p1} gives $\mathcal{P}_1$, and
Eq.~\eqref{eq:general-class-hard-edge-o1} then gives $\mathcal{O}_1$.
The results for the seven classes are collected in
Table~\ref{tab:replica-results-summary}.

\subsection{Overlaps at Fixed Eigenvalues and Their Asymptotics}
\label{subsec:conditional-overlaps}

The preceding subsections describe the density-weighted overlaps
$\mathcal O_1$ and $\mathcal O_2$, defined in
Eqs.~\eqref{eq:cm_o1} and~\eqref{eq:cm_o2}. They combine the individual
overlaps $O_{aa}$ and $O_{ab}$ in
Eq.~\eqref{eq:general-class-rank-normalized-overlap} with the eigenvalue
densities $R_1(z)$ and $R_2(z_1,z_2)$ in
Eqs.~\eqref{eq:conditional-r1-definition} and~\eqref{eq:conditional-r2-definition}.
To isolate eigenvector nonorthogonality from the eigenvalue densities,
we consider the ensemble-averaged overlaps at fixed eigenvalues
\begin{equation}
  \langle O_{aa}\rangle_z
  :=\left\langle O_{aa}\mid\lambda_a=z\right\rangle
  =\frac{N\mathcal O_1(z)}{R_1(z)},
  \label{eq:conditional-o1-ratio}
\end{equation}
\begin{equation}
  \begin{aligned}
    \langle O_{ab}\rangle_{z_1,z_2}
    &:=\left\langle O_{ab}\mid
      \lambda_a=z_1,\lambda_b=z_2\right\rangle\\
    &=\frac{N\mathcal O_2(z_1,z_2)}{R_2(z_1,z_2)},
    \qquad z_1\ne z_2.
  \end{aligned}
  \label{eq:conditional-o2-ratio}
\end{equation}
Here, $N$ cancels the factor $1/N$ in Eqs.~\eqref{eq:cm_o1}
and~\eqref{eq:cm_o2}. The average diagonal overlap satisfies
$\langle O_{aa}\rangle_z\ge1$.
The individual overlaps quantify eigenvector nonorthogonality: $O_{aa}$
is the diagonal overlap (the Petermann factor for a nondegenerate
eigenvalue), while $O_{ab}$ is the off-diagonal overlap.
Divergences of these average overlaps at fixed eigenvalues describe
an enhancement of eigenvector nonorthogonality.

In the bulk, this distinction is apparent as two eigenvalues approach
each other. For class A, using $\pi^2R_2(z_1,z_2)=1-e^{-|\omega|^2}$ for $|\omega|>0$ and
the class-A overlap in Table~\ref{tab:replica-bulk-summary} gives
\begin{align}
    \langle O_{ab}\rangle_{z_1,z_2}
    &=-N\pi\mathcal O_1(z)
      \frac{1-(1+|\omega|^2)e^{-|\omega|^2}}{|\omega|^4(1-e^{-|\omega|^2})} \notag \\
    &\sim-\frac{N\pi\mathcal O_1(z)}{2|\omega|^2},
      \quad |\omega|\to0.
  \label{eq:conditional-class-a-contact}
\end{align}
Although $\mathcal O_2\to-\mathcal O_1(z)/(2\pi)$ is finite,
the average off-diagonal overlap at fixed eigenvalues diverges as
$-|\omega|^{-2}$ because
$R_2\sim |\omega|^2/\pi^2$. This is consistent with two-level perturbation
theory, in which the spectral projectors near a generic non-normal
coalescence scale as $(\lambda_a-\lambda_b)^{-1}$.
For classes $\mathrm A$, $\mathrm{AI}^{\dagger}$, and
$\mathrm{AII}^{\dagger}$, the corresponding leading behaviors at fixed
midpoint are, with positive numerical prefactors suppressed,
\begin{align}
    &\langle O_{ab}\rangle_{z_1,z_2} \notag \\
    &\quad\propto-N\pi\mathcal O_1(z)
    \begin{cases}
      |\omega|^{-2}, &\text{(Class }\mathrm A\text{)},\\[5pt]
      [|\omega|^2\log(1/|\omega|)]^{-1},
        &\text{(Class }\mathrm{AI}^{\dagger}\text{)},\\[5pt]
      |\omega|^{-2}, &\text{(Class }\mathrm{AII}^{\dagger}\text{)}.
    \end{cases}
  \label{eq:conditional-three-bulk-contact-limits}
\end{align}
Table~\ref{tab:conditional-bulk-asymptotics} collects these divergences
together with the eigenvalue densities and density-weighted overlaps.
The logarithmic factor in class $\mathrm{AI}^{\dagger}$ comes from
$R_2\sim|\omega|^2\log(1/|\omega|)$.

\begin{table*}[t]

  \caption{Leading bulk scalings as \(|\omega|\to0\), omitting
  prefactors but retaining signs; \(-1\) denotes a finite negative limit.
  The third column gives density-weighted overlaps; the last column
  gives average overlaps at fixed eigenvalues.}
  \label{tab:conditional-bulk-asymptotics}
  \vspace{2pt}
  \centering
  \footnotesize
  \renewcommand{\arraystretch}{1.30}
  \setlength{\tabcolsep}{2.5pt}
  \begin{tabular}{@{}
    c
    c
    c
    c@{}}
    \toprule
    \bfseries Class
      & \bfseries \(R_2(z_1,z_2)\)
      & \bfseries \(\mathcal O_2/[\pi\mathcal O_1(z)]\)
      & \bfseries \(\langle O_{ab}\rangle_{z_1,z_2}/[N\pi\mathcal O_1(z)]\) \\
    \midrule
    \(\mathrm A\)
      & \(|\omega|^2\)
      & \(-1\)
      & \(-|\omega|^{-2}\) \\[2pt]
    \(\mathrm{AI}^{\dagger}\)
      & \(|\omega|^2\log(1/|\omega|)\)
      & \(-1\)
      & \(-[|\omega|^2\log(1/|\omega|)]^{-1}\) \\[2pt]
    \(\mathrm{AII}^{\dagger}\)
      & \(|\omega|^2\)
      & \(-1\)
      & \(-|\omega|^{-2}\) \\
    \bottomrule
  \end{tabular}
\end{table*}

Near the origin, the same separation of overlap and density gives
\begin{equation}
  \frac{1}{N}
  \langle O_{aa}\rangle_z
  =\frac{\pi\mathcal O_1(z)}
         {\pi R_1(z)}
  =
  \frac{\beta
    \left[1+\mathcal P_1(|z|^2)\right]}
       {4\pi R_1(z)}.
  \label{eq:conditional-hard-edge-general}
\end{equation}
For example, the class-C and topologically trivial class-D results are
\begin{equation}
    \frac{1}{N}
      \langle O_{aa}\rangle_{z,\mathrm C}
    =\frac{1-(1-e^{-2|z|^2})/(4|z|^2)}{1-e^{-2|z|^2}},
  \label{eq:conditional-class-c-hard-edge}
\end{equation}
\begin{equation}
    \frac{1}{N}
      \langle O_{aa}\rangle_{z,\mathrm D,\nu=0}
    =\frac{1+(1-e^{-2|z|^2})/(4|z|^2)}{1+e^{-2|z|^2}}.
  \label{eq:conditional-class-d-hard-edge}
\end{equation}
Thus, $\langle O_{aa}\rangle_z\sim N/(4|z|^2)$ in class C, whereas it
tends to $3N/4$ in class D at $\nu=0$. For $\nu=1$ in class D, the
zero-mode-induced $1/|z|^2$ term in $\mathcal O_1$ combines with $R_1\sim |z|^2$
to give $\langle O_{aa}\rangle_z\propto N/|z|^4$.
More generally, Table~\ref{tab:conditional-hard-edge-asymptotics} shows
that the $\nu=0$ average diagonal overlaps at fixed eigenvalues scale as $N/[|z|^2\log(1/|z|^2)]$ in
classes $\mathrm{AIII}^{\dagger}$, $\mathrm{CI}_0$, and
$\mathrm{DIII}_0$, as $N/|z|^2$ in class $\mathrm{CII}_0$, and remain
finite in class $\mathrm{BDI}_0$.
The nonzero sectors have $N/|z|^4$ scaling, except for
$|\nu|=1$ in class $\mathrm{BDI}_0$, where the behavior is
$N/[|z|^4\log(1/|z|^2)]$. Numerical prefactors in these scalings are suppressed.

\begin{table*}[t]

  \caption{Leading hard-edge scalings as \(|z|\to0\), omitting prefactors.
  For \(|z|>0\), \(R_1(z)\) is the eigenvalue density;
  \(\mathcal O_1(z)\) includes the protected-zero-mode contribution
  proportional to \(1/|z|^2\).}
  \label{tab:conditional-hard-edge-asymptotics}
  \vspace{2pt}
  \centering
  \footnotesize
  \renewcommand{\arraystretch}{1.30}
  \setlength{\tabcolsep}{2.5pt}
  \begin{tabular}{@{}
    c
    c
    c
    c
    c
    c@{}}
    \toprule
    \bfseries Class
      & \bfseries Sector set
      & \bfseries Sector
      & \bfseries \(R_1(z)\)
      & \bfseries \(\mathcal O_1(z)\)
      & \bfseries
        \(N^{-1}\langle O_{aa}\rangle_z\) \\
    \midrule
    \multirow{2}{*}{\(\mathrm{AIII}^{\dagger}\)}
      & \multirow{2}{*}{\(\mathbb Z\)} & \(\nu=0\)
      & \(|z|^2\log(1/|z|^2)\) & \(1\)
      & \([|z|^2\log(1/|z|^2)]^{-1}\) \\
      & & \(|\nu|\ge1\)
      & \(|z|^2\) & \(|z|^{-2}\) & \(|z|^{-4}\) \\
    \cmidrule(lr){1-6}
    \multirow{3}{*}{\(\mathrm{BDI}_0\)}
      & \multirow{3}{*}{\(\mathbb Z\)} & \(\nu=0\)
      & \(1\) & \(1\) & \(1\) \\
      & & \(|\nu|=1\)
      & \(|z|^2\log(1/|z|^2)\) & \(|z|^{-2}\)
      & \([|z|^4\log(1/|z|^2)]^{-1}\) \\
      & & \(|\nu|\ge2\)
      & \(|z|^2\) & \(|z|^{-2}\) & \(|z|^{-4}\) \\
    \cmidrule(lr){1-6}
    \multirow{2}{*}{\(\mathrm{CII}_0\)}
      & \multirow{2}{*}{\(\mathbb Z\)} & \(\nu=0\)
      & \(|z|^2\) & \(1\) & \(|z|^{-2}\) \\
      & & \(|\nu|\ge1\)
      & \(|z|^2\) & \(|z|^{-2}\) & \(|z|^{-4}\) \\
    \cmidrule(lr){1-6}
    \multirow{2}{*}{\(\mathrm D\)}
      & \multirow{2}{*}{\(\mathbb Z_2\)} & \(\nu=0\)
      & \(1\) & \(1\) & \(1\) \\
      & & \(\nu=1\)
      & \(|z|^2\) & \(|z|^{-2}\) & \(|z|^{-4}\) \\
    \cmidrule(lr){1-6}
    \(\mathrm C\) & --- & ---
      & \(|z|^2\) & \(1\) & \(|z|^{-2}\) \\
    \cmidrule(lr){1-6}
    \(\mathrm{CI}_0\) & --- & ---
      & \(|z|^2\log(1/|z|^2)\) & \(1\) & \([|z|^2\log(1/|z|^2)]^{-1}\) \\
    \cmidrule(lr){1-6}
    \multirow{2}{*}{\(\mathrm{DIII}_0\)}
      & \multirow{2}{*}{\(\mathbb Z_2\)} & \(\nu=0\)
      & \(|z|^2\log(1/|z|^2)\) & \(1\) & \([|z|^2\log(1/|z|^2)]^{-1}\) \\
      & & \(\nu=1\)
      & \(|z|^2\) & \(|z|^{-2}\) & \(|z|^{-4}\) \\
    \bottomrule
  \end{tabular}
\end{table*}

At large separations, the off-diagonal overlap instead decays.
At fixed bulk midpoint in the mesoscopic window $1\ll |\omega|^2\ll N$,
expanding Table~\ref{tab:replica-bulk-summary} gives
\begin{align}
    &\frac{\pi\mathcal O_2(z_1,z_2)}{\mathcal O_1(z)}
    =-\frac{1}{|\omega|^4}\times \notag \\
    &\begin{cases}
      1-(1+|\omega|^2)e^{-|\omega|^2},
        &\text{(Class }\mathrm A\text{)},\\[5pt]
      2-\dfrac{8}{|\omega|^2}+\dfrac{36}{|\omega|^4}+\bigO(|\omega|^{-6}),
        &\text{(Class }\mathrm{AI}^{\dagger}\text{)},\\[5pt]
      \dfrac12+\dfrac{1}{|\omega|^2}+\dfrac{9}{4|\omega|^4}+\bigO(|\omega|^{-6}),
        &\text{(Class }\mathrm{AII}^{\dagger}\text{)}.
    \end{cases}
  \label{eq:large-separation-three-class-expansion}
\end{align}
Thus, classes $\mathrm A$, $\mathrm{AI}^{\dagger}$, and
$\mathrm{AII}^{\dagger}$ share the leading tail
\begin{equation}
  \mathcal O_2(z_1,z_2)
  \sim-\frac{2\mathcal O_1(z)}{\beta\pi|\omega|^4}.
  \label{eq:large-separation-universal-o2}
\end{equation}
Using the Gaussian amplitude in Eq.~\eqref{eq:general-class-bulk-o1}
gives $\mathcal O_2\sim-(1-|z|^2/N)/(\pi^2|\omega|^4)$ in all three classes.
For class A, this leading tail agrees with the result
in Ref.~\cite{chalker1998eigenvector} for $1\ll|\omega|^2\ll N$ at fixed
bulk midpoint,
\begin{equation}
  \mathcal O_{2}^{\mathrm{A}}(z_1,z_2)
  =-\frac{1-z_1\overline z_2/N}
          {\pi^2|\omega|^4}.
  \label{eq:large-separation-chalker-mehlig}
\end{equation}
Its full complex dependence on $z_1$ and $z_2$ lies outside our
fixed-midpoint scaling regime.

\section{Replica Method}
\label{sec:derivations}

We first construct the $n$-replica partition function $Z_{1,n}$ for the diagonal overlap $\mathcal O_1$ and calculate its bulk form,
then evaluate a similar partition function $Z_{2,n}$ for the off-diagonal overlap $\mathcal O_2$.
At the large-$N$ saddle
point, the source derivatives can be expressed in terms of the angular
integrals that determine the eigenvalue statistics.
This relates the
overlaps to the Poisson transforms $\mathcal P_1$ and $\mathcal P_2$.
Finally, we apply the one-point construction near the spectral origin,
i.e., to the hard-edge diagonal overlap.
\subsection{Bulk Diagonal Overlap}
\label{sec:brief-class-a-o1}
\label{subsec:brief-class-a-o1-summary}

We consider the class-A Gaussian ensemble, i.e., the complex Ginibre
ensemble, with $H$ an unconstrained $N\times N$ complex matrix and
$P(H)\propto e^{-\tr(H^\dagger H)}$. Its large-$N$ spectrum occupies
a disk of radius $\sqrt N$. The bulk diagonal overlap was obtained by
Schur decomposition in Ref.~\cite{chalker1998eigenvector}; here we
derive it using the replica nonlinear $\sigma$
model~\cite{Nishigaki2002,Kanzieper2005}.
We introduce the one-point partition function
\begin{equation}
 \begin{aligned}
 Z_{1,n}^{\mathrm A}(z,\eta) &:= \left\langle \det\left[\begin{pmatrix}
     z - H & \eta I \\ -\eta I & \overline{z} - H^\dagger
 \end{pmatrix}\right]^n \right\rangle\\
 &=
 \left\langle
 \det[(z-H)(\overline{z}-H^\dagger)+\eta^2 I]^n\right\rangle,
 \end{aligned}
 \label{eq:a-brief-partition}
\end{equation}
where $\eta$ is a real source parameter.
We denote the resolvent by $\mathcal R(z)=(z-H)^{-1}$.
Because the determinant is even in $\eta$, its first source derivative
vanishes at $\eta=0$. The second derivative gives
\begin{align}
 &\left.\frac12\partial_\eta^2
 \det[(z-H)(\bar z-H^\dagger)+\eta^2 I]^n
 \right|_{\eta=0}\notag\\*
 &\quad=n|\det(z-H)|^{2n}\tr[\mathcal R(z)\mathcal R(z)^\dagger].
 \label{eq:main-contact-source-derivative}
\end{align}
Near a nondegenerate eigenvalue $\lambda_a$, the resolvent has the pole
$\mathcal R(z)=\Pi_a/(z-\lambda_a)+\bigO(1)$.
Since $\tr(\Pi_a\Pi_a^\dagger)=O_{aa}$, the trace in
Eq.~\eqref{eq:main-contact-source-derivative} behaves as
\begin{equation}
 \tr[\mathcal R(z)\mathcal R(z)^\dagger]
 =\frac{O_{aa}}{|z-\lambda_a|^2}
   +\bigO(|z-\lambda_a|^{-1}).
 \label{eq:main-trace-overlap-residue}
\end{equation}
Thus, the most singular local contribution to Eq.~\eqref{eq:main-contact-source-derivative} is
\begin{equation}
 nO_{aa}|z-\lambda_a|^{2n-2}
 \prod_{b\ne a}|z-\lambda_b|^{2n}.
 \label{eq:main-contact-local-weight}
\end{equation}
The product over $b\ne a$ tends to unity as $n\to0$.
Writing $n|z-\lambda_a|^{2n-2}$ as
$n^{-1}\partial_z\partial_{\bar z}|z-\lambda_a|^{2n}$ and using
$\partial_z\partial_{\bar z}\log|z-\lambda_a|^2
=\pi\delta^{(2)}(z-\lambda_a)$ gives the distributional limit
\begin{equation}
 \lim_{n\to0}n|z-\lambda_a|^{2n-2}
 =\pi\delta^{(2)}(z-\lambda_a).
 \label{eq:main-contact-distributional-limit}
\end{equation}
The less singular terms vanish in this limit. Summing the contributions
from all eigenvalues and taking the ensemble average therefore yields
\begin{equation}
 \begin{aligned}
 &\frac1{2N}\lim_{n\to0}
 \left.\partial_\eta^2 Z_{1,n}^{\mathrm A}(z,\eta)\right|_{\eta=0}\\
 &\quad=\frac{\pi}{N}\left\langle\sum_a O_{aa}\delta^{(2)}(z-\lambda_a)\right\rangle
 =\pi\mathcal O_1(z).
 \end{aligned}
 \label{eq:a-brief-replica}
\end{equation}
This derivation uses only the pole structure of the resolvent. The same
identity therefore holds for any ensemble with nondegenerate
eigenvalues, with $Z_{1,n}$ defined by the corresponding average.

A Hubbard--Stratonovich transformation of the Gaussian average
introduces an $n\times n$ complex matrix field $Q$. At $\eta=0$,
its bulk saddle is $Q=qU$, with $U\in\mathrm U(n)$ and
$q^2=N-|z|^2$. Restricting the integral to this saddle-point manifold
gives the nonlinear $\sigma$ model
(see Appendix~\ref{subsec:nlsm_and_replica_limit}),
\begin{equation}
  Z_{1,n}^{\mathrm A}(z,\eta)\simeq e^{-nq^2-n\eta^2}\int_{\mathrm U(n)}dU\,e^{q\eta\tr(U+U^\dagger)},
  \label{eq:main-self-bulk-source-action}
\end{equation}
where the Haar measure is normalized to unit volume.

The nonlinear $\sigma$ model in
Eq.~\eqref{eq:main-self-bulk-source-action} can be understood through
Hermitization~\cite{Feinberg1997,Kawabata2023SVD,Chen2025}.
We introduce the Hermitized matrix
\begin{equation}
 \mathcal H_z=\begin{pmatrix}
  0&z-H\\ \bar z-H^\dagger&0
 \end{pmatrix},\qquad \{\mathcal H_z,\tau_3\}=0,
 \label{eq:bulk-hermitized-hamiltonian}
\end{equation}
where $\tau_3=\operatorname{diag}(I_N,-I_N)$. Since $H$ has no
additional symmetry constraint, $\mathcal H_z$ belongs to the Hermitian
chiral unitary class AIII. In terms of $\mathcal H_z$, the replica partition
function in Eq.~\eqref{eq:a-brief-partition} reads
$Z_{1,n}^{\mathrm A}=\langle\det^n(\eta+i\mathcal H_z)\rangle$. Since
\begin{equation}
 \det(\eta+i\mathcal H_z)=(-1)^N\det(i\eta-\mathcal H_z),
 \label{eq:bulk-hermitized-source}
\end{equation}
the source $\eta$ plays the role of an imaginary energy, while $z$
enters as a parameter of $\mathcal H_z$. The field $U\in\mathrm U(n)$
and the linear source coupling $\eta\tr(U+U^\dagger)$ in
Eq.~\eqref{eq:main-self-bulk-source-action} thus have the structure
of the zero-dimensional class-AIII nonlinear $\sigma$ model of
Anderson localization~\cite{Gade1993,Karcher2023}. The Gaussian
calculation above fixes the coupling coefficient and the prefactor.

Expanding $Z_{1,n}^{\mathrm A}$ to second order in $\eta$ and using the Schur
orthogonality relation [Eq.~\eqref{eq:buub}],
\begin{equation}
 \int dU\,\tr U\tr U^\dagger=1,
 \qquad \int dU\,(\tr U)^2=0,
 \label{eq:main-self-unitary-contraction}
\end{equation}
we obtain
\begin{equation}
 Z_{1,n}^{\mathrm A}(z,\eta)\simeq e^{-nq^2}[1+\eta^2(q^2-n)+\bigO(\eta^4)].
 \label{eq:main-self-source-expansion}
\end{equation}
Equation~\eqref{eq:a-brief-replica} then gives
\begin{equation}
 \pi\mathcal O_1(z)=\frac{q^2}{N}=1-\frac{|z|^2}{N},
 \label{eq:draft-a-bulk-self}
\end{equation}
which recovers the result in
Ref.~\cite{chalker1998eigenvector}.

We now turn to the other two bulk families. The same source
construction applies to class $\mathrm{AI}^{\dagger}$, whereas class
$\mathrm{AII}^{\dagger}$ requires a Pfaffian to count each Kramers pair
once. We keep the spectral radius $\sqrt N$ and count
$N$ distinct eigenvalues in all three classes. The changes in the
source coupling and the angular integral determine the coefficient
$\beta/2$ in Eq.~\eqref{eq:general-class-bulk-o1}.

For class $\mathrm{AI}^{\dagger}$, we take \(H=H^T\) with
\(P(H)\propto\exp\,\left[-\tr\,(H^\dagger H)/2\right]\), and define
\begin{equation}
  Z_{1,n}^{\mathrm{AI}^{\dagger}}(z,\eta)
  :=\left\langle\det\!\left[
    (z-H)(z-H)^\dagger+\eta^2 I
  \right]^n\right\rangle.
  \label{eq:aidag-brief-petermann-partition}
\end{equation}
At the bulk saddle point, \(q^2=N-|z|^2\) as in class A, and the
saddle-point manifold is \(\mathrm{Sp}(n)\). The nonlinear $\sigma$
model reads
\begin{equation}
  Z_{1,n}^{\mathrm{AI}^{\dagger}}(z,\eta)
  \simeq e^{-nq^2-n\eta^2}
    \int_{\mathrm{Sp}(n)}\!\mathrm dU\,
    e^{\frac{\eta q}{2}\tr(U+U^\dagger)}.
  \label{eq:aidag-brief-petermann-sigma-model}
\end{equation}
Here $U\in\mathrm{Sp}(n)\subset\mathrm U(2n)$. The quaternion-real
Hubbard--Stratonovich field has Gaussian weight
$e^{-\tr(QQ^\dagger)/2}$~\cite{Kulkarni2025}. Shifting $Q$ by the
scalar source and setting $Q=qU$ gives the coupling
$(q\eta/2)\tr(U+U^\dagger)$ and the factor $e^{-n\eta^2}$.
The fundamental representation of \(\mathrm{Sp}(n)\) is irreducible and
has a real character, so that \(\tr\,U^\dagger=\tr\,U\). The Schur
orthogonality relation [Eq.~\eqref{eq:schur_orthogonality}] then gives
\(\int\mathrm dU\,\tr\,U=0\) and \(\int\mathrm dU\,(\tr\,U)^2=1\).
Hence,
\begin{equation}
  Z_{1,n}^{\mathrm{AI}^{\dagger}}(z,\eta)
  \simeq e^{-nq^2}\left[
    1+\eta^2\left(\frac{q^2}{2}-n\right)+\bigO(\eta^4)
  \right].
  \label{eq:aidag-brief-petermann-expansion}
\end{equation}
Equation~\eqref{eq:a-brief-replica} extracts the coefficient of $\eta^2$
and divides it by $N$ after taking the replica limit $n\to0$.
In this limit, $e^{-nq^2}\to1$ and $q^2/2-n\to q^2/2$, giving
$\pi\mathcal O_1=q^2/(2N)=\tfrac12(1-|z|^2/N)$.
Thus the bulk amplitude is $\beta/2=1/2$, in agreement with the
bulk mean overlap obtained from the Petermann-factor distribution
in Ref.~\cite{Akemann2025b}.

For classes with Kramers degeneracy, the replica partition function must
be modified so that each Kramers pair is counted once. For these classes,
we choose the TRS${}^{\dagger}$ matrix $\mathcal{C}_+ = J$, with
$J=\left(\begin{smallmatrix}0&I_N\\-I_N&0\end{smallmatrix}\right)$.
The corresponding one-point replica partition function is
\begin{align}
 Z_{1,n}(z,\eta)&=\Biggl\langle\Biggl\{\operatorname{Pf}\!\begin{pmatrix}
 (z-H)J&\eta I\\
 -\eta I&[(z-H)J]^\dagger
 \end{pmatrix}\Biggr\}^{n}\Biggr\rangle.
 \label{eq:general-class-contact-partition}
\end{align}
With this $Z_{1,n}$ and the rank-normalized overlap
$O_{aa}=\tr(\Pi_a\Pi_a^\dagger)/2$
[Eq.~\eqref{eq:general-class-rank-normalized-overlap}], the
source-extraction identity in Eq.~\eqref{eq:a-brief-replica} retains
its form. Together with the determinant construction for the classes
without Kramers degeneracy, this identity covers all classes considered
here.

For class $\mathrm{AII}^{\dagger}$, $H$ is a $2N\times2N$ complex
self-dual matrix, $H=JH^TJ^{-1}$, with
$P(H)\propto e^{-\tr(H^\dagger H)}$. The auxiliary field is
real~\cite{Kulkarni2025}. With $n$ Pfaffian replicas, it is an
$n\times n$ matrix with Gaussian weight $e^{-\tr(QQ^T)}$.
At the bulk saddle $Q=qO$, $O\in\mathrm O(n)$, the scalar source
shift gives $q\eta\tr(O+O^T)=2q\eta\tr O$. Hence
\begin{equation}
 Z_{1,n}^{\mathrm{AII}^{\dagger}}(z,\eta)
 \simeq e^{-nq^2-n\eta^2}\int_{\mathrm O(n)}dO\,e^{2q\eta\tr O}.
 \label{eq:aiidag-brief-petermann-sigma-model}
\end{equation}
The normalized Haar measure includes both components of
$\mathrm O(n)$. Schur orthogonality gives
$\int dO\,\tr O=0$ and $\int dO\,(\tr O)^2=1$, so the coefficient
of $\eta^2$ is $2q^2-n$. Equation~\eqref{eq:a-brief-replica} therefore
gives $\pi\mathcal O_1=2q^2/N$, or $\beta/2=2$.
The coefficients $1$, $1/2$, and $2$ for classes A,
$\mathrm{AI}^{\dagger}$, and $\mathrm{AII}^{\dagger}$ thus follow from
their source couplings and group averages, for a common normalization
of the overlap density.

The same Hermitization also explains the other two target manifolds.
For $H$ in classes $\mathrm{AI}^{\dagger}$ and
$\mathrm{AII}^{\dagger}$, the Hermitized matrix $\mathcal H_z$ in
Eq.~\eqref{eq:bulk-hermitized-hamiltonian} belongs to Hermitian classes
CI and DIII, respectively~\cite{Chen2025}.
Their compact fermionic-replica target manifolds are
$\mathrm{Sp}(n)$ and $\mathrm O(n)$, with $n$
counting Pfaffian replicas in the latter case.
Equations~\eqref{eq:aidag-brief-petermann-sigma-model}
and~\eqref{eq:aiidag-brief-petermann-sigma-model} are therefore the
corresponding zero-dimensional nonlinear $\sigma$ models, with $\eta$
again acting as an imaginary energy.

\subsection{Bulk Off-Diagonal Overlap}
\label{sec:brief-class-a-o2}

For the same class-A Gaussian ensemble, we now calculate the
off-diagonal overlap $\mathcal O_2(z_1,z_2)$. We keep the
midpoint in the bulk and the separation between distinct eigenvalues
microscopic,
\begin{equation}
 z=\frac{z_1+z_2}{2},\qquad \omega=z_1-z_2.
 \label{eq:a-o2-variables}
\end{equation}
To generate this two-point function, we introduce the mixed resolvent
\begin{equation}
 \mathcal G(z_1,z_2)=\frac1{N}\left\langle
 \tr[\mathcal R(z_1)\mathcal R(z_2)^\dagger]\right\rangle.
 \label{eq:main-mixed-resolvent}
\end{equation}
Its spectral expansion is
\begin{equation}
 \mathcal G(z_1,z_2)=\frac1N\left\langle
 \sum_{a,b}\frac{O_{ab}}
 {(z_1-\lambda_a)(\bar z_2-\bar\lambda_b)}
 \right\rangle.
 \label{eq:g2_spectral}
\end{equation}
Its derivatives give
\begin{equation}
 \frac1{\pi^2}\partial_{\bar z_1}\partial_{z_2}
 \mathcal G(z_1,z_2)
 =\mathcal O_1(z_1)\delta^{(2)}(\omega)+\mathcal O_2(z_1,z_2).
 \label{eq:main-mixed-resolvent-contact}
\end{equation}
We introduce a single source $\eta$ through the two-copy partition
function $Z_{2,n}^{\mathrm A}(\eta)=\langle\det[D(\eta)]^n\rangle$, where
\begin{equation}
 D(\eta)=
 \begin{pmatrix}
 z_1-H&\eta I&0&0\\
 -\eta I&\bar z_2-H^\dagger&0&0\\
 0&0&z_2-H&0\\
 0&0&0&\bar z_1-H^\dagger
 \end{pmatrix}.
 \label{eq:main-pair-source-matrix}
\end{equation}
Only the first two blocks are coupled by $\eta$. Taking the Schur
complement of this $2\times2$ block gives
\begin{equation}
 \det D(\eta)=\det D(0)\det[I+\eta^2\mathcal R_1\mathcal R_2^\dagger],
 \label{eq:main-pair-determinant-factorization}
\end{equation}
where $\mathcal R_a=\mathcal R(z_a)$. The determinant is even in
$\eta$, and differentiating twice yields
\begin{equation}
 \left.\frac1{2n}\partial_\eta^2[\det D(\eta)]^n\right|_{\eta=0}
 =[\det D(0)]^n\tr(\mathcal R_1\mathcal R_2^\dagger).
 \label{eq:main-pair-determinant-derivative}
\end{equation}
Since $[\det D(0)]^n\to1$ in the replica limit, the ensemble average
recovers the mixed resolvent,
\begin{equation}
 \mathcal G(z_1,z_2)=\frac1{2N}\lim_{n\to0}\frac1n
 \left.\partial_\eta^2 Z_{2,n}^{\mathrm A}(\eta)\right|_{\eta=0}.
 \label{eq:main-pair-source-extraction}
\end{equation}

The Hubbard--Stratonovich transformation introduces a $2n\times 2n$
complex matrix field $Q$. At $\eta=\omega=0$, its bulk saddle is
$Q=qU$, with $U\in\mathrm U(2n)$ and $q^2=N-|z|^2$. Defining
\begin{equation}
 \Lambda=\operatorname{diag}(I_n,-I_n),
\end{equation}
we expand about this saddle to second order in $\omega$ and integrate
out the radial fluctuations. Shifting the source blocks and evaluating
the resulting source insertion at $Q=qU$ then gives
(see Appendix~\ref{subsec:a-o2-nlsm-derivation})
\begin{equation}
 \begin{aligned}
 &Z_{2,n}^{\mathrm A}(\eta)\simeq\int_{\mathrm U(2n)}dU\,
 \exp\!\Bigl\{\frac {|\omega|^2}4\tr(U^\dagger\Lambda U\Lambda)\\
 &-\tr\!\Bigl[\left(qU+
 \left(\begin{smallmatrix}0&\eta I_n\\0&0\end{smallmatrix}\right)\right)
 \left(qU^\dagger+
 \left(\begin{smallmatrix}0&0\\\eta I_n&0\end{smallmatrix}\right)\right)
 -q^2 I_{2n}\Bigr]\Bigr\}.
 \end{aligned}
 \label{eq:main-pair-bulk-source-action}
\end{equation}
We omit a source-independent radial factor that tends to unity in the
replica limit. Subtracting $q^2I_{2n}$ makes the second trace vanish at
$\eta=0$.

At $\eta=0$, Hermitization gives the Hermitian chiral unitary
class-AIII matrices $\mathcal H_{z_1}$ and $\mathcal H_{z_2}$
[Eq.~\eqref{eq:bulk-hermitized-hamiltonian}], each with $n$ replicas.
When $z_1=z_2$, the two matrices coincide, giving
$U\in\mathrm U(2n)$. The spectral splitting $\omega=z_1-z_2$ generates the first term
$|\omega|^2\tr(U^\dagger\Lambda U\Lambda)/4$ in
Eq.~\eqref{eq:main-pair-bulk-source-action}, which suppresses mixing
between the two replica sectors. The source $\eta$ in
Eq.~\eqref{eq:main-pair-source-matrix} couples these sectors.
Shifting the $(1,2)$ block of $Q$ by $\eta I_n$, together
with its conjugate, transfers the source from the determinant to
the Gaussian weight $e^{-\tr(QQ^\dagger)}$. At $Q=qU$, this gives
\begin{equation}
 -q\eta\tr\!\left[
 \begin{pmatrix}0&I_n\\0&0\end{pmatrix}U^\dagger
 +U\begin{pmatrix}0&0\\I_n&0\end{pmatrix}\right]
 -n\eta^2.
 \label{eq:main-pair-source-exponent}
\end{equation}
Here, the linear term probes the off-diagonal blocks of $U$, while
$-n\eta^2$ comes from the product of the two source blocks.

At $\eta=0$, the normalized angular integral is
\begin{equation}
 Y_n^{\mathrm A}(|\omega|^2)=\int_{\mathrm U(2n)}dU\,
 \exp\!\left[\frac {|\omega|^2}4\tr(U^\dagger\Lambda U\Lambda)\right],
 \label{eq:main-pair-angular-integral}
\end{equation}
with $Y_n^{\mathrm A}(0)=1$.
The second derivative of $Z_{2,n}^{\mathrm A}$ with respect to the source $\eta$ can be
replaced by a derivative with respect to the squared microscopic separation $|\omega|^2$ (Appendix~\ref{subsec:a-o2-nlsm-derivation}),
\begin{equation}
 \left.\frac12\partial_\eta^2 Z_{2,n}^{\mathrm A}(\eta)\right|_{\eta=0}
 =\left(\frac{q^2}{2}-n-\frac{q^2}{n}\frac{\mathrm d}{\mathrm d|\omega|^2}\right)
 Y_n^{\mathrm A}(|\omega|^2).
 \label{eq:main-pair-source-insertion}
\end{equation}
At fixed midpoint and for $z_1\ne z_2$,
$\partial_{\bar z_1}\partial_{z_2}f(|\omega|^2)=-\partial_{|\omega|^2}[|\omega|^2f'(|\omega|^2)]$, so that
\begin{align}
 \pi^2\mathcal O_2
 &=\frac{q^2}{N}\frac{\mathrm d}{\mathrm d|\omega|^2}\biggl\{
 |\omega|^2\notag\\
 &\quad\times\lim_{n\to0}\left[
 \frac1{n^2}\frac{\mathrm d^2Y_n^{\mathrm A}(|\omega|^2)}{\mathrm d(|\omega|^2)^2}
 -\frac1{2n}\frac{\mathrm dY_n^{\mathrm A}(|\omega|^2)}{\mathrm d|\omega|^2}
 \right]\biggr\}.
 \label{eq:main-pair-differentiated-replica}
\end{align}
To take the replica limit $n\to 0$, we use the known replica limit of
$Y_n^{\mathrm A}$~\cite{ChenXiaoLiuRyu2026}. With $\mathcal{P}_2$ defined in
Eq.~\eqref{eq:def_p2}, the replica limit of the second derivative is
given by
\begin{equation}
 \mathcal P_2(|\omega|^2)=\frac12+2\lim_{n\to0}\frac1{n^2}
 \frac{\mathrm d^2Y_n^{\mathrm A}(|\omega|^2)}{\mathrm d(|\omega|^2)^2}.
 \label{eq:main-pair-poisson-matching}
\end{equation}
This gives the first term in
Eq.~\eqref{eq:main-pair-differentiated-replica}. For the second term,
the finiteness of the replica limit and the normalization
$Y_n^{\mathrm A}(0)=1$ give
\begin{equation}
 Y_n^{\mathrm A}(|\omega|^2)=1+cn|\omega|^2+\bigO(n^2),
 \label{eq:main-pair-linear-replica}
\end{equation}
with a constant $c$.
The overlap $\mathcal{O}_2$ is then proportional to
$\mathrm d[|\omega|^2\mathcal P_2(|\omega|^2)]/\mathrm d|\omega|^2-1/2-c$. Since
$\mathrm d[|\omega|^2\mathcal P_2(|\omega|^2)]/\mathrm d|\omega|^2\to1$ at large $|\omega|$, the condition
$\mathcal O_2\to0$ fixes $c=1/2$ (see
Appendix~\ref{subsec:pair_correlation_replica_limit} for details).

The resulting $\mathcal{O}_2$ for class A is
\begin{equation}
 \pi^2\mathcal O_2(z_1,z_2)
 =\frac{1-|z|^2/N}{2}
 \left\{-1+\frac{\mathrm d}{\mathrm d|\omega|^2}[|\omega|^2\mathcal P_2(|\omega|^2)]\right\}.
 \label{eq:a-o2-poisson-relation}
\end{equation}
For class A and $|\omega|>0$, $\mathcal{P}_2$ reads~\cite{ChenXiaoLiuRyu2026}
\begin{equation}
 \mathcal P_2(|\omega|^2)
 =1+\frac{2(1-e^{-|\omega|^2})}{|\omega|^4}.
 \label{eq:a-o2-poisson-class-a}
\end{equation}
Substituting Eq.~\eqref{eq:a-o2-poisson-class-a} into
Eq.~\eqref{eq:a-o2-poisson-relation} yields
\begin{equation}
 \pi^2\mathcal O_2(z_1,z_2)
 =-\left(1-\frac{|z|^2}{N}\right)
 \frac{1-(1+|\omega|^2)e^{-|\omega|^2}}{|\omega|^4},
 \label{eq:a-o2-final}
\end{equation}
recovering the result of Ref.~\cite{chalker1998eigenvector}.
The off-diagonal overlap is finite as $|z_1 - z_2| \to 0$, where it approaches
$-(1-|z|^2/N)/(2\pi^2)$. At large microscopic separation, it has a
$|z_1 - z_2|^{-4}$ tail.

For the other two bulk families, only the ensemble average and the
generating function change. For class $\mathrm{AI}^{\dagger}$, we use
the same determinant partition function with the complex-symmetric
Gaussian average.
For class $\mathrm{AII}^{\dagger}$ and the other Kramers-degenerate
classes, we instead use a Pfaffian partition
function, with $J$ as in
Eq.~\eqref{eq:general-class-contact-partition}. We write
$K_a=(z_a-H)J$ and define
$Z_{2,n}(\eta)=\langle\operatorname{Pf}[\widetilde D(\eta)]^n\rangle$, with
\begin{equation}
 \widetilde D(\eta)=
 \begin{pmatrix}
 K_1&0&0&\eta I\\
 0&K_2&0&0\\
 0&0&K_1^\dagger&0\\
 -\eta I&0&0&K_2^\dagger
 \end{pmatrix}.
 \label{eq:general-class-pair-partition}
\end{equation}
Replacing $Z_{2,n}^{\mathrm A}$ by this $Z_{2,n}$ in
Eq.~\eqref{eq:main-pair-source-extraction} gives the mixed resolvent
\begin{equation}
   \mathcal G(z_1,z_2)=\frac1{d N}\left\langle
 \tr[\mathcal R(z_1)\mathcal R(z_2)^\dagger]\right\rangle.
\end{equation}
Equations~\eqref{eq:g2_spectral} and~\eqref{eq:main-mixed-resolvent-contact}
still hold, so $\mathcal{O}_2$ follows from $Z_{2,n}$ in the same way.

For all three bulk classes, the source shift and the block-subgroup
contractions are given explicitly in
Appendices~\ref{subsec:pair-source-three-classes} and~\ref{app:schur}.
Equation~\eqref{eq:pair-three-class-source-reduction} converts the
second source derivative into the operator
$\beta q^2/4-n-\beta q^2\partial_{|\omega|^2}/(2n)$ acting on the
corresponding source-free angular integral $Y_n^X$.
Combining this identity with the class-dependent spectral replica
limit and the large-separation boundary condition gives
\begin{equation}
 \pi^2\mathcal O_2(z_1,z_2)
 =\frac{1}{2}\pi \mathcal{O}_1(z)
 \left\{-1+\frac{\mathrm d}{\mathrm d|\omega|^2}[|\omega|^2\mathcal P_2(|\omega|^2)]\right\},
\end{equation}
which is Eq.~\eqref{eq:general-class-bulk-o2-ratio}. Substituting
$\mathcal{P}_2$ for classes $\mathrm{AI}^{\dagger}$ and
$\mathrm{AII}^{\dagger}$ gives the other two bulk rows in
Table~\ref{tab:replica-bulk-summary}.

\subsection{Hard-Edge Diagonal Overlap}
\label{sec:class-d-hard-edge-resolvent}

Finally, we calculate $\mathcal{O}_1$ near the origin for the seven
hard-edge classes. As an example, we consider class D, whose Gaussian
ensemble consists of complex antisymmetric matrices,
\begin{equation}
 H^T=-H,\quad P(H)\propto e^{-\tr(H^\dagger H)/2},
 \quad N\text{ even},
 \label{eq:d-resolvent-ensemble}
\end{equation}
normalized such that $\pi R_1=1$ in the bulk.
We follow Sec.~\ref{subsec:brief-class-a-o1-summary} and use the
class-D replica partition function $Z_{1,n}^{\mathrm D}$ obtained by
replacing the ensemble average in Eq.~\eqref{eq:a-brief-partition}
with the class-D average.

At fixed $|z|^2=\bigO(1)$ as $N\to\infty$, the saddle point has
$q^2=N$. With
$J_n=\left(\begin{smallmatrix}0&I_n\\-I_n&0\end{smallmatrix}\right)$, we
define
\begin{equation}
 Y_n^{\mathrm D}(|z|^2)=\int_{\mathrm O(2n)}dO\,
 \exp\!\left[-\frac {|z|^2}2\tr(O^TJ_nOJ_n)\right].
 \label{eq:main-hard-edge-angular-integral}
\end{equation}
The normalized Haar measure includes both connected components of
$\mathrm O(2n)$. The saddle-point expression with the source becomes
\begin{align}
 Z_{1,n}^{\mathrm D}(z,\eta)&\simeq e^{-n\eta^2}
 \int_{\mathrm O(2n)}dO\,
 \exp\!\left[-\frac {|z|^2}2\tr(O^TJ_nOJ_n)\right]\notag\\*
 &\quad\times\exp\!\left[
 \frac{\eta\sqrt N}2\tr(O+O^T)\right].
 \label{eq:main-hard-edge-source-action}
\end{align}
A Ward identity for the group integral converts the second derivative
with respect to the source $\eta$ into a derivative with respect to $|z|^2$,
\begin{equation}
 \left.\frac12\partial_\eta^2 Z_{1,n}^{\mathrm D}(z,\eta)\right|_{\eta=0}
 =\left[\frac N2\left(1+\frac1n\partial_{|z|^2}\right)-n\right]Y_n^{\mathrm D}(|z|^2).
 \label{eq:main-hard-edge-source-ward}
\end{equation}
This is proved in Appendix~\ref{subsec:class_d_replica_limit}.
With $\mathcal{P}_1$ defined in Eq.~\eqref{eq:def_p1}, the replica limits
of $Y^{\mathrm{D}}_n$ are known~\cite{ChenXiaoLiuRyu2026},
\begin{equation}
 \lim_{n\to0}Y_n^{\mathrm D}(|z|^2)=1,\quad
 \lim_{n\to0}\frac1n\partial_{|z|^2}Y_n^{\mathrm D}(|z|^2)=\mathcal P_1(|z|^2).
 \label{eq:main-hard-edge-poisson-matching}
\end{equation}
Equation~\eqref{eq:a-brief-replica} then gives
$\pi\mathcal O_1(z)=\tfrac12[1+\mathcal P_1(|z|^2)]$, i.e., the explicit
expression in Eq.~\eqref{eq:draft-d-hard-edge-result}.

Appendix~\ref{subsec:d-topological-source} extends this derivation to
odd matrix dimension. Retaining the Pfaffian factor $(\det O)^\nu$,
where $\nu=N\bmod2$, leaves the source differential operator unchanged
and replaces $Y_n^{\mathrm D}$ and $\mathcal P_1$ by their
sector-resolved counterparts.
The class-D calculation is an example of the general hard-edge relation in
Eq.~\eqref{eq:general-class-hard-edge-o1}. The same calculation for the
other six classes gives that relation with the corresponding $\beta$
from Table~\ref{tab:symmetry-classification}, i.e., $\beta=2$ for
classes $\mathrm{AIII}^{\dagger}$, $\mathrm D$, and $\mathrm C$,
$\beta=1$ for classes $\mathrm{BDI}_0$ and $\mathrm{CI}_0$, and
$\beta=4$ for classes $\mathrm{CII}_0$ and $\mathrm{DIII}_0$. The
prefactor thus depends only on the bulk family, whereas
$\mathcal P_1$, the Poisson transform of the Hermitian hard-edge
density of the partner class, carries the dependence on the full
symmetry class and topological sector. In the matching regime
$1\ll |z|^2\ll N$, $\mathcal P_1\to1$, which recovers the central bulk
value $\pi\mathcal O_1\to\beta/2$.

Protected zero modes are included in the Hermitian density entering
$\mathcal P_1$. Their $\delta$ functions produce $1/|z|^2$ terms in this
transform and hence modify the overlap density of the nonzero
eigenvalues. Equation~\eqref{eq:general-class-hard-edge-o1} describes
this density at $|z|>0$. The overlap weight carried by the zero modes at
$z=0$ is separate. Substituting the Poisson transforms of the seven
classes and their topological sectors gives the expressions in
Table~\ref{tab:replica-results-summary}.

\section{Numerical Verification}
\label{sec:numerics}

We test the bulk and hard-edge predictions in
Tables~\ref{tab:replica-bulk-summary} and~\ref{tab:replica-results-summary}
using Gaussian ensembles with bulk normalization $\pi R_1=1$.

\begin{figure}[!tp]
  \centering
  \includegraphics[width=\columnwidth]
    {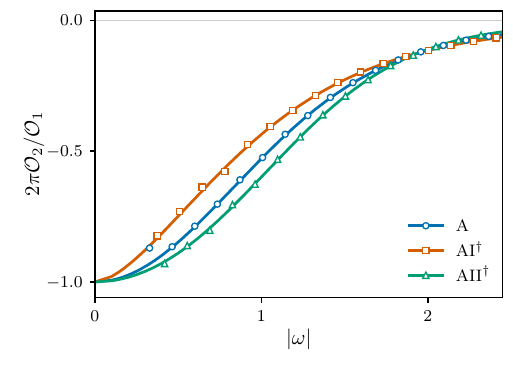}
  \caption{Off-diagonal overlaps in classes \(\mathrm A\),
  \(\mathrm{AI}^{\dagger}\), and \(\mathrm{AII}^{\dagger}\), using
  \(100{,}000\) matrices of effective size \(N=512\) per class.
  Lines are the replica predictions {in} Eq.~\eqref{eq:general-class-bulk-o2-ratio}.}
  \label{fig:numerical-three-class-o2}
\end{figure}

\begin{figure*}[t]
  \centering
  \includegraphics[width=5.10in]
    {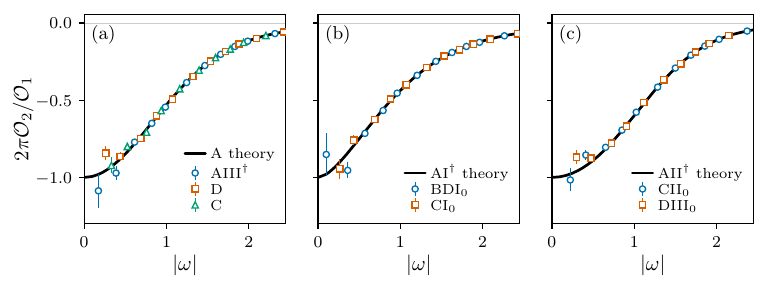}
  \caption{Bulk off-diagonal overlaps at \(N=256\) and
  \(\nu=0\), using \(100{,}000\) matrices per class.
  (a) Classes \(\mathrm{AIII}^{\dagger}\), \(\mathrm D\), and
  \(\mathrm C\) compared with class \(\mathrm A\).
  (b) Classes \(\mathrm{BDI}_0\) and \(\mathrm{CI}_0\) compared with
  class \(\mathrm{AI}^{\dagger}\).
  (c) Classes \(\mathrm{CII}_0\) and \(\mathrm{DIII}_0\) compared with
  class \(\mathrm{AII}^{\dagger}\).
  Midpoints lie between radii \(0.25\sqrt N\) and
  \(0.60\sqrt N\).
  Lines are the replica predictions {in} Eq.~\eqref{eq:general-class-bulk-o2-ratio}.}
  \label{fig:numerical-seven-class-o2}
\end{figure*}

We first test the bulk prediction, Eq.~\eqref{eq:general-class-bulk-o2-ratio},
in all ten AZ$_0$ classes, with $N=512$ for classes
$\mathrm A$, $\mathrm{AI}^{\dagger}$, and $\mathrm{AII}^{\dagger}$,
and $N=256$, $\nu=0$ for the other seven classes.
We select eigenvalues away from the spectral edge and, for the seven
hard-edge classes, away from the origin. Pairs are binned by midpoint
$z=(z_1+z_2)/2$ and squared separation $|\omega|^2$, and the overlap
density is normalized by $\mathcal O_1(z)$. Combining both orderings gives
$O_{ab}+O_{ba}=2\operatorname{Re}O_{ab}$, so the plotted averages are real.

We find good agreement between the numerical bulk off-diagonal overlaps and
the replica predictions for classes $\mathrm A$, $\mathrm{AI}^{\dagger}$,
and $\mathrm{AII}^{\dagger}$ (Fig.~\ref{fig:numerical-three-class-o2}).
Among the remaining classes, classes $\mathrm{AIII}^{\dagger}$,
$\mathrm D$, and $\mathrm C$ follow the class-$\mathrm A$ prediction,
classes $\mathrm{BDI}_0$ and $\mathrm{CI}_0$ follow the
class-$\mathrm{AI}^{\dagger}$ prediction, and classes $\mathrm{CII}_0$
and $\mathrm{DIII}_0$ follow the class-$\mathrm{AII}^{\dagger}$
prediction (Fig.~\ref{fig:numerical-seven-class-o2}). These results support the
prediction that the universal form of the normalized off-diagonal
overlap in the spectral bulk is determined solely by TRS${}^{\dagger}$.

\begin{figure*}[t]
  \centering
  \includegraphics[width=\textwidth]
    {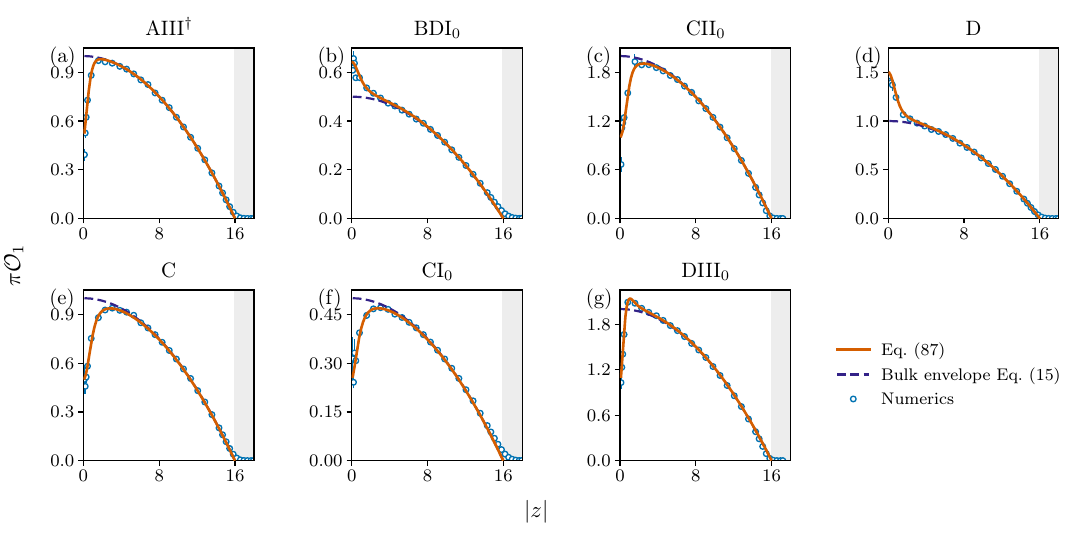}
  \caption{Diagonal overlaps at \(N=256\) and \(\nu=0\),
  from \(600{,}000\) matrices per class: (a) \(\mathrm{AIII}^{\dagger}\),
  (b) \(\mathrm{BDI}_0\), (c) \(\mathrm{CII}_0\), (d) \(\mathrm D\),
  (e) \(\mathrm C\), (f) \(\mathrm{CI}_0\), and (g) \(\mathrm{DIII}_0\).
  Solid orange curves show the additive interpolation
  in Eq.~\eqref{eq:numerical-o1-additive}, while dashed curves retain only the
  bulk envelope.
  The shaded region lies beyond \(\sqrt N=16\).
  Each displayed bin contains at least \(100\) spectral multiplets.}
  \label{fig:numerical-seven-class-o1}
\end{figure*}

\begin{figure*}[t]
  \centering
  \includegraphics[width=\textwidth]
    {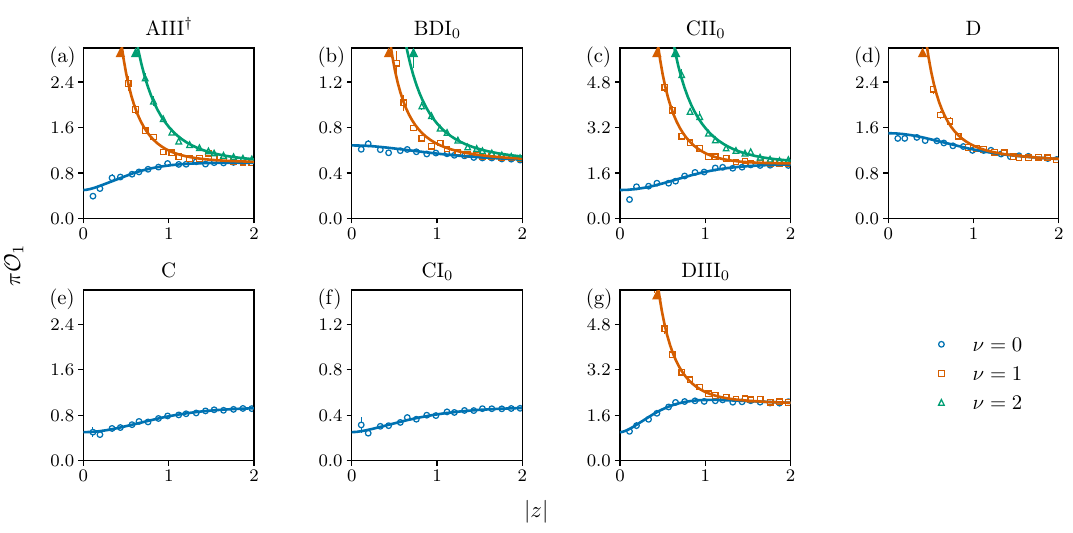}
  \caption{Diagonal overlaps, in
  (a) \(\mathrm{AIII}^{\dagger}\), (b) \(\mathrm{BDI}_0\),
  (c) \(\mathrm{CII}_0\), (d) \(\mathrm D\), (e) \(\mathrm C\),
  (f) \(\mathrm{CI}_0\), and (g) \(\mathrm{DIII}_0\).
  Colors and symbols identify \(\nu\). Each sector uses \(600{,}000\)
  matrices, with \(N=257\) for \(\nu=1\) and \(N=256\) otherwise.
  Lines show the additive interpolation in Eq.~\eqref{eq:numerical-o1-additive}.
  Top-edge triangles indicate the outermost displayed off-scale mean in
  each series.
  Protected zero modes, origin-touching bins, and bins with fewer than \(100\) multiplets are omitted.
  }
  \label{fig:numerical-nu-overlay}
\end{figure*}

We next test the diagonal overlaps in the seven hard-edge classes,
with $N=256$ for $\nu=0,2$ and $N=257$ for $\nu=1$.
Equation~\eqref{eq:general-class-hard-edge-o1} applies at
$|z|\ll\sqrt N$. To compare over a wider range, we include the bulk
envelope from Table~\ref{tab:replica-bulk-summary} through the additive
interpolation
\begin{equation}
  \label{eq:numerical-o1-additive}
  \pi \mathcal{O}_1(z)
  = \frac{\beta}{2}\left[
      1-\frac{|z|^2}{N}
      +\frac{\mathcal P_{1}(|z|^2)-1}{2}
    \right],
\end{equation}
with $\beta$ fixed by Table~\ref{tab:symmetry-classification}.
We find that this interpolation captures the $\nu=0$ diagonal-overlap
profiles across the spectral interior in all seven classes
(Fig.~\ref{fig:numerical-seven-class-o1}). Near the origin, the numerical
results agree with the predicted class and sector dependence in
Table~\ref{tab:replica-results-summary}, including the $|\nu|/|z|^2$
divergence in nonzero sectors (Fig.~\ref{fig:numerical-nu-overlay}).

\section{Universality in Physical Models}
\label{sec:physical-models}

We now test universality in the non-Hermitian SYK and
Anderson models, and in a dissipative fermion model on a bipartite
lattice. For bulk off-diagonal overlaps, we compare
$\mathcal O_2(z_1,z_2)/[R_1(z)\mathcal O_1(z)]$ as a function of the
unfolded separation $\omega=\sqrt{\pi R_1(z)}\,(z_1-z_2)$ with
Table~\ref{tab:replica-bulk-summary}. In the normalization $\pi R_1=1$
used in Secs.~\ref{sec:main-results}--\ref{sec:numerics}, this
$\omega$ reduces to $z_1-z_2$. For the hard edge, we compare the
profile in Table~\ref{tab:replica-results-summary} after fixing the
spectral scale and overlap amplitude. The unfolding and normalization
formulas are given in Appendix~\ref{app:unfolding-normalization}.

\subsection{SYK Model}

We first consider the non-Hermitian SYK
model with four-body interaction~\cite{GarciaGarcia2022,ChenXiaoLiuRyu2026},
\begin{equation}
  H=\sum_{1\leq i<j<k<\ell\leq N_\gamma}
  K_{ijk\ell}\,\gamma_i\gamma_j\gamma_k\gamma_\ell,
  \label{eq:physical-syk-hamiltonian}
\end{equation}
where the $N_\gamma$ Majorana operators satisfy
$\{\gamma_i,\gamma_j\}=2\delta_{ij}$, and the couplings $K_{ijk\ell}$
are independent, centered complex Gaussian variables with variance
$\langle|K_{ijk\ell}|^{2}\rangle=\binom{N_\gamma}{4}^{-1}$.
We diagonalize the even-fermion-parity block.
The symmetry class
is determined by $N_\gamma\bmod8$~\cite{GarciaGarcia2022}:
we use $N_\gamma=18$, $16$, and $20$, realizing classes $\mathrm{A}$,
$\mathrm{AI}^{\dagger}$, and $\mathrm{AII}^{\dagger}$, respectively.
We find good agreement between the normalized bulk off-diagonal overlaps
and Eq.~\eqref{eq:general-class-bulk-o2-ratio} in all three classes
(Fig.~\ref{fig:numerical-physical-o2}(a)).

\subsection{Anderson Model}

\begin{figure*}[t]
  \centering
  \includegraphics[width=2.62in]{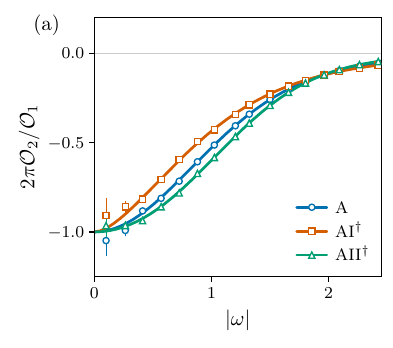}\hspace{0.16in}\includegraphics[width=2.62in]{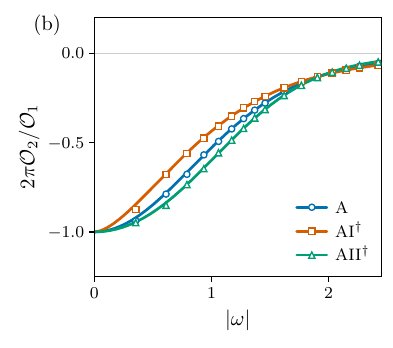}
  \caption{Off-diagonal overlaps in physical models.
  \textbf{(a)} The \(q=4\) non-Hermitian SYK models in classes
  \(\mathrm A\), \(\mathrm{AI}^{\dagger}\), and \(\mathrm{AII}^{\dagger}\)
  have \(18\), \(16\), and \(20\) Majorana fermions and parity-block
  dimensions \(256\), \(128\), and \(512\), respectively. Midpoints lie in
  the central \(30\%\) of the spectrum.
  \textbf{(b)} The Anderson data in classes \(\mathrm A\) and
  \(\mathrm{AI}^{\dagger}\) use periodic three-dimensional lattices with
  \(L=8\) and \(W=3\). The data in class \(\mathrm{AII}^{\dagger}\) use {a periodic
  two-dimensional square lattice with} \(L=16\),
  \(W=3.8\), and \(\mu=2\). Each Anderson matrix has dimension \(512\).
  {Midpoints lie in the central \(30\%\) of the spectrum.} In
  class \(\mathrm{AII}^{\dagger}\), the spectrum contains \(256\) distinct
  rank-two eigenspaces.
  Both panels use \(100{,}000\) realizations per
  class and local unfolding to \(\pi R_1=1\).
  Lines are the random-matrix predictions {in} Eq.~\eqref{eq:general-class-bulk-o2-ratio}.
  }
  \label{fig:numerical-physical-o2}
\end{figure*}

To test the bulk correlations in a spatially local system, we next
consider the non-Hermitian Anderson
model~\cite{LuoOhtsukiShindou2021,LuoXiao2022}.
Let \(c_{\boldsymbol r}\) denote a one-component fermionic operator
in classes \(\mathrm A\) and \(\mathrm{AI}^{\dagger}\) and a
two-component spinor in class \(\mathrm{AII}^{\dagger}\).
We denote Hermitian conjugation by H.c. For the three symmetry
classes, the Hamiltonians read~\cite{Tzortzakakis2020,HuangShklovskii2020Spectral,Yan2022}
\begin{align}
\label{eq:physical-anderson-hamiltonians}
    H
    &=\sum_{\boldsymbol r}\epsilon_{\boldsymbol r}
      c_{\boldsymbol r}^{\dagger}c_{\boldsymbol r}+\sum_{\langle\boldsymbol r,\boldsymbol r'\rangle}
      \left(c_{\boldsymbol r}^{\dagger}
      T_{\boldsymbol r\boldsymbol r'}c_{\boldsymbol r'}
      +\mathrm{H.c.}\right),\\
    \epsilon_{\boldsymbol r}&=x_{\boldsymbol r}+iy_{\boldsymbol r},\\
    T_{\boldsymbol r\boldsymbol r'}
    &=\begin{cases}
      e^{i\phi_{\boldsymbol r\boldsymbol r'}},
        & \mathrm A,\\
      1,
        & \mathrm{AI}^{\dagger},\\
      I_2+i\mu\,\boldsymbol v_{\boldsymbol r\boldsymbol r'}
        \cdot\boldsymbol\sigma,
        & \mathrm{AII}^{\dagger}.
    \end{cases}
\end{align}
The real and imaginary parts $x_{\boldsymbol r}$ and $y_{\boldsymbol r}$ of the on-site energies $\epsilon_{\boldsymbol r}$ are independent
and uniformly distributed in $[-W/2,W/2]$.
The phase $\phi$ is uniformly
distributed in $[0,2\pi)$, and the three components of
\(\boldsymbol v\) are independent and uniformly distributed in
\([-1/2,1/2]\).
We find that the normalized bulk off-diagonal overlaps follow the
class-dependent predictions of Eq.~\eqref{eq:general-class-bulk-o2-ratio}
overall [Fig.~\ref{fig:numerical-physical-o2}(b)].

\subsection{Dissipative Fermions on a Bipartite Lattice}

Having tested bulk off-diagonal overlaps, we turn to the hard-edge
diagonal overlap in a bipartite lattice model with linear loss~\cite{Xiao2024}.
As a physical realization of class $\mathrm{BDI}_{0}$, we consider
spinless fermions on a periodic bipartite cubic lattice with random real
hopping,
\begin{equation}
  \mathcal H
  =\sum_{\langle i,j\rangle}t_{ij}c_i^\dagger c_j+\mathrm{H.c.},
  \label{eq:physical-bdi0-loss-model}
\end{equation}
where $t_{ij}$ is drawn uniformly from $[t-W/2,\,t+W/2]$. We further
introduce the linear loss operators
$L_i=\sqrt{\gamma}\,(c_i+c_{i+e_z})$~\cite{Xiao2024}.
Since the Lindbladian is quadratic, its spectrum is determined by the
effective one-particle Hamiltonian~\cite{ChenXiaoLiuRyu2026}, which,
up to a uniform decay shift, reads
\begin{equation}
  H_{\mathrm{eff}}
  =\mathcal H
  -\mathrm{i}\gamma\sum_i
  \left(c_{i+e_z}^\dagger c_i+c_i^\dagger c_{i+e_z}\right).
  \label{eq:physical-bdi0-hamiltonian}
\end{equation}
With the sublattice operator $\mathcal S$, diagonal with entries
$\mathcal{S}_{ii}=(-1)^{i_x+i_y+i_z}$, the
effective Hamiltonian obeys
$\mathcal{S}H_{\mathrm{eff}}\mathcal{S}=-H_{\mathrm{eff}}$ and
$H_{\mathrm{eff}}^{\mathrm{T}}=H_{\mathrm{eff}}$, placing it in class
$\mathrm{BDI}_{0}$~\cite{Xiao2024,ChenXiaoLiuRyu2026}.
We choose $t=1$, $W=1.6$, and $\gamma=0.1$.
\par
\begin{figure}[!htbp]
  \centering
  \includegraphics[width=\columnwidth]
    {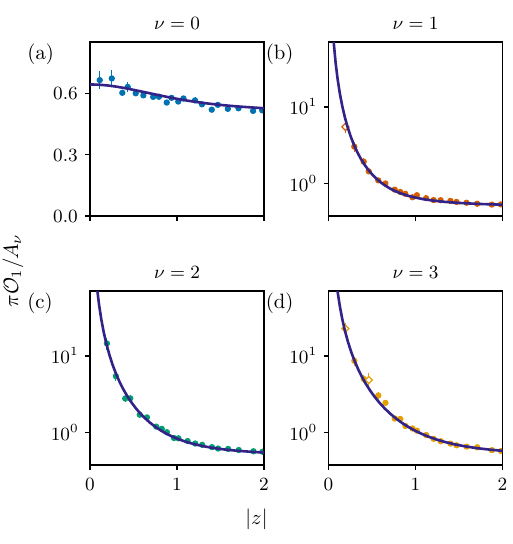}
  \caption{Hard-edge diagonal overlap in the class-\(\mathrm{BDI}_0\)
  model defined by Eq.~\eqref{eq:physical-bdi0-hamiltonian}.
  Panels (a)--(d) show \(\nu=0,1,2,3\), with
  \(500{,}000\) matrices per sector.
  {See Appendix~\ref{app:unfolding-normalization} for details of the
  unfolding and overlap normalization}.
  Panel (a) is linear; panels (b)--(d) are logarithmic.
  Dark solid lines show the hard-edge predictions in
  Eq.~\eqref{eq:general-class-hard-edge-o1} for \(\beta=1\).
  Colored markers are numerical data.
  Open diamonds flag bins where a single nonzero
  \(\pm z\) pair contributes at least \(10\%\) of the overlap sum.
  Protected zero modes, origin-touching bins for \(\nu>0\), and bins with fewer
  than \(100\) multiplets are omitted.}
  \label{fig:numerical-bdi-vacancy}
\end{figure}

To realize sectors with $\nu>0$, we remove $\nu$ sites from one
sublattice of the periodic $L=18$ cubic lattice.
The loss operators $L_i$ become $\sqrt{\gamma}\,c_i$ if $i+e_z$ is a
vacancy and $\sqrt{\gamma}\,c_{i+e_z}$ if $i$ is a vacancy.
After removing the uniform decay shift, only hoppings between retained
sites remain in the effective Hamiltonian $H_{\mathrm{eff}}$.
For $\nu=1$, $2$, and $3$, we sequentially remove the sites at
$(i_x,i_y,i_z)=(1,0,0)$, $(10,9,8)$, and $(10,0,13)$.
The uniform decay shift removed above does not change the spectral
projectors. We retain the
\(16\) nonzero \(\pm z\) pairs closest to the origin.
{See Appendix~\ref{app:unfolding-normalization} for details of the
unfolding and overlap normalization}. This comparison tests the normalized shape, not a universal
absolute amplitude. We find that the normalized profiles follow the hard-edge predictions in
Eq.~\eqref{eq:general-class-hard-edge-o1}, including the
\(|z|^{-2}\) enhancement near the origin
(Fig.~\ref{fig:numerical-bdi-vacancy}).

\section{Conclusion}
\label{sec:conclusion}
We obtain analytical expressions for eigenvector
overlaps $\mathcal{O}_1(z)$ and $\mathcal{O}_2(z_1,z_2)$ in the large-$N$ limit across the tenfold
Altland--Zirnbauer$_0$ (AZ$_0$) classification. These ten classes exhaust
the non-Hermitian internal symmetry classes whose defining constraints
involve neither complex conjugation nor Hermitian conjugation of the
matrix. In the spectral bulk, the diagonal overlap has the
symmetry-dependent amplitude in Eq.~\eqref{eq:general-class-bulk-o1},
while the normalized off-diagonal overlap takes three universal forms
selected by transposition-based time-reversal symmetry
(TRS${}^{\dagger}$), Eq.~\eqref{eq:general-class-bulk-o2-ratio}.
The class-A results recover those of
Refs.~\cite{chalker1998eigenvector,MehligChalker2000}.
Near the origin, Eq.~\eqref{eq:general-class-hard-edge-o1} produces the
full symmetry-class and topological-sector dependence of the diagonal
overlap, including its enhancement by protected zero modes.
Because $\mathcal O_1$ and $\mathcal O_2$ include eigenvalue-density
factors, they can remain finite even when the average overlaps at fixed
eigenvalues diverge. In the bulk, this occurs as two eigenvalues
approach each other: $\mathcal O_2$ remains finite, while the average
off-diagonal overlap at fixed eigenvalues diverges
[Eq.~\eqref{eq:conditional-three-bulk-contact-limits}].
The same distinction appears near the origin in class C, where
$\mathcal O_1$ remains finite but the average diagonal overlap at a
fixed eigenvalue diverges [Eq.~\eqref{eq:conditional-class-c-hard-edge}].

These relations follow from fermionic replica {nonlinear} $\sigma$
models, in which overlap-source insertions reduce to differential
operators acting on spectral partition functions. Together with the
spectral duality established in Ref.~\cite{ChenXiaoLiuRyu2026}, they
connect Hermitian level statistics, non-Hermitian level statistics,
and non-Hermitian eigenvector nonorthogonality through the same
Poisson transforms. The Hermitian spectral correlations thus determine
both non-Hermitian eigenvalue and eigenvector correlations in the
random-matrix scaling limit. This shared structure suggests a deeper
connection between spectral correlations and eigenvector geometry
beyond the individual ensembles considered here.

Our results extend the random-matrix description of quantum chaos from
level statistics to eigenvector nonorthogonality. Numerical comparisons
with Gaussian ensembles, non-Hermitian Sachdev--Ye--Kitaev models,
Anderson metals, and dissipative fermions support this extension.
After spectral unfolding and overlap normalization, the numerical
overlap profiles in these physical models agree with the
symmetry-dependent theoretical predictions.
These results provide eigenvector-based benchmarks for quantum chaos
that complement the established spectral diagnostics.

A natural next step is to extend the relation between level statistics
and eigenvector overlaps to more general physical {scenarios}.
Equation~\eqref{eq:bulk-overlap-from-eigenvalue-correlation} expresses
the normalized off-diagonal overlap directly in terms of the
non-Hermitian eigenvalue pair correlation, suggesting a possible
universal relation even when the level statistics {deviate} from
random{-}matrix {statistics}.
Anderson localization offers a concrete setting: in both Hermitian
and non-Hermitian systems, ergodic extended and localized regimes can
be distinguished by random-matrix and Poisson spectral statistics,
respectively~\cite{Evers2008,Xiao2022}.
Whether the overlap--level-correlation relation survives at localization
transitions and in intermediate regimes is an open question.
The source construction in {nonlinear} $\sigma$ models provides a
route to investigating such questions and the associated eigenvector
correlations~\cite{Ghosh2023}.

\section*{Acknowledgment}

We thank
Gernot Akemann,
Yan Fyodorov,
Kohei Kawabata and Yifei Liu for helpful discussions.
Z.C. and Z.X. thank Shuo Liu for helpful discussions.
The computations reported in this paper were performed using Princeton Research Computing resources.
Z.X. is supported by the Princeton
Quantum Initiative Fellowship. S.R. is supported by the
National Science Foundation under Award No.\ DMR-2409412.

The authors used OpenAI Codex (GPT-6 Astra) and Claude Code (Opus 5) for code development and debugging, checks of analytical steps, and manuscript revision.
The authors critically reviewed and independently verified all outputs through analytical and numerical checks, and retain full responsibility for this work.

\textup{Note added.}---
While we were finalizing this work,
Ref.~\cite{Fyodorov2026Parametric} appeared on arXiv, studying a closely
related but complementary quantity, the Petermann factor density
correlation. It also predicts the same bulk diagonal overlap for
class $\mathrm{AII}^{\dagger}$ as obtained here.
In Appendix~\ref{app:petermann_factor_density_correlation}, we compare
its results with ours and discuss the connection between the two works.

\onecolumngrid
\appendix
\section{Replica Derivation of the Class-A Diagonal Overlap}
\label{app:class-a-self}

The source partition function in Eq.~\eqref{eq:a-brief-partition} gives
the diagonal overlap through Eq.~\eqref{eq:a-brief-replica}. In this
appendix, we establish this distributional identity and evaluate it at
the leading bulk saddle point.

\subsection{Diagonal Overlap from the Resolvent}
\label{subsec:petermann_resolvent}

Let
\begin{equation}
  Z_{1,n}(z,\eta)
  =\left\langle
    \det\!\left[(z-H)(z-H)^\dagger+\eta^2\right]^n
  \right\rangle.
\end{equation}
We show that
\begin{equation}
  \pi\mathcal{O}_1(z)
  =\frac{1}{2N}\lim_{n\to0}
    \left.\partial_\eta^2 Z_{1,n}(z,\eta)\right|_{\eta=0}.
\end{equation}
There is no factor $1/n$, because the source derivative is taken before
the distributional replica limit. The first source derivative reads
\begin{equation}
  \partial_\eta Z_{1,n}(z,\eta)
  =2n\eta\left\langle
    \tr\,\left[((z-H)(z-H)^\dagger+\eta^2)^{-1}\right]
    \det\!\left[(z-H)(z-H)^\dagger+\eta^2\right]^n
  \right\rangle.
\end{equation}
The first derivative vanishes at $\eta=0$. Differentiating once more
and then setting the source to zero gives
\begin{align}
  &\left.\frac12\partial_\eta^2 Z_{1,n}(z,\eta)\right|_{\eta=0}
  =n\left\langle
    \tr\,\left[((z-H)(z-H)^\dagger)^{-1}\right]
    \det\!\left[(z-H)(z-H)^\dagger\right]^n
  \right\rangle
  \nonumber \\
  &\quad=n\Biggl\langle
    \sum_a\tr\,(\Pi_a\Pi_a^\dagger)
      |z-\lambda_a|^{2n-2}
      \prod_{c\ne a}|z-\lambda_c|^{2n}
    +\sum_{a\ne b}\tr\,(\Pi_a\Pi_b^\dagger)
      \frac{|z-\lambda_a|^{2n}|z-\lambda_b|^{2n}}
           {(z-\lambda_a)(\overline z-\overline\lambda_b)}
      \prod_{c\ne a,b}|z-\lambda_c|^{2n}
  \Biggr\rangle.
\end{align}
The products over the remaining eigenvalues tend to unity as $n\to 0$.
The diagonal term is selected by the distributional limits
\begin{equation}
  \lim_{n\to0}n|z-\lambda_a|^{2n-2}
  =\pi\delta^{(2)}(z-\lambda_a), \qquad \lim_{n\to0}n
    \frac{|z-\lambda_a|^{2n}|z-\lambda_b|^{2n}}
         {(z-\lambda_a)(\overline z-\overline\lambda_b)}
  =0,  a\ne b.
\end{equation}
Here, we assume $\lambda_a \neq \lambda_b$ for $a\neq b$, i.e., there
are no degeneracies other than those contained in the projectors.
{The exponents displayed above are those of rank-one eigenspaces. For
an eigenspace of rank $d$, the determinant contributes
$|z-\lambda_a|^{2dn}$ instead of $|z-\lambda_a|^{2n}$, and the
distributional limit becomes
$\lim_{n\to0}n|z-\lambda_a|^{2dn-2}=(\pi/d)\,\delta^{(2)}(z-\lambda_a)$.
The factor $1/d$ is precisely the normalization of $O_{aa}$ in
Eq.~\eqref{eq:general-class-rank-normalized-overlap}, so the identity below holds for
Kramers-degenerate classes as well.}
Consequently,
\begin{equation}
  \frac1{2N}\lim_{n\to0}
    \left.\partial_\eta^2 Z_{1,n}(z,\eta)\right|_{\eta=0}
  =\frac\pi N\left\langle
    \sum_a\tr\,(\Pi_a\Pi_a^\dagger)
      \delta^{(2)}(z-\lambda_a)
  \right\rangle
  =\pi\mathcal O_1(z).
\end{equation}

\subsection{\texorpdfstring{Nonlinear $\sigma$ Model}{Nonlinear sigma model}}
\label{subsec:nlsm_and_replica_limit}

We write $Z_{1,n}^{\mathrm A}(z,\eta)$ as a block determinant,
\begin{equation}
  Z_{1,n}^{\mathrm A}(z,\eta)
  :=\left\langle
    \det\!\left[\begin{pmatrix}
      z-H&-\eta\\
      \eta&\overline z-H^\dagger
    \end{pmatrix}\right]^n
  \right\rangle.
\end{equation}
The Hubbard--Stratonovich transformation gives
\begin{equation}
  Z_{1,n}^{\mathrm A}(z,\eta)
  =\int_{\mathbb C^{n\times n}}\,\mathrm dQ\,
    e^{-\tr\,(QQ^\dagger)}
    \det\!\left[\begin{pmatrix}
      z&-Q-\eta\\
      Q^\dagger+\eta&\overline z
    \end{pmatrix}\right]^N.
  \label{eq:a-brief-hs-after-grassmann}
\end{equation}
After the shift $Q\mapsto Q-\eta$, this becomes
\begin{equation}
  Z_{1,n}^{\mathrm A}(z,\eta)
  =\int_{\mathbb C^{n\times n}}\,\mathrm dQ\,
    e^{-\tr\,(QQ^\dagger)+\eta\,\tr\,(Q+Q^\dagger)-n\eta^2}
    \det\!\left[\begin{pmatrix}
      z&-Q\\
      Q^\dagger&\overline z
    \end{pmatrix}\right]^N.
\end{equation}
At the saddle point for $\eta=0$, $Q=qU$ with $U\in\mathrm U(n)$ and
$q^2=N-|z|^2$. Restricting the integral to the saddle-point manifold, we
obtain{, to leading order in large $N$ and up to source-independent
radial factors that tend to unity as $n\to0$,}
\begin{equation}
  Z_{1,n}^{\mathrm A}(z,\eta){\simeq}\int_{\mathrm{U}(n)}\,\mathrm dU\,
    \exp\,\left[-nq^2+\eta\,q\,\tr\,(U+U^\dagger)-n\eta^2\right].
\end{equation}

\subsection{Replica Limit}
\label{subsec:a-replica-limit}

To evaluate the second source derivative, we expand through order
$\eta^2$. Phase invariance under $U\mapsto e^{i\phi}U$ removes
$\tr\,U$, $\tr\,U^\dagger$, and their individual squares, while
$U\mapsto-U$ removes all odd powers of $\eta$. Thus,
\begin{equation}
  Z_{1,n}^{\mathrm A}(z,\eta)
  =e^{-nq^2}\int_{\mathrm{U}(n)}\,\mathrm dU\,\Bigl[
    1+\eta^2 q^2\tr\,(U)\tr\,(U^\dagger)-n\eta^2+\bigO(\eta^4)
  \Bigr].
\end{equation}
Using Eq.~\eqref{eq:buub}, we obtain
\begin{equation}
  Z_{1,n}^{\mathrm A}(z,\eta)=e^{-nq^2}\left[1+\eta^2(q^2-n)+\bigO(\eta^4)\right].
\end{equation}
This gives
\begin{equation}
  \pi\mathcal O_1(z)
  =\frac1{2N}\lim_{n\to0}
  \left.\partial_\eta^2 Z_{1,n}^{\mathrm A}(z,\eta)\right|_{\eta=0}
  =\frac{q^2}{N}=1-\frac{|z|^2}{N}.
\end{equation}

\section{Replica Derivation of the Bulk Off-Diagonal Overlap}
\label{app:class-a-pair}

In this appendix, we derive the bulk relation stated in
Sec.~\ref{sec:brief-class-a-o2}. The midpoint \(z\) and separation
\(\omega\) are defined in
Eq.~\eqref{eq:a-o2-variables}. We work to leading order in large \(N\),
with \(z\) in the bulk and \(\omega=\mathcal O(1)\). The replica
calculation first treats distinct points. The contribution at $z_1 = z_2$ is
restored in Sec.~\ref{subsubsec:recover_self_overlap}.

\subsection{Source Construction and Angular Integral}

We define the two-copy replica partition function
\begin{equation}
  Z_{2,n}^{\mathrm A}(\eta)
  :=\left\langle\det\!\left[D(\eta)\right]^n\right\rangle,
  \qquad
  D(\eta):=
  \begin{pmatrix}
    z_1-H & \eta & 0 & 0\\
    -\eta & \bar z_2-H^\dagger & 0 & 0\\
    0 & 0 & z_2-H & 0\\
    0 & 0 & 0 & \bar z_1-H^\dagger
  \end{pmatrix}.
  \label{eq:a-o2-replica-partition}
\end{equation}
Each block acts on the space of \(H\). The real source \(\eta\)
is set to zero after taking its second derivative.
The replica partition function generates the mixed resolvent
\begin{equation}
  \mathcal G(z_1,z_2)
  :=\frac1N\left\langle
    \tr\,\left[(z_1-H)^{-1}
      (\bar z_2-H^\dagger)^{-1}\right]\right\rangle =\frac1N\left\langle
    \sum_{a,b}
    \frac{O_{ab}}
    {(z_1-\lambda_a)(\bar z_2-\bar\lambda_b)}
    \right\rangle,
  \label{eq:a-o2-mixed-resolvent}
\end{equation}
through
\begin{equation}
  \mathcal G(z_1,z_2)
  =\frac1{2N}\lim_{n\to0}\frac1n
    \left.
    \partial_\eta^2
    Z_{2,n}^{\mathrm A}(\eta)
    \right|_{\eta=0}.
  \label{eq:a-o2-k-from-replica}
\end{equation}
The determinant identity is derived in Sec.~\ref{subsec:resolvent_from_replica_limit}.
Applying the two
derivatives to Eq.~\eqref{eq:a-o2-mixed-resolvent} gives
\begin{equation}
  \frac1{\pi^2}\partial_{\bar z_1}\partial_{z_2}
    \mathcal G(z_1,z_2)
  =\mathcal O_1(z_1)\delta^{(2)}(z_1-z_2)
   +\mathcal O_2(z_1,z_2).
  \label{eq:a-o2-k-contact}
\end{equation}
Consequently, away from the diagonal,
\begin{equation}
  \pi^2 \mathcal O_2(z_1,z_2)
  =
    \partial_{\bar z_1}\partial_{z_2}\mathcal G(z_1,z_2),
  \qquad z_1\ne z_2.
  \label{eq:a-o2-from-k}
\end{equation}

The saddle-point integral could be written in terms of the following group integral
\begin{equation}
  Y_n^{\mathrm A}(|\omega|^2)=\int_{\mathrm U(2n)}\mathrm dU\,
  \exp\,\left[
    \frac{|\omega|^2}{4}\tr\,(U^\dagger\Lambda U\Lambda)
  \right].
  \label{eq:a-o2-nlsm}
\end{equation}
Here, \(\mathrm dU\) is the normalized Haar measure, with
\(\int_{\mathrm U(2n)}\mathrm dU=1\) and hence \(Y_n^{\mathrm A}(0)=1\), and
$\Lambda = \begin{pmatrix}
  I_{n} & \\ & -I_n
\end{pmatrix}$.
The saddle calculation in Sec.~\ref{subsec:a-o2-nlsm-derivation} gives
\begin{align}
  \left.\frac12\partial_\eta^2 Z_{2,n}^{\mathrm A}(\eta)\right|_{\eta=0}
  &= \left(\frac{q^2}{2}-n-\frac{q^2}{n}\frac{\mathrm d}{\mathrm d|\omega|^2}\right)Y_n^{\mathrm A}(|\omega|^2), \label{eq:a-o2-source-insertion}
\end{align}
where $q^2 = N - |z|^2$.

\subsection{Self-Correlation Term and Overlap Sum Rule}
\label{subsubsec:recover_self_overlap}

Equation~\eqref{eq:a-o2-final} is obtained from the full Poisson
transform in Eq.~\eqref{eq:a-o2-poisson-class-a} for $|\omega|>0$. The same
separated-point result follows from its regular part, since the
self-correlation contributes $2/|\omega|^2$ and
$\mathrm d[|\omega|^2(2/|\omega|^2)]/\mathrm d|\omega|^2=0$ for $|\omega|>0$. To recover the self-correlation term
at $|\omega|=0$, we retain the contribution of the $\delta$ function in $R_2^{\mathrm{H}}$ in the form
\begin{equation}
  \label{eq:full_p2}
  \mathcal P_2(|\omega|^2)
  =1-\frac2{|\omega|^2}+\frac{2(1-e^{-|\omega|^2})}{|\omega|^4} + \frac{2}{|\omega|^2}\Theta(|\omega|^2) = \mathcal{P}_2^{\mathrm{regular}}(|\omega|^2) + \frac{2}{|\omega|^2}\Theta(|\omega|^2),
\end{equation}
where $\Theta(|\omega|^2)$ is the Heaviside step function.
The term \(2\Theta(|\omega|^2)/|\omega|^2\) comes from the Poisson transform of the
self-correlation \(\delta(x)\) in \(R_2^{\mathrm H}\).
While the Poisson integral in Eq.~\eqref{eq:def_p2} fixes the term to be \(2/|\omega|^2\) for \(|\omega|>0\), its value for $s\le 0$, where $s$ is the radial argument, is determined by the following argument which leads to Eq.~\eqref{eq:r_2_with_delta}.
\par
We use $\Theta'(s)=\delta_+(s)$, where $\delta_+$ is the one-sided
$\delta$ function defined by
\begin{equation}
  \int_0^\infty \delta_+(s) f(s)\,\mathrm ds=f(0).
\end{equation}
For the squared radius, $\delta_+(|\omega|^2)=\pi\delta^{(2)}(\omega)$.
Eq.~\eqref{eq:full_p2} then
gives the eigenvalue self-correlation,
\begin{equation}
  \pi^2 R_2(\omega)
  =\frac12\frac{\mathrm d^2}{\mathrm d(|\omega|^2)^2}
    [|\omega|^4\mathcal P_2(|\omega|^2)]
  =\pi^2R_2^{\mathrm{regular}}(\omega)+\pi\delta^{(2)}(\omega).
  \label{eq:r_2_with_delta}
\end{equation}
The first term is the contribution of distinct eigenspaces to $R_2$, and
the second term is the $\delta$ function at $\omega=0$. Here, we use the
normalization $\pi R_1 = 1$. Before unfolding, the exact
$\delta$-function term is \(R_1(z_1)\delta^{(2)}(z_1-z_2)\). In the same
way, we recover the $\delta$-function term in the eigenvector correlation
by inserting the full $\mathcal P_2$ [Eq.~\eqref{eq:full_p2}] into
Eq.~\eqref{eq:a-o2-poisson-relation},
\begin{align}
   &\phantom{{}={}} \pi^2 \mathcal O_2(z_1,z_2) + \pi^2 \mathcal O_1(z_1)\delta^{(2)}(z_1-z_2)
   \nonumber \\
   &= \frac{1}{2}\left(1-\frac{|z|^2}{N}\right)
  \left\{-1+\frac{\mathrm d}{\mathrm d|\omega|^2}
    \bigl[|\omega|^2\mathcal P_2(|\omega|^2)\bigr]\right\}
    \nonumber \\
    &= \pi^2 \mathcal O_2(z_1,z_2) + \left(1-\frac{|z|^2}{N}\right)\pi \delta^{(2)}(z_1 - z_2).
  \label{eq:full_o2_from_p2}
\end{align}
The coefficient of the self-correlation term is therefore
\begin{equation}
  \pi \mathcal O_1(z) = 1-\frac{|z|^2}{N}.
\end{equation}

\subsubsection{Sum Rule}

The self-correlation coefficient could also be determined by the overlap sum rule.
The sum rule for $R_2$ is given by
\begin{equation}
  \int_{\mathbb C}\mathrm d^2\omega\,
    [\pi^2 R_2(\omega)-1] =\pi^2R_1+
    \int_{\mathbb C}\mathrm d^2\omega\,
    [\pi^2R_2^{\mathrm{regular}}(\omega)-1]=0.
\end{equation}
Thus, the regular part determines the weight of the self-correlation.
The same mechanism relates \(\mathcal O_2\) to \(\mathcal O_1\). We first
establish a sum rule for $\mathcal{P}_2$. The argument uses the sum rule for $R_2$ but not the explicit class-A form of $\mathcal P_2$.
It therefore applies to all three bulk families.
\par
We assume the asymptotic behavior
\(\mathcal P_2(|\omega|^2)-1=a/|\omega|^2+o(|\omega|^{-2})\) at large $|\omega|$ and no boundary
contribution at \(|\omega|^2=0\) beyond the step-function term. The sum rule for
the full \(R_2\) then implies
\begin{equation}
\int_0^\infty \mathrm ds\,
\left\{-1+\frac{\mathrm d}{\mathrm ds}[s\mathcal P_2(s)]\right\}
=0,
\label{eq:p_sum_rule}
\end{equation}
where $\mathcal{P}_2$ contains the step-function term. Indeed,
\begin{equation}
  \pi^2R_2^{\mathrm{regular}}(\omega)-1
=\frac12\,\frac{\mathrm d^2}{\mathrm d(|\omega|^2)^2}
[|\omega|^4(\mathcal P_2^{\mathrm{regular}}(|\omega|^2)-1)]
\end{equation}
together with the screening identity implies
\(|\omega|^2(\mathcal P_2^{\mathrm{regular}}-1)\to-2\). Thus, the regular part of
the integrand in Eq.~\eqref{eq:p_sum_rule} contributes \(-2\), whereas
the step-function term \(2\delta_+(s)\) contributes \(+2\). This proves
Eq.~\eqref{eq:p_sum_rule} without using the explicit formula for
$\mathcal{P}_2$.
\par
We next apply Eq.~\eqref{eq:p_sum_rule} to the overlaps.
The sum rule for $\mathcal{O}_2$ is given by
Eq.~\eqref{eq:definitions-overlap-sum-rule},
\begin{equation}
  \int_{\mathbb C}\mathrm d^2z_2\,\mathcal O_2(z_1,z_2)
  =\frac{1}{N}R_1(z_1)-\mathcal O_1(z_1).
  \label{eq:a-o2-exact-sum-rule}
\end{equation}
With our normalization, the sum rule in the large-$N$ limit becomes
\begin{equation}
  \label{eq:o2-large-N-sum-rule}
  \int_{\mathbb C}\mathrm d^2z_2\, \left( \mathcal O_2(z_1,z_2) + \mathcal O_1(z_1) \delta^{(2)}(z_1 - z_2)\right) = \bigO\!\left(\frac{1}{N}\right).
\end{equation}
The integrand is given by Eq.~\eqref{eq:full_o2_from_p2},
\begin{align}
    \phantom{{}={}} \pi^2 \mathcal O_2(z_1,z_2) + \pi^2 \mathcal O_1(z_1)\delta^{(2)}(z_1-z_2) &= \frac{1}{2}\left(1-\frac{|z_1|^2}{N} + \bigO\!\left(\frac{1}{N}\right)\right)
  \left\{-1+\frac{\mathrm d}{\mathrm d|\omega|^2}
    \bigl[|\omega|^2\mathcal P_2(|\omega|^2)\bigr]\right\}.
\end{align}
{Here, we have replaced the midpoint prefactor $1-|z|^2/N$ by
$1-|z_1|^2/N$. Pointwise the difference is $\mathrm{Re}(z_1\bar\omega)/N=\bigO(N^{-1/2})$
for $|z_1|=\bigO(\sqrt N)$, but this term is odd under $\omega\to-\omega$ and
vanishes upon integration over $z_2$ at fixed $z_1$, since the bracket depends
only on $|\omega|$.}
Applying the sum rule for $\mathcal{P}_2$ [Eq.~\eqref{eq:p_sum_rule}],
we find that Eq.~\eqref{eq:full_o2_from_p2} is consistent with the sum
rule in Eq.~\eqref{eq:o2-large-N-sum-rule} up to $\bigO(1/N)$ terms.

\subsection{Resolvent from the Replica Limit}
\label{subsec:resolvent_from_replica_limit}

Equation~\eqref{eq:a-o2-k-from-replica} follows from the block structure
of $D(\eta)$ in Eq.~\eqref{eq:a-o2-replica-partition}:
\begin{align}
  \det D(\eta)
  &=\det\!\begin{pmatrix}
    z_1-H&\eta\\
    -\eta&\bar z_2-H^\dagger
  \end{pmatrix}
  \det(z_2-H)\det(\bar z_1-H^\dagger)
  \label{eq:a-o2-source-block-form} \\
  &={}\det(z_2-H)\det(\bar z_1-H^\dagger)\times\det\!\left[
    (\bar z_2-H^\dagger)(z_1-H)+\eta^2
  \right]\nonumber\\
  &={}\det(z_1-H)\det(z_2-H)
    \det(\bar z_1-H^\dagger)\det(\bar z_2-H^\dagger)\times\det\!\left[
    I+\eta^2(z_1-H)^{-1}
      (\bar z_2-H^\dagger)^{-1}
  \right].
  \label{eq:a-o2-source-block-determinant}
\end{align}
Thus, the determinant is an even function of the source, depending only
on \(\eta^2\), and
\begin{equation}
  \left.\frac12\partial_\eta^2
    [\det D(\eta)]^n\right|_{\eta=0}
  =n[\det D(0)]^n
  \tr\,\left[(z_1-H)^{-1}
    (\bar z_2-H^\dagger)^{-1}\right].
  \label{eq:a-o2-source-derivative}
\end{equation}
It follows that
\begin{equation}
  \frac1{2N}\lim_{n\to0}\frac1n
    \left.\partial_\eta^2
      Z_{2,n}^{\mathrm A}(\eta)\right|_{\eta=0} =\frac1N\lim_{n\to0}\left\langle
    [\det D(0)]^n
    \tr\,\left[(z_1-H)^{-1}
      (\bar z_2-H^\dagger)^{-1}\right]
    \right\rangle
  =\mathcal G(z_1,z_2).
  \label{eq:a-o2-direct-replica-limit}
\end{equation}
Here, \([\det D(0)]^n\to1\) as \(n\to0\), so no normalization by
\(Z_{2,n}^{\mathrm A}(0)\) is needed.  This proves
Eq.~\eqref{eq:a-o2-k-from-replica}.

\subsection{\texorpdfstring{Nonlinear $\sigma$ Model}{Nonlinear sigma model}}
\label{subsec:a-o2-nlsm-derivation}

Grouping the $z_1-H$ and $z_2-H$ blocks together, and likewise their
conjugate blocks, the Hubbard--Stratonovich transformation introduces
a $2n\times 2n$ complex matrix field $Q$ and gives
\begin{equation}
  Z_{2,n}^{\mathrm A}(\eta)
  =\int_{\mathbb{C}^{2n\times 2n}} \mathrm dQ\,e^{-\tr\,(QQ^\dagger)}
    \det\!\left[\begin{pmatrix}
      \begin{pmatrix}
        z_1&0\\0&z_2
      \end{pmatrix}
      &
      \begin{pmatrix}
        0&\eta\\0&0
      \end{pmatrix}-Q\\[1.2ex]
      Q^\dagger-
      \begin{pmatrix}
        0&0\\\eta&0
      \end{pmatrix}
      &
      \begin{pmatrix}
        \bar z_1&0\\0&\bar z_2
      \end{pmatrix}
    \end{pmatrix}\right]^N.
  \label{eq:a-o2-hs-representation}
\end{equation}
We need the expansion through second order in $\eta$. Shifting $Q$
and $Q^\dagger$ together gives
\begin{equation}
  Z_{2,n}^{\mathrm A}(\eta)
  ={}\int \mathrm dQ\,
    \exp\,\Biggl\{-\tr\,\left[
      \left(Q+\eta\begin{pmatrix}0&I_n\\0&0\end{pmatrix}\right)
      \left(Q^\dagger+\eta\begin{pmatrix}0&0\\I_n&0\end{pmatrix}\right)
    \right]\Biggr\} \times
    \det\!\left[\begin{pmatrix}
      \begin{pmatrix}
        z_1&0\\0&z_2
      \end{pmatrix}
      &
      -Q\\[1.2ex]
      Q^\dagger
      &
      \begin{pmatrix}
        \bar z_1&0\\0&\bar z_2
      \end{pmatrix}
    \end{pmatrix}\right]^N.
  \label{eq:a-o2-hs-shifted-representation}
\end{equation}
The shifted Gaussian factor is $e^{-\tr(QQ^\dagger)}F(Q)$, where
\begin{equation}
  F(Q)=\exp\,\Biggl\{
    -\eta\tr\,\left[
      \begin{pmatrix}0&I_n\\0&0\end{pmatrix}Q^\dagger
      +Q\begin{pmatrix}0&0\\I_n&0\end{pmatrix}
    \right]-n\eta^2\Biggr\}.
  \label{eq:a-o2-shifted-gaussian-eta}
\end{equation}
The partition function
$Z_{2,n}^{\mathrm A}(\eta)$ then takes the form of an expectation value,
\begin{equation}
  Z_{2,n}^{\mathrm A}(\eta)
  ={}\int \mathrm dQ\,e^{-\tr\,(QQ^\dagger)}F(Q) \times\det\!\left[\begin{pmatrix}
      \begin{pmatrix}
        z_1&0\\0&z_2
      \end{pmatrix}
      &
      -Q\\[1.2ex]
      Q^\dagger
      &
      \begin{pmatrix}
        \bar z_1&0\\0&\bar z_2
      \end{pmatrix}
    \end{pmatrix}\right]^N.
  \label{eq:z_n_as_exp_val}
\end{equation}
This is equal to the partition function for level correlations with the
insertion of $F(Q)$. The technique is therefore the same. The saddle
point is determined by the action without the source insertion $F(Q)$.
We then write $Z_{2,n}^{\mathrm A}$ as an integral over the saddle-point manifold, with
$F(Q)$ evaluated there.
\par
At \(\eta=\omega=0\), the bulk saddle point of the radial action is
\begin{equation}
  Q=qU,
  \qquad
  q^2=N-|z|^2,
  \qquad
  U\in \mathrm U(2n).
  \label{eq:a-o2-bulk-saddle}
\end{equation}

For eigenvector correlations, the radial saddle point must retain its
dependence on the midpoint. In the replica calculation of level
correlations, one can set $q^2 = N$, since the $|z|^2$ correction does
not matter in the large-$N$ limit. For eigenvector correlations, by
contrast, the source insertion retains \(q^2=N-|z|^2\), which gives the
prefactor \(1-|z|^2/N\). The radial calculation below preserves this
dependence.
\par
We first set $\eta=0$, so that $F(Q)=1$. The source insertion is
restored through Eq.~\eqref{eq:z_n_restore} after writing the integral on the saddle-point manifold.
We perform the polar decomposition
$Q = (q+R)U$ with Hermitian $R$ and integrate over $R$.
{At leading order in large $N$, the Jacobian of this decomposition
and the restriction $q+R>0$ contribute only source-independent factors,
so we replace the measure by $\mathrm dR\,\mathrm dU$ and extend the
$R$ integral to all Hermitian matrices.}
\par
After the polar decomposition, $Z_{2,n}^{\mathrm A}$ becomes
\begin{equation}
  Z_{2,n}^{\mathrm A}(0)
  =
  \int_{\mathrm U(2n)}\mathrm dU
  \int_{\operatorname{Herm}(2n)}\mathrm dR\,\mathscr{I}(U,R),
\end{equation}
where the integrand is
\begin{align}
  \mathscr{I}(U,R) &:= e^{-\tr\,(QQ^\dagger)} \det\!\left[\begin{pmatrix}
    z+\omega\Lambda/2 & -(q+R)U \\
    U^\dagger(q+R) & \bar z+\bar\omega\Lambda/2
  \end{pmatrix}\right]^N \notag\\*
  &\quad=
  e^{-\tr\,\left[(q+R)^2\right]} \det\!\left[
    \begin{pmatrix}
      z&-q\\q&\bar z
    \end{pmatrix}
    +
    \begin{pmatrix}
      \omega\Lambda/2&-R\\
      R&\bar\omega U\Lambda U^\dagger/2
    \end{pmatrix}
  \right]^N.
\end{align}
Expanding the logarithm to quadratic order in $R$ and $\omega$, and
using $q^2+|z|^2=N$, gives
\begin{align}
  \mathscr{I}(U,R)
  &={}
  \exp\,\Biggl\{
    -2nq^2+N\tr\,\log\!\begin{pmatrix}z&-q\\q&\bar z\end{pmatrix}
    -\frac{n}{4N}(\bar z^2\omega^2+z^2\bar\omega^2)
  \Biggr\}\nonumber\\
  &\quad\times\exp\,\Biggl\{
    -\frac{2q^2}{N}\tr\,(R^2)
    -\frac{q}{N}\tr\,\left[
      R(\bar z\omega\Lambda+z\bar\omega U\Lambda U^\dagger)\right]
    +\frac{q^2|\omega|^2}{4N}\tr\,(U\Lambda U^\dagger\Lambda)
  \Biggr\}.
  \label{eq:integrand}
\end{align}
Completing the square in the radial terms,
\begin{align}
  &-\frac{2q^2}{N}\tr\,(R^2)
   -\frac{q}{N}\tr\,[R(\bar z\omega\Lambda+z\bar\omega U\Lambda U^\dagger)]
   \nonumber\\
  &\quad=-\frac{2q^2}{N}\tr\,\left[
    \left(R+\frac{\bar z\omega\Lambda+z\bar\omega U\Lambda U^\dagger}{4q}\right)^2
  \right]
  +\frac{1}{8N}\tr\,[(\bar z\omega\Lambda+z\bar\omega U\Lambda U^\dagger)^2],
\end{align}
we integrate over $R$. The remaining square satisfies
\begin{equation}
  \tr\,[(\bar z\omega\Lambda+z\bar\omega U\Lambda U^\dagger)^2]
  =2n(\bar z^2\omega^2+z^2\bar\omega^2)
   +2|z|^2|\omega|^2\tr\,(U\Lambda U^\dagger\Lambda).
\end{equation}
The first term cancels the scalar $\omega$ dependence in
Eq.~\eqref{eq:integrand}, while the second changes the angular coefficient
from $q^2|\omega|^2/(4N)$ to $|\omega|^2/4$. Thus,
\begin{equation}
  Z_{2,n}^{\mathrm A}(0)
  =\exp\,\Biggl\{
    -2nq^2+N\tr\,\log\!\begin{pmatrix}z&-q\\q&\bar z\end{pmatrix}
  \Biggr\}
  \int_{\mathrm U(2n)}\mathrm dU\,
    e^{\frac{|\omega|^2}{4}\tr(U\Lambda U^\dagger\Lambda)}.
\end{equation}
Factors that are independent of the source and tend to unity as
$n\to0$ can be omitted at this order. This differs from the calculation
of level correlations, because we take derivatives with respect to
$\eta$ rather than $\omega$. This leaves $Z_{2,n}^{\mathrm A}(0)=Y_n^{\mathrm A}(|\omega|^2)$.
The angular action is the same as for \(q^2=N\), and the dependence on
\(q\) remains only in the source insertion. From
Eq.~\eqref{eq:z_n_as_exp_val}, we obtain
\begin{align}
  Z_{2,n}^{\mathrm A}(\eta)
  &=
  \int_{\mathrm U(2n)}\mathrm dU\,
  \exp\,\left\{
    \frac{|\omega|^2}{4}\tr\,(\Lambda U\Lambda U^\dagger)
  \right\}F(qU). \label{eq:z_n_restore}
\end{align}
To extract the source derivative, we expand $F(qU)$ through order
$\eta^2$.
The Haar measure and angular weight are invariant under the independent
block-diagonal phase rotations
\begin{equation}
  U\longmapsto
  \begin{pmatrix}e^{i\phi}I_n&0\\0&I_n\end{pmatrix}
  U
  \begin{pmatrix}I_n&0\\0&e^{i\psi}I_n\end{pmatrix}.
\end{equation}
The two source traces acquire opposite phases
$e^{-i(\phi+\psi)}$ and $e^{i(\phi+\psi)}$. Hence, the linear terms
and the squares of either trace integrate to zero, while their product
survives. The symmetry $U\mapsto-U$ also removes all odd powers of
$\eta$. Therefore,
\begin{align}
  \left.\frac12\partial_\eta^2 Z_{2,n}^{\mathrm A}(\eta)\right|_{\eta=0}
  &=
  \int_{\mathrm U(2n)}\mathrm dU\,
  \exp\,\left\{
    \frac{|\omega|^2}{4}\tr\,(\Lambda U\Lambda U^\dagger)
  \right\}
    \times\left\{
      q^2\tr\,\left[
        \begin{pmatrix}0&I_n\\0&0\end{pmatrix}U^\dagger
      \right]
      \tr\,\left[
        U\begin{pmatrix}0&0\\I_n&0\end{pmatrix}
      \right]-n\right\}.
\end{align}
The source insertion thus supplies the required prefactor
$q^2=N-|z|^2$. The remaining angular integral is fixed by a Ward
identity. In Sec.~\ref{sec:a-o2-y-branch-and-angular-identity}, we show
that
\begin{equation}
  \left.\frac12\partial_\eta^2 Z_{2,n}^{\mathrm A}(\eta)\right|_{\eta=0}
  =\left(\frac{q^2}{2}-n-\frac{q^2}{n}\frac{\mathrm d}{\mathrm d|\omega|^2}\right)
    Y_n^{\mathrm A}(|\omega|^2).
  \label{eq:a-o2-unitary-insertion-identity}
\end{equation}
This gives Eq.~\eqref{eq:a-o2-source-insertion}.

\subsection{Replica Limit}
\label{subsec:pair_correlation_replica_limit}

Combining Eqs.~\eqref{eq:a-o2-k-from-replica}
and~\eqref{eq:a-o2-source-insertion} gives
\begin{equation}
  \mathcal G(z_1,z_2)
  =\left(1-\frac{|z|^2}{N}\right)
   \lim_{n\to0}\frac1{n^2}
   \left[\frac n2Y_n^{\mathrm A}(|\omega|^2)
     -\frac{\mathrm dY_n^{\mathrm A}(|\omega|^2)}{\mathrm d|\omega|^2}\right]
   +\bigO(N^{-1}).
\end{equation}
Here, the term $-Y_n^{\mathrm A}/N$ is subleading. The replica limit
can be taken only after the derivatives
$\partial_{\bar z_1}\partial_{z_2}$.

We hold the midpoint \(z\) fixed. For any
function \(f(|\omega|^2)\),
\(\partial_{\bar z_1}\partial_{z_2}f(|\omega|^2)
=-\frac{\mathrm d}{\mathrm d|\omega|^2}[|\omega|^2 f'(|\omega|^2)]\).  Carrying out the differentiation in
Eq.~\eqref{eq:a-o2-from-k} therefore gives
\begin{equation}
  \pi^2 \mathcal O_2(z_1,z_2)
  =\left(1-\frac{|z|^2}{N}\right)\frac{\mathrm d}{\mathrm d|\omega|^2}\!\left[
    |\omega|^2\lim_{n\to0}\left(
      \frac1{n^2}\frac{\mathrm d^2Y_n^{\mathrm A}(|\omega|^2)}{\mathrm d(|\omega|^2)^2}
      -\frac1{2n}\frac{\mathrm dY_n^{\mathrm A}(|\omega|^2)}{\mathrm d|\omega|^2}
    \right)
  \right],
  \qquad z_1\ne z_2.
  \label{eq:a-o2-direct-y}
\end{equation}
The differentiated expression can be evaluated with the known replica limits for $Y_n^{\mathrm A}(|\omega|^2)$. In Ref.~\cite{ChenXiaoLiuRyu2026}, the
replica limit of the same \(Y_n^{\mathrm A}\) is related to
\(\mathcal P_2\) by
\begin{equation}
  \mathcal P_2(|\omega|^2)
  =\frac12+2\lim_{n\to0}\frac1{n^2}
    \frac{\mathrm d^2Y_n^{\mathrm A}(|\omega|^2)}{\mathrm d(|\omega|^2)^2}.
  \label{eq:a-o2-poisson-replica}
\end{equation}
Equation~\eqref{eq:a-o2-direct-y} also requires the term linear in $n$ in $Y_n^{\mathrm A}(|\omega|^2)$.
The argument in Sec.~\ref{subsubsec:linear_term} shows that
\(Y_n^{\mathrm A}(|\omega|^2)=1+cn|\omega|^2+\bigO(n^2)\) for \(|\omega|>0\), where \(c\) is a constant
not fixed by that argument.
Consequently
\begin{equation}
  \lim_{n\to0}\left[
    \frac1{n^2}\frac{\mathrm d^2Y_n^{\mathrm A}(|\omega|^2)}{\mathrm d(|\omega|^2)^2}
    -\frac1{2n}\frac{\mathrm dY_n^{\mathrm A}(|\omega|^2)}{\mathrm d|\omega|^2}
  \right]
  =\frac12\left[\mathcal P_2(|\omega|^2)-\frac12-c\right].
  \label{eq:a-o2-unlogged-replica}
\end{equation}
Substituting Eq.~\eqref{eq:a-o2-unlogged-replica} into
Eq.~\eqref{eq:a-o2-direct-y} gives
\begin{equation}
  \pi^2\mathcal O_2(z_1,z_2)
  =\frac{1}{2}\left(1-\frac{|z|^2}{N}\right)
    \left[\frac{\mathrm d}{\mathrm d|\omega|^2}\!\left[ |\omega|^2 \mathcal{P}_2(|\omega|^2) \right]-\frac12-c\right],
  \qquad z_1\ne z_2.
  \label{eq:a-o2-clustering-with-c}
\end{equation}
The asymptotic behavior of $\mathcal{P}_2$ gives \(\frac{\mathrm d}{\mathrm d|\omega|^2}\!\left[ |\omega|^2 \mathcal{P}_2(|\omega|^2) \right]\to1\) as \(|\omega|^2\to\infty\).
Since the off-diagonal overlap must also vanish at large
separation, \(\mathcal O_2(z_1,z_2)\to0\), and \(z\) lies in the bulk,
Eq.~\eqref{eq:a-o2-clustering-with-c} fixes
\begin{equation}
  c=\frac12.
  \label{eq:a-o2-clustering-fixes-c}
\end{equation}
Therefore, we obtain
\begin{align}
  \pi^2 \mathcal O_2(z_1,z_2) &=\frac{1}{2}\left(1-\frac{|z|^2}{N}\right)
  \left\{-1+\frac{\mathrm d}{\mathrm d|\omega|^2}
    \bigl[|\omega|^2\mathcal P_2(|\omega|^2)\bigr]\right\},
  \qquad z_1\ne z_2.
\end{align}

\subsubsection{Linear Replica Term from the Hermitian Level Correlation}
\label{subsubsec:linear_term}

Here, we justify the expansion \(Y_n^{\mathrm A}(|\omega|^2)=1+cn|\omega|^2+\bigO(n^2)\) for some constant $c$.
The
integral \(Y_n^{\mathrm A}(|\omega|^2)\) is the analytic continuation of the compact
two-level replica partition function of the Gaussian unitary ensemble, and its replica limit
for level correlations is Eq.~\eqref{eq:a-o2-poisson-replica}.
For $|\omega|>0$, we write
\begin{equation}
  Y_n^{\mathrm A}(|\omega|^2)=1+n g(|\omega|^2)+\bigO(n^2).
  \label{eq:a-o2-proof-general-replica-expansion}
\end{equation}
Then,
\begin{equation}
  \frac1{n^2}\frac{\mathrm d^2Y_n^{\mathrm A}(|\omega|^2)}{\mathrm d(|\omega|^2)^2}
  =\frac{g''(|\omega|^2)}{n}+\bigO(1).
  \label{eq:a-o2-proof-gue-finiteness}
\end{equation}
Since the replica limit in
Eq.~\eqref{eq:a-o2-poisson-replica} is finite at fixed \(|\omega|>0\), we need
\(g''(|\omega|^2)=0\). The normalized Haar measure gives \(Y_n^{\mathrm A}(0)=1\), so
\(g(0)=0\), and therefore \(g(|\omega|^2)=c|\omega|^2\) with a constant \(c\). Thus,
\begin{equation}
  Y_n^{\mathrm A}(|\omega|^2)=1+cn|\omega|^2+\bigO(n^2).
  \label{eq:a-o2-proof-linear-replica-term}
\end{equation}

\subsection{Angular Ward Identity}
\label{sec:a-o2-y-branch-and-angular-identity}

Let $B$ denote the upper-right $n\times n$ block of $U$. The two source
traces in Sec.~\ref{subsec:a-o2-nlsm-derivation} are $\tr\,B^\dagger$
and $\tr\,B$. The Haar measure and angular weight are invariant under
$U\mapsto\operatorname{diag}(V,I_n)U$ for any $V\in\mathrm U(n)$,
which sends $B\mapsto VB$. Averaging over $V$ and applying
Eq.~\eqref{eq:buub} gives
\begin{equation}
  \int_{\mathrm U(n)}\mathrm dV\,
    \tr\,(B^\dagger V^\dagger)\tr\,(VB)
  =\frac1n\tr\,(B^\dagger B).
  \label{eq:u_n_orthogonal_decomposition}
\end{equation}
Since
\begin{equation}
  \tr\,(B^\dagger B)
  =\frac14\tr\,[(I_{2n}-\Lambda)U^\dagger(I_{2n}+\Lambda)U]
  =\frac14\tr\,[I_{2n}-\Lambda U^\dagger\Lambda U],
\end{equation}
the source contraction reduces to
\begin{align}
  &\int_{\mathrm U(2n)}\mathrm dU\,
    e^{\frac{|\omega|^2}{4}\tr(\Lambda U\Lambda U^\dagger)}
    \tr\,(B^\dagger)\tr\,B\nonumber\\
  &\quad=\frac1n\int_{\mathrm U(2n)}\mathrm dU\,
    e^{\frac{|\omega|^2}{4}\tr(\Lambda U\Lambda U^\dagger)}
    \tr\,(B^\dagger B)
  =\left(\frac12-\frac1n\frac{\mathrm d}{\mathrm d|\omega|^2}\right)
    Y_n^{\mathrm A}(|\omega|^2).
\end{align}
This proves Eq.~\eqref{eq:a-o2-unitary-insertion-identity}.

\subsection{Extension to the Other Two Bulk Classes}
\label{subsec:pair-source-three-classes}

For class $\mathrm{AI}^{\dagger}$ we average the determinant in
Eq.~\eqref{eq:a-o2-replica-partition} over complex-symmetric matrices;
for class $\mathrm{AII}^{\dagger}$ we use the Pfaffian in
Eq.~\eqref{eq:general-class-pair-partition}. In both cases, the source
couples the first spectral sector to the conjugate of the second.
We keep $s=|\omega|^2$ and $q^2=N-|z|^2$.
Define the parameters for the three classes as follows.
\begin{equation}
\begin{array}{c|ccccc}
 X&G_X&m_X&c_X&a_X&p_X\\ \hline
 \mathrm A&\mathrm U(2n)&n&1&1/4&N\\
 \mathrm{AI}^{\dagger}&\mathrm{Sp}(2n)&2n&1/2&1/8&N/2\\
 \mathrm{AII}^{\dagger}&\mathrm O(2n)&n&1&1/4&N
\end{array}
 \label{eq:pair-three-class-parameters}
\end{equation}
Let $\Lambda_X=\operatorname{diag}(I_{m_X},-I_{m_X})$.
The source-free angular
integrals are~\cite{ChenXiaoLiuRyu2026}
\begin{equation}
 Y_n^X(s)=\int_{G_X}dU\,e^{a_XsT_X(U)},\qquad
 T_X(U)=\tr(\Lambda_XU\Lambda_XU^\dagger).
 \label{eq:pair-three-class-angular}
\end{equation}
In the symplectic case we order the basis by spectral sector,
so each sector contains a quaternion space of dimension $n$ and its
complex representation has dimension $m_X=2n$.

The Gaussian average introduces a complex, quaternion-real, or real
auxiliary field $Q$, respectively~\cite{Kulkarni2025}.
Write $\mathsf Z_X=\operatorname{diag}(z_1I_{m_X},z_2I_{m_X})$ and
$E_X=\left(\begin{smallmatrix}0&I_{m_X}\\0&0\end{smallmatrix}\right)$.
The source-dependent auxiliary-field integral is
\begin{equation}
 Z_{2,n}^X(\eta)\propto\int dQ\,e^{-c_X\tr(QQ^\dagger)}
 \det{}^{p_X}\!\begin{pmatrix}
 \mathsf Z_X&\eta E_X-Q\\
 Q^\dagger-\eta E_X^\dagger&\overline{\mathsf Z}_X
 \end{pmatrix}.
 \label{eq:pair-three-class-hs-source}
\end{equation}
The proportionality constant is source independent. In the symplectic
case, the half-integer determinant power denotes the Pfaffian branch
fixed by the original fermionic integral. Translating
$Q\mapsto Q+\eta E_X$ removes the source from the determinant and
stays within each of the three auxiliary-field spaces.
The resulting Gaussian factor is
\begin{align}
 F_X(Q,\eta)
 &=\exp\!\left\{-c_X\eta\tr(E_XQ^\dagger+QE_X^\dagger)
              -c_X\eta^2\tr(E_XE_X^\dagger)\right\},
 \label{eq:pair-three-class-source-shift}\\
 F_X(qU,\eta)
 &=\exp\!\left[-c_Xq\eta\tr(B+B^\dagger)-n\eta^2\right],
 \qquad B=U_{12}.
 \label{eq:pair-three-class-source-factor}
\end{align}
In particular, $c_X\tr(E_XE_X^\dagger)=c_Xm_X=n$ in every class.
The reduced Pfaffian construction in class $\mathrm{AII}^{\dagger}$
uses a real $2n\times2n$ field for the two spectral sectors, rather
than the $4n\times4n$ field of full determinant replicas.
As in class A, source-independent radial factors are omitted and the
source insertion is evaluated at the leading bulk saddle.

For class A, the block phase average removes $(\tr B)^2$ and its
complex conjugate. For the other two classes, $B$ is quaternion-real
or real, respectively, so $\tr B^\dagger=\tr B$. Expanding
Eq.~\eqref{eq:pair-three-class-source-factor} thus gives
\begin{equation}
 \left.\frac12\partial_\eta^2Z_{2,n}^X\right|_0
 \simeq\int_{G_X}dU\,e^{a_XsT_X}
 \left[q^2\mathcal S_X(B)-n\right],\qquad
 \mathcal S_X(B)=
 \begin{cases}
 |\tr B|^2,&X=\mathrm A,\\
 \tfrac12(\tr B)^2,&X=\mathrm{AI}^{\dagger},\\
 2(\tr B)^2,&X=\mathrm{AII}^{\dagger}.
 \end{cases}
 \label{eq:pair-three-class-insertion}
\end{equation}
The coefficients here follow from the Gaussian source shift, before
performing any angular average.

The weight in Eq.~\eqref{eq:pair-three-class-angular} is invariant
under left multiplication by $\operatorname{diag}(V,I_{m_X})$, with
$V$ in $\mathrm U(n)$, $\mathrm{Sp}(n)$, or $\mathrm O(n)$,
respectively. The contractions in Appendix~\ref{app:schur} give
\begin{equation}
 \int dV\,\mathcal S_X(VB)=
 \begin{cases}
 \tr(BB^\dagger)/n,&X=\mathrm A,\\
 \tr(BB^\dagger)/(4n),&X=\mathrm{AI}^{\dagger},\\
 2\tr(BB^T)/n,&X=\mathrm{AII}^{\dagger}.
 \end{cases}
 \label{eq:pair-three-class-block-average}
\end{equation}
Unitarity, with the same block dimensions as above, implies
\begin{equation}
 \tr(BB^\dagger)=\frac{m_X}{2}-\frac14T_X(U),\qquad
 \partial_sY_n^X=a_X\int_{G_X}dU\,T_Xe^{a_XsT_X}.
 \label{eq:pair-three-class-trace-derivative}
\end{equation}
Combining these equations yields the source reduction, with
$(\beta_{\mathrm A},\beta_{\mathrm{AI}^{\dagger}},
\beta_{\mathrm{AII}^{\dagger}})=(2,1,4)$,
\begin{equation}
 \left.\frac12\partial_\eta^2Z_{2,n}^X\right|_0
 \simeq\left[\frac{\beta_Xq^2}{4}-n
       -\frac{\beta_Xq^2}{2n}\partial_s\right]Y_n^X(s).
 \label{eq:pair-three-class-source-reduction}
\end{equation}
For example, the symplectic row gives
$q^2[1/4-T_X/(16n)]-n$ before integration, and
$a_X=1/8$ converts this to $q^2/4-n-q^2\partial_s/(2n)$.
The group contractions are exact for positive integer $n$; the symbol
$\simeq$ refers to the leading large-$N$ saddle approximation.

With one replica per eigenvalue or Kramers pair, the spectral replica
matching in all three classes reads~\cite{ChenXiaoLiuRyu2026}
\begin{equation}
 \mathcal P_2^X(s)=\frac12+
 2\lim_{n\to0}\frac{\partial_s^2Y_n^X(s)}{n^2}.
 \label{eq:pair-three-class-poisson-matching}
\end{equation}
Using Eq.~\eqref{eq:main-pair-source-extraction}, taking the spectral
derivatives before the replica limit, and retaining the leading
$q^2/N$ term gives
\begin{equation}
 \pi^2\mathcal O_2^X
 =\frac{\beta_Xq^2}{2N}\partial_s\!\left\{
 s\lim_{n\to0}\left[
 \frac{\partial_s^2Y_n^X}{n^2}-\frac{\partial_sY_n^X}{2n}
 \right]\right\}.
 \label{eq:pair-three-class-replica-overlap}
\end{equation}
The finiteness of Eq.~\eqref{eq:pair-three-class-poisson-matching}
makes the order-$n$ term in $Y_n^X$ linear in $s$. As in
Sec.~\ref{subsec:pair_correlation_replica_limit}, the condition
$\mathcal O_2^X\to0$ at large separation fixes its slope to $1/2$.
Consequently, for $s>0$,
\begin{equation}
 \pi^2\mathcal O_2^X(z_1,z_2)
 =\frac{\beta_Xq^2}{4N}
 \left\{-1+\partial_s[s\mathcal P_2^X(s)]\right\}
 =\frac{\pi\mathcal O_1^X(z)}2
 \left\{-1+\partial_s[s\mathcal P_2^X(s)]\right\}.
 \label{eq:pair-three-class-final}
\end{equation}
This establishes Eq.~\eqref{eq:general-class-bulk-o2-ratio} for all three classes.

\section{Replica Derivation of the Class-D Hard-Edge Diagonal Overlap}
\label{app:class-d-hard-edge}

We consider the even-dimensional Gaussian ensemble of class D in
Eq.~\eqref{eq:d-resolvent-ensemble}, with \(|z|^2=\mathcal O(1)\) fixed
as \(N\to\infty\). All saddle-point expressions below are understood at
leading order for large $N$ in this hard-edge limit.
Section~\ref{subsec:d-topological-source} extends the source reduction
to odd dimension by retaining the Pfaffian sign.

\subsection{Source Partition Function and Hard-Edge Limit}

We proceed as in class A. With a small source $\eta$ that is set to zero
at the end, we define the partition function
\begin{equation}
  Z_{1,n}^{\mathrm D}(z,\eta)
  :=\left\langle
    \det\!\left[
      (z-H)(z-H)^\dagger+\eta^2
    \right]^n
  \right\rangle.
\end{equation}
The distributional identity derived in Sec.~\ref{subsec:petermann_resolvent} gives
\begin{equation}
  \pi\mathcal O_{1}(z)
  =\frac1{2N}\lim_{n \to 0}
    \left.\partial_\eta^2 Z_{1,n}^{\mathrm D}(z,\eta)\right|_{\eta=0}.
  \label{eq:d-resolvent-replica-contact}
\end{equation}
In Sec.~\ref{sec:d-contact-nlsm-derivation}, we derive the nonlinear
$\sigma$ model, with $q = \sqrt{N}$,
\begin{align}
Z_{1,n}^{\mathrm D}(z,\eta) &= e^{-n \eta^2}\int_{\mathrm{O}(2n)}\,\mathrm dO\,
  \exp\,\left[
    -\frac {|z|^2}2\tr\,(O^TJ_nOJ_n)
    +\frac{q\eta}{2}\tr\,(O+O^T)
  \right].
  \label{eq:d-resolvent-nlsm}
\end{align}
Here, \(\mathrm dO\) denotes the normalized Haar measure on both
connected components of \(\mathrm{O}(2n)\). In
Sec.~\ref{subsec:class_d_replica_limit}, we take the replica limit and
find
\begin{equation}
  \pi \mathcal{O}_1(z) = \frac{1}{2N} \lim_{n\to 0}
      \left.\partial_\eta^2 Z_{1,n}^{\mathrm D}(z,\eta)\right|_{\eta=0} = \lim_{n\to 0} \frac{1}{2}\left(1+\frac{1}{n}\frac{\mathrm d}{\mathrm d|z|^2}\right)
      \int_{\mathrm{O}(2n)}\,\mathrm dO\,
      \exp\,\left[-\frac{|z|^2}{2}\tr\,(J_nO^TJ_nO)\right].
  \label{eq:class_d_replica_limit}
\end{equation}
The integral is the same as that in the replica partition function for
the spectral density of non-Hermitian random matrices in class D,
\begin{equation}
  \left\langle
    \det\!\left[
      (z-H)(\overline{z}-H^\dagger)
    \right]^n
  \right\rangle = \int_{\mathrm{O}(2n)}\,\mathrm dO\,
    \exp\,\left[-\frac{|z|^2}{2}\tr\,(J_nO^TJ_nO)\right].
\end{equation}
Its replica limit is known from Ref.~\cite{ChenXiaoLiuRyu2026},
\begin{equation}
  \lim_{n\to 0}\frac{1}{n}\frac{\mathrm d}{\mathrm d|z|^2}
    \int_{\mathrm{O}(2n)}\,\mathrm dO\,
    \exp\,\left[-\frac{|z|^2}{2}\tr\,(J_nO^TJ_nO)\right]
  = \mathcal P_1(|z|^2).
\end{equation}
Consequently,
\begin{equation}
  \pi \mathcal{O}_1(z)
  =\frac{1}{2}\left[1+\mathcal P_1(|z|^2)\right].
\end{equation}
The following subsections derive the source action and the Ward identity.

\subsection{\texorpdfstring{Nonlinear $\sigma$ Model}{Nonlinear sigma model}}
\label{sec:d-contact-nlsm-derivation}

With
\(\Sigma_x=\left(\begin{smallmatrix}0&I_n\\ I_n&0\end{smallmatrix}\right)\),
we apply the Hubbard--Stratonovich transformation to
\begin{equation}
  \det\!\left[
    (z-H)^\dagger(z-H)+\eta^2
  \right]
  =\det\!\begin{pmatrix}
    z-H&-\eta\\
    \eta&\bar z-H^\dagger
  \end{pmatrix}.
  \label{eq:d-contact-block-linearization}
\end{equation}
Following Ref.~\cite{Chen2025} and changing variables as
$Q\mapsto Q - \eta\,\Sigma_x$, we obtain
\begingroup
\begin{equation}
  Z_{1,n}^{\mathrm D}(z,\eta)
  =e^{-n\eta^2}
    \int_{\Sigma_xQ^*\Sigma_x=Q}\!\mathrm dQ\,
    \exp\,\left[
      -\frac12\tr\,(Q^\dagger Q)
      +\frac{1}{2}\eta\,\tr\,\bigl[(Q+Q^\dagger)\Sigma_x\bigr]
    \right] \times
    \left\{
      \operatorname{Pf}\!\begin{pmatrix}
        zJ_n&-Q\\
        Q^T&\bar zJ_n
      \end{pmatrix}
    \right\}^{N}.
  \label{eq:d-contact-exact-hs}
\end{equation}
\endgroup
At \(z=\eta=0\), the saddle-point manifold $Q^\dagger Q = N I_{2n}$ is
parametrized by
\begin{equation}
  Q=q\,S O S,
    \qquad O\in \mathrm{O}(2n),
  \label{eq:d-contact-hs-saddle}
\end{equation}
where $S = S^T = \Sigma_x^{1/2}$ and $S^\dagger S = I_{2n}$.
On the saddle-point manifold, the source term becomes
\begin{equation}
  \frac{1}{2}\eta\,\tr\,\bigl[(Q+Q^\dagger)\Sigma_x\bigr]
  =\frac{q\eta}{2}\tr\,(O+O^T).
  \label{eq:d-contact-transverse-source}
\end{equation}
The Pfaffian factor becomes
\begin{align}
  &\left[
    {\operatorname{Pf}\!\begin{pmatrix}
      zJ_n&-Q\\
      Q^T&\bar zJ_n
    \end{pmatrix}}
    \right]^{N}
  = \det\!\left[Q Q^T\right]^{N/2} \det\!\left[
    I_{2n}+|z|^2J_n(Q^T)^{-1}J_nQ^{-1}
  \right]^{N/2}
  \nonumber \\
  &\quad
  =q^{2nN}\exp\,\left[
    -\frac {|z|^2}2\tr\,(J_nO^TJ_nO)+\bigO(N^{-1})
  \right],
\end{align}
where we used $q^2=N$ and expanded the logarithm at fixed $|z|^2$.
The saddle-point expression is therefore
\begin{equation}
  Z_{1,n}^{\mathrm D}(z,\eta) = q^{2nN} e^{-n \eta^2 - n q^2}
    \int_{\mathrm{O}(2n)}\,\mathrm dO\,
    \exp\,\left[-\frac{|z|^2}{2}\tr\,(J_nO^TJ_nO)
      +\frac{q\eta}{2}\tr\,(O+O^T)\right].
\end{equation}
Dropping the factors that are independent of $\eta$ and tend to unity as
$n\to0$ gives Eq.~\eqref{eq:d-resolvent-nlsm}.

\subsection{Replica Limit}
\label{subsec:class_d_replica_limit}

We first establish the Ward identity
\begin{align}
\int_{\mathrm{O}(2n)}\,\mathrm dO\,
  e^{-\frac {|z|^2}2\tr\,(O^TJ_nOJ_n)}
  \left[\frac12\tr\,(O+O^T)\right]^2
&=\left(1+\frac{1}{n}\frac{\mathrm d}{\mathrm d|z|^2}\right)
  \int_{\mathrm{O}(2n)}\,\mathrm dO\,
  e^{-\frac {|z|^2}2\tr\,(O^TJ_nOJ_n)}.
    \label{eq:d-replica-limit-ward}
\end{align}
Since $\tr\,O^T=\tr\,O$, the source insertion is
$\tr\,(O^T)\tr\,O$. The weight and Haar measure are invariant under
$O\mapsto UO$ for $U\in\mathrm U(n)\subset\mathrm O(2n)$.
We therefore average the insertion over $U$
and use Eq.~\eqref{eq:jojo},
\begin{align}
  \int_{\mathrm{U}(n)\subset\mathrm{O}(2n)}
    \,\mathrm dU\,
    \tr\,\bigl(O^TU^T\bigr)\tr\,(UO)
  &=1-\frac{1}{2n}\tr\,\bigl(J_nO^TJ_nO\bigr),
\end{align}
inside the orthogonal-group integral. The trace term is generated by
differentiating the weight with respect to $|z|^2$, which establishes
Eq.~\eqref{eq:d-replica-limit-ward}.

Taking the second source derivative at $\eta=0$ and applying this
identity gives
\begingroup
\begin{align}
  \left.\frac12\partial_\eta^2 Z_{1,n}^{\mathrm D}(z,\eta)\right|_{\eta=0}
  &=\int_{\mathrm{O}(2n)}\!\mathrm dO\,
    \exp\,\left[-\frac{|z|^2}{2}\tr\,(J_nO^TJ_nO)\right]
    \times\left\{
    \frac{q^2}{2}\left[\frac12\tr\,(O+O^T)\right]^2-n
    \right\}
    \nonumber \\
  &=\left[\frac{q^2}{2}\left(1+\frac{1}{n}\frac{\mathrm d}{\mathrm d|z|^2}\right)-n\right]
    \int_{\mathrm{O}(2n)}\!\mathrm dO\,
    \exp\,\left[-\frac{|z|^2}{2}\tr\,(J_nO^TJ_nO)\right].
\end{align}
\endgroup
Taking $n\to0$ with $q^2=N$ gives
Eq.~\eqref{eq:class_d_replica_limit} and hence the hard-edge
diagonal overlap.

\subsection{Odd Matrix Dimension}
\label{subsec:d-topological-source}

Now let $N$ have either parity and write $\nu=N\bmod2$.
The Hubbard--Stratonovich representation
[Eq.~\eqref{eq:d-contact-exact-hs}] is unchanged, but its Pfaffian
must retain its sign. At $z=0$,
\begin{equation}
 \operatorname{Pf}\!\begin{pmatrix}0&-Q\\Q^T&0\end{pmatrix}
 =(-1)^n\det Q=q^{2n}\det O,
 \qquad Q=qSOS,
 \label{eq:d-topological-pfaffian-sign}
\end{equation}
since $\det(S)^2=\det\Sigma_x=(-1)^n$.
Raising this Pfaffian to the physical matrix dimension gives
$(\det O)^N=(\det O)^\nu$. For odd $N$, replacing the Pfaffian
by an unsigned square root of its determinant would lose this factor.
Expanding its ratio to the $z=0$ value gives the same angular action
as in the even-dimensional calculation~\cite{ChenXiaoLiuRyu2026}.
Thus, with $s=|z|^2$ and the same source-independent normalization
as above,
\begin{align}
 Y_{n,\nu}^{\mathrm D}(s)
 &=\int_{\mathrm O(2n)}dO\,(\det O)^\nu
 e^{-\frac s2\tr(O^TJ_nOJ_n)},
 \label{eq:d-topological-angular}\\
 Z_{1,n,\nu}^{\mathrm D}(z,\eta)
 &\simeq e^{-n\eta^2}\int_{\mathrm O(2n)}dO\,(\det O)^\nu
 e^{-\frac s2\tr(O^TJ_nOJ_n)+q\eta\tr O},\qquad q^2=N.
 \label{eq:d-topological-source-action}
\end{align}
The Haar measure has unit volume on the full group; it is not
renormalized after inserting the character. In particular,
$Y_{n,1}^{\mathrm D}(0)=0$ for positive integer $n$, reflecting the
protected zero mode. Replica limits below are taken at fixed $s>0$.

Under $O\mapsto kO$ with $k\in K\simeq\mathrm U(n)$,
$k^TJ_nk=J_n$ preserves the angular action and $\det k=1$ preserves
the topological weight [Eq.~\eqref{eq:schur-topological-invariance}].
We may therefore perform the same subgroup average as for $\nu=0$:
\begin{align}
 &\int_{\mathrm O(2n)}dO\,(\det O)^\nu
 e^{-\frac s2\tr(O^TJ_nOJ_n)}(\tr O)^2\notag\\
 &\quad=\int_{\mathrm O(2n)}dO\,(\det O)^\nu
 e^{-\frac s2\tr(O^TJ_nOJ_n)}
 \left[1-\frac1{2n}\tr(J_nO^TJ_nO)\right]\notag\\
 &\quad=\left(1+\frac1n\partial_s\right)Y_{n,\nu}^{\mathrm D}(s).
 \label{eq:d-topological-ward}
\end{align}
The second source derivative consequently satisfies
\begin{equation}
 \left.\frac12\partial_\eta^2Z_{1,n,\nu}^{\mathrm D}\right|_0
 \simeq\left[\frac N2\left(1+\frac1n\partial_s\right)-n\right]
 Y_{n,\nu}^{\mathrm D}(s),\qquad\nu=0,1.
 \label{eq:d-topological-source-reduction}
\end{equation}
This establishes the source identity with the topological weight
present throughout the calculation. Using the sector-resolved
spectral replica limits~\cite{ChenXiaoLiuRyu2026},
$\lim_{n\to0}Y_{n,\nu}^{\mathrm D}=1$ and
$\lim_{n\to0}n^{-1}\partial_sY_{n,\nu}^{\mathrm D}
=\mathcal P_{1,\nu}^{\mathrm D}$, gives
\begin{equation}
 \pi\mathcal O_{1,\nu}^{\mathrm D}(z)
 =\frac12[1+\mathcal P_{1,\nu}^{\mathrm D}(s)],\qquad s>0.
 \label{eq:d-topological-overlap}
\end{equation}
For example, in the odd-dimensional sector,
\begin{equation}
 \mathcal P_{1,1}^{\mathrm D}(s)
 =1-\frac{1-e^{-2s}}{2s}+\frac1s,\qquad
 \pi\mathcal O_{1,1}^{\mathrm D}(z)
 =1+\frac{1+e^{-2s}}{4s}.
 \label{eq:d-odd-explicit-overlap}
\end{equation}
The $1/s$ term in $\mathcal P_{1,1}^{\mathrm D}$ comes from the
protected zero eigenvalue of the Hermitian partner. The resulting
enhancement concerns the overlap density of nonzero eigenvalues;
the atomic overlap weight at $z=0$ is a separate observable.

\section{Schur Orthogonality}\label{app:schur}

All compact group measures below are normalized to unit volume.
We use
$\mathrm{Sp}(n)\subset\mathrm U(2n)$ and the matrices
\begin{equation}
 J_n=\begin{pmatrix}0&I_n\\-I_n&0\end{pmatrix},\qquad
 \Lambda_n=\begin{pmatrix}I_n&0\\0&-I_n\end{pmatrix}.
\end{equation}
We record the group contractions used in taking the replica limit. In
the calculations of $R_2$ and $R_1$ for non-Hermitian random matrices,
the replica partition functions are identified with their Hermitian
counterparts, so no explicit group integration is needed. For
$\mathcal O_2$ and $\mathcal O_1$, however, the partition functions
contain a source $\eta$. Both calculations use the second
derivative with respect to $\eta$. The following result allows
us to replace these source derivatives by derivatives with respect to
{$|\omega|^2$ for $\mathcal O_2$ and $|z|^2$ for $\mathcal O_1$}. After the source is set to zero, we recover the replica partition
functions for $R_2$ and $R_1$, whose replica limits are known.
\par
Let $G$ be a compact Lie group, and let
\begin{equation}
  \rho,\sigma\colon G \to \mathrm U(V)
\end{equation}
be irreducible unitary representations of $G$ on a complex vector space
$V$, with $d_\rho:=\dim V$. The Schur orthogonality
relation~\cite{Helgason1984} reads
\begin{equation}
  \label{eq:schur_orthogonality}
  \int_G \mathrm d g\, \tr\,(A\rho(g))
  \tr\,(\sigma(g)^\dagger B)
  = \frac{1}{d_\rho} \delta_{\rho \sigma} \tr\,(A B),
\end{equation}
where $\delta_{\rho\sigma} = 1$ if $\rho$ and $\sigma$ are equivalent and $0$ otherwise.
When the representations are equivalent, the bases are chosen so that $\rho(g)=\sigma(g)$.
\par
We first give the unitary contractions.
For $G=\mathrm U(n)$, we have
\begin{equation}
  \int_{\mathrm U(n)}\!\mathrm dU\,
  \tr\,(B^\dagger U)\tr\,(U^\dagger B)
  =\frac{1}{n}\tr\,(B^\dagger B).
  \label{eq:buub}
\end{equation}
For $\mathrm{U}(n)\subset\mathrm O(2n)$,
\begin{equation}
  \int_{\mathrm U(n)\subset\mathrm O(2n)}\!\mathrm dU\,
  \tr\,(O^T U^T)\tr\,(U O)
  = 1 - \frac{1}{2n}\tr\,(J_n O^T J_n O).
  \label{eq:jojo}
\end{equation}
\par
For a quaternion-real $2n\times2n$ matrix $B$,
$B=J_n\overline B J_n^{-1}$, the trace of $VB$ is real for
$V\in\mathrm{Sp}(n)$. Applying
Eq.~\eqref{eq:schur_orthogonality} to the fundamental complex
representation, whose dimension is $2n$, therefore gives
\begin{equation}
 \int_{\mathrm{Sp}(n)}dV\,[\tr(VB)]^2
 =\int_{\mathrm{Sp}(n)}dV\,|\tr(VB)|^2
 =\frac1{2n}\tr(BB^\dagger).
 \label{eq:schur-symplectic-source}
\end{equation}
Likewise, for a real $n\times n$ matrix $B$ and the full orthogonal
group, including both connected components,
\begin{equation}
 \int_{\mathrm O(n)}dV\,[\tr(VB)]^2
 =\frac1n\tr(BB^T).
 \label{eq:schur-orthogonal-source}
\end{equation}
These are the contractions used in the three-class bulk reduction in
Sec.~\ref{subsec:pair-source-three-classes}.

For the class-D hard edge, the subgroup in Eq.~\eqref{eq:jojo} is
$K=\{k\in\mathrm O(2n):k^TJ_nk=J_n\}\simeq\mathrm U(n)$.
If $k$ is the real representation of $V\in\mathrm U(n)$, then
\begin{equation}
 \det_{\mathbb R}k=|\det_{\mathbb C}V|^2=1,\qquad
 [\det(kO)]^\nu=(\det O)^\nu.
 \label{eq:schur-topological-invariance}
\end{equation}
Thus Eq.~\eqref{eq:jojo} also applies inside the class-D integral
weighted by $(\det O)^\nu$ for either $\nu=0$ or $1$, as used in
Sec.~\ref{subsec:d-topological-source}.

\section{Unfolding and Normalization}
\label{app:unfolding-normalization}

For the comparisons in Sec.~\ref{sec:physical-models}, we unfold the
spectrum and normalize the overlap amplitude.

\subsection{Bulk Off-Diagonal Overlaps}

For bulk eigenvalues $z_1,z_2$ with midpoint $z=(z_1+z_2)/2$,
we normalize the off-diagonal overlap as
\begin{align}
  \label{eq:o2o1}
  \frac{\mathcal{O}_2(z_1,z_2)}{R_1(z)\mathcal{O}_1(z)}
  &= \dfrac{\left\langle
      \sum_{\substack{a,b\\{a\ne b}}} O_{ab}\,
      \delta^{(2)}(z_1-\lambda_a)
      \delta^{(2)}(z_2-\lambda_b)\right\rangle}{
      R_1(z) \left\langle
      \sum_{a} O_{aa}\,
      \delta^{(2)}(z-\lambda_a)\right\rangle},
\end{align}
where the sums run over distinct eigenspaces. The local mean spacing
sets the unfolded separation,
\begin{equation}
  \label{eq:unfolded_t}
  \omega = \sqrt{\pi R_1(z)} (z_1 - z_2),
\end{equation}
where $R_1(z)$ is the local eigenvalue density. For the Gaussian
normalization $\pi R_1=1$, this reduces to $\omega=z_1-z_2$.
The same unfolding applies to eigenvalue correlations~\cite{ChenXiaoLiuRyu2026}.
Changing variables in the $\delta$ functions at fixed midpoint gives
\begin{align}
  &\frac{\mathcal{O}_2(z_1,z_2)}{R_1(z)\mathcal{O}_1(z)}
    =\pi\left\langle\sum_a O_{aa}\,
      \delta^{(2)}(z-\lambda_a)\right\rangle^{-1}
  \times\Biggl\langle\sum_{\substack{a,b\\a\ne b}}O_{ab}\,
      \delta^{(2)}\!\left(z-\frac{\lambda_a+\lambda_b}{2}\right)
  \times\delta^{(2)}\!\left(
      \omega-\sqrt{\pi R_1(z)}(\lambda_a-\lambda_b)\right)\Biggr\rangle.
  \label{eq:bulk_universal_operational}
\end{align}
The universality claim to be tested numerically is
\begin{equation}
  \label{eq:bulk_universal_unfolded}
  \frac{\mathcal{O}_2(z_1,z_2)}{R_1(z)\mathcal{O}_1(z)}
  =\frac{1}{2}\left[ -1+\frac{\mathrm d}{\mathrm d|\omega|^2}
    \left[|\omega|^2\mathcal P_2(|\omega|^2)\right] \right].
\end{equation}
The right-hand side is given in Table~\ref{tab:replica-bulk-summary};
Eq.~\eqref{eq:bulk_universal_operational} supplies the numerical estimator
on the left. Both orderings of each pair are combined, as in
Sec.~\ref{sec:numerics}. At $\pi R_1=1$, this relation reduces to
Eq.~\eqref{eq:general-class-bulk-o2-ratio}.

\subsection{Hard-Edge Overlaps}

For hard-edge statistics, denote the physical spectral coordinate by
$z_{\mathrm{phys}}$ and its unfolded value by
$z=\sqrt{\pi R_1^{\mathrm{bulk}}}\,z_{\mathrm{phys}}$, where
$R_1^{\mathrm{bulk}}$ is the bulk eigenvalue density near the origin.
The Gaussian ensembles already have $\pi R_1^{\mathrm{bulk}}=1$.
In physical systems, estimating this density is difficult: measurements
too close to the origin are affected by the hard edge, whereas those
too far away may no longer represent the local bulk density.
We therefore unfold using the mean modulus of the smallest nonzero
eigenvalue~\cite{Xiao2024},
\begin{equation}
  z={c_\nu z_{\mathrm{phys}},\qquad c_\nu:=}\frac{\langle|\lambda_{\mathrm{min}}|\rangle_{\mathrm{RM}}}
         {\langle|\lambda_{\mathrm{min}}|\rangle_{\mathrm{Phys}}}{.}
  \label{eq:hard-edge-unfolding-scale}
\end{equation}
Here, the averages refer to the reference random matrix ensemble and
the physical system, respectively. The reference ensemble has the same
symmetry class and topological sector, with the Gaussian normalization
used throughout the paper{, so $c_\nu$ is determined separately for
each topological sector. Since $\mathcal O_1$ is a density, the unfolded
profile is
$\mathcal O_1^{\mathrm{unf}}(z)=c_\nu^{-2}\mathcal O_1^{\mathrm{phys}}(z/c_\nu)$;
the Jacobian cancels in the ratio in
Eq.~\eqref{eq:hard_edge_universal_unfolded} below but enters the
amplitude $A_\nu$ defined in Eq.~\eqref{eq:hard-edge-amplitude}}.
The overlap amplitude remains system dependent, so we normalize it at
the mean smallest-eigenvalue modulus. Equation~\eqref{eq:general-class-hard-edge-o1}
then predicts
\begin{equation}
  \label{eq:hard_edge_universal_unfolded}
  \frac{\mathcal O_1(z_{\mathrm{phys}})}
       {\mathcal O_1(\langle|\lambda_{\mathrm{min}}|\rangle_{\mathrm{Phys}})}
  =\frac{1+\mathcal P_1(|z|^2)}
        {1+\mathcal P_1(\langle|\lambda_{\mathrm{min}}|\rangle_{\mathrm{RM}}^2)}.
\end{equation}
{In practice, the numerical data are binned in $u=|z|^2$, so the
pointwise normalization in Eq.~\eqref{eq:hard_edge_universal_unfolded}
is implemented over the radial bin $B_\nu=[u_-,u_+]$ that contains
$\langle|\lambda_{\mathrm{min}}|\rangle_{\mathrm{RM}}^2$. With
$F_\nu(u)=\frac{\beta}{4}[1+\mathcal P_{1,\nu}(u)]$ and
$\overline{g}_{B_\nu}=(u_+-u_-)^{-1}\int_{u_-}^{u_+}g(u)\,\mathrm du$,
the amplitude of sector $\nu$ is
\begin{equation}
  A_\nu:=\frac{\overline{\pi\mathcal O_{1,\nu}^{\mathrm{unf}}}_{B_\nu}}
              {\overline{F_\nu}_{B_\nu}},
  \label{eq:hard-edge-amplitude}
\end{equation}
so that $\pi\mathcal O_{1,\nu}^{\mathrm{unf}}/A_\nu$, the ordinate of
Fig.~\ref{fig:numerical-bdi-vacancy}, has the theoretical mean over
the calibration bin by construction. The solid lines in
Fig.~\ref{fig:numerical-bdi-vacancy} are $F_\nu(u)$ without a bulk
envelope. The calibration bin therefore does not constitute an
independent test; the comparison concerns the shape of the profile
outside it.}

\section{Diagonal Overlap Correlation}
\label{app:petermann_factor_density_correlation}

\begin{table}[b]
  \centering
  \caption{Bulk-pair relations through \(\mathcal P_2\) and
    hard-edge one-point relations through \(\mathcal P_1\).}
  \label{tab:p-relations}
  \small
  \renewcommand{\arraystretch}{1.3}
  \setlength{\tabcolsep}{3pt}
  \begin{tabular}{@{}l r c l l@{}}
    \toprule
    \bfseries Quantity
      & \multicolumn{3}{c}{\bfseries Relation}
      & \bfseries Source \\
    \midrule
    \multicolumn{5}{@{}l}{\bfseries Bulk pair \qquad $\omega=z_1-z_2,\ z = (z_1 + z_2)/2$} \\[3pt]
    \(\mathcal W_2\)
      & \(\displaystyle \frac{\mathcal{W}_2(z_1,z_2)}{\mathcal{O}_1(z)^2}\)
      & \(=\)
      & \(\displaystyle \frac12[1+\mathcal P_2(|\omega|^2)]\)
      & Ref.~\cite{Fyodorov2026Parametric} \\[5pt]
    \(\mathcal O_2\)
      & \(\displaystyle \frac{\pi\mathcal O_2(z_1,z_2)}{\mathcal O_1(z)}
          \)
      & \(=\)
      & \(\displaystyle \frac12\left[-1+\frac{\mathrm d}{\mathrm d|\omega|^2}
          [|\omega|^2\mathcal P_2(|\omega|^2)]\right]\)
      & Eq.~\eqref{eq:general-class-bulk-o2-ratio} \\[6pt]
    \(R_2\)
      & \(\displaystyle \pi^2R_2(\omega)\)
      & \(=\)
      & \(\displaystyle \frac12\frac{\mathrm d^2}{\mathrm d(|\omega|^2)^2}
          [|\omega|^4\mathcal P_2(|\omega|^2)]\)
      & Ref.~\cite{ChenXiaoLiuRyu2026} \\[5pt]
    \midrule
    \multicolumn{5}{@{}l}{\bfseries Hard-edge one-point \qquad $|z|>0$} \\[3pt]
    \(\mathcal O_1\)
      & \(\displaystyle \pi\mathcal O_1(z)\)
      & \(=\)
      & \(\displaystyle \frac{\beta}{4}[1+\mathcal P_1(|z|^2)]\)
      & Eq.~\eqref{eq:general-class-hard-edge-o1} \\[5pt]
    \(R_1\)
      & \(\displaystyle \pi R_1(z)\)
      & \(=\)
      & \(\displaystyle \frac{\mathrm d}{\mathrm d|z|^2}[|z|^2\mathcal P_1(|z|^2)]\)
      & Ref.~\cite{ChenXiaoLiuRyu2026} \\
    \bottomrule
  \end{tabular}
\end{table}

The diagonal overlap correlation studied in
Ref.~\cite{Fyodorov2026Parametric} is defined, in a similar way to
$\mathcal{O}_2$, by
\begin{equation}
    \mathcal{W}_2(z_1,z_2)
    =\frac{1}{N^2}\Biggl\langle\sum_{\substack{a,b=1\\a\ne b}}^N
      O_{aa}O_{bb}\,
    \delta^{(2)}(z_1-\lambda_a)
      \delta^{(2)}(z_2-\lambda_b)\Biggr\rangle.
\end{equation}
While $\mathcal{W}_2$ is not itself universal (like $\mathcal{O}_2$),
it could be made universal after normalization by the local $\mathcal{O}_1$ value
and by unfolding such that $\pi R_1 = 1$.
We therefore study the normalized ${\mathcal{W}_2(z_1,z_2)}/{\mathcal{O}_1(z)^2}$.
Ref.~\cite{Fyodorov2026Parametric} conjectured the following relation
between $\mathcal W_2$ and $\mathcal P_2$, which has been proved for
class A~\cite{BourgadeDubach2020}.
\begin{equation}
  \label{eq:w_by_p}
  \frac{\mathcal{W}_2(z_1,z_2)}{\mathcal{O}_1(z)^2}
  =\frac12\left[1+\mathcal P_2(|\omega|^2)\right],\qquad |\omega|>0.
\end{equation}
This, together with the results in this article and in Ref.~\cite{ChenXiaoLiuRyu2026}, allows us to identify $\mathcal{P}_2$ as the generating function of universal statistical quantities of non-Hermitian random matrices.
We collect these results, including those of hard-edge statistics, in {Table}~\ref{tab:p-relations}.
\par
We now derive Eq.~\eqref{eq:w_by_p} in class A to leading order in
large $N$. We keep the midpoint $z=(z_1+z_2)/2$ fixed in the spectral
bulk, with $\omega=z_1-z_2$ and $|\omega|>0$.
\par
Define the following replica partition function which contains two sources $\eta_1$ and $\eta_2${, distinguished by the superscript $\mathcal W$ from the mixed-source $Z_{2,n}^{\mathrm A}(\eta)$ of Appendix~\ref{app:class-a-pair}},
\begin{equation}
  Z_{2,n}^{\mathrm A{,\mathcal W}}(\eta_1,\eta_2)
  :=\left\langle
    \det\!\left[
      (z_1-H)(\overline z_1-H^\dagger)+\eta_1^2
    \right]^n
    \det\!\left[
      (z_2-H)(\overline z_2-H^\dagger)+\eta_2^2
    \right]^n
  \right\rangle.
  \label{eq:w2-mass-source-partition}
\end{equation}
With the resolvent result in Appendix~\ref{subsec:petermann_resolvent},
$\mathcal{W}_2$ is calculated from $Z^{\mathrm A}_{2,n}$ by
\begin{equation}
  \mathcal W_2(z_1,z_2)
  =\frac1{4\pi^2N^2}\lim_{n\to0}
  \left.
  \partial_{\eta_1}^2\partial_{\eta_2}^2
  Z_{2,n}^{\mathrm A{,\mathcal W}}(\eta_1,\eta_2)
  \right|_{\eta_1=\eta_2=0},
  \qquad z_1\ne z_2.
  \label{eq:w2-replica-extraction}
\end{equation}
The same Hubbard{--}Stratonovich transform as in Appendix~\ref{app:class-a-pair} gives
\begin{align}
  Z_{2,n}^{\mathrm A{,\mathcal W}}(\eta_1,\eta_2)
  ={}&\int_{\mathrm U(2n)}\!\mathrm dU\,
  \exp\,\left[\frac{|\omega|^2}{4}\tr\,(\Lambda U\Lambda U^\dagger)\right]
  \notag\\
  &\quad\times
  \left\{1+\eta_1^2
    \left[q^2\tr\,(U_{11})\tr\,(U_{11}^\dagger)-n\right]
    +\bigO(\eta_1^4)\right\}
  \notag\\
  &\quad\times
  \left\{1+\eta_2^2
    \left[q^2\tr\,(U_{22})\tr\,(U_{22}^\dagger)-n\right]
    +\bigO(\eta_2^4)\right\}.
  \label{eq:w2-source-coupled-nlsm}
\end{align}
Here, $q^2=N-|z|^2$, $\Lambda=\operatorname{diag}(I_n,-I_n)$, and $U\in \mathrm{U}(2n)$ has the block structure
\begin{equation}
  U=\begin{pmatrix}U_{11}&U_{12}\\U_{21}&U_{22}\end{pmatrix}.
\end{equation}
The zero-source compact integral is the same \(Y_n^{\mathrm A}(|\omega|^2)\) defined in
Eq.~\eqref{eq:a-o2-nlsm}:
\begin{equation}
  Y_n^{\mathrm A}(|\omega|^2)=\int_{\mathrm U(2n)}\!\mathrm dU\,
  \exp\,\left[\frac{|\omega|^2}{4}\tr\,(\Lambda U\Lambda U^\dagger)\right].
  \label{eq:w2-yn}
\end{equation}
It follows from Eq.~\eqref{eq:w2-source-coupled-nlsm} that
\begin{align}
  &\frac14\left.
  \partial_{\eta_1}^2\partial_{\eta_2}^2
  Z_{2,n}^{\mathrm A{,\mathcal W}}(\eta_1,\eta_2)
  \right|_{\eta_1=\eta_2=0}
  \notag\\
  &\quad=\int_{\mathrm U(2n)}\!\mathrm dU\,
  \exp\,\left[\frac{|\omega|^2}{4}\tr\,(\Lambda U\Lambda U^\dagger)\right]
  \left[q^2\tr\,(U_{11})\tr\,(U_{11}^\dagger)-n\right]
  \left[q^2\tr\,(U_{22})\tr\,(U_{22}^\dagger)-n\right].
  \label{eq:w2-source-insertion-saddle}
\end{align}
\par
As before, we write Eq.~\eqref{eq:w2-source-insertion-saddle} as a linear combination of Eq.~\eqref{eq:w2-yn} and its derivatives with respect to $|\omega|^2$ using the Schur orthogonality relations in Appendix~\ref{app:schur}.
Eq.~\eqref{eq:buub} gives
\begin{align}
  \int_{\mathrm U(n)}\!\mathrm dV_1\,
    \tr\,(U_{11}V_1)\tr\,(V_1^\dagger U_{11}^\dagger)
    \label{eq:w2-fiber-average-1}
  &=\frac1n\tr\,(U_{11}U_{11}^\dagger),\\
  \int_{\mathrm U(n)}\!\mathrm dV_1
    \int_{\mathrm U(n)}\!\mathrm dV_2\,
    \tr\,(U_{11}V_1)\tr\,(V_1^\dagger U_{11}^\dagger)
    \tr\,(U_{22}V_2)\tr\,(V_2^\dagger U_{22}^\dagger)
  &=\frac1{n^2}
    \tr\,(U_{11}U_{11}^\dagger)
    \tr\,(U_{22}U_{22}^\dagger).
  \label{eq:w2-fiber-average-2}
\end{align}
Unitarity implies
\begin{equation}
  \tr\,(U_{11}U_{11}^\dagger)
  =\tr\,(U_{22}U_{22}^\dagger)
  =\frac14\left[2n+\tr\,(\Lambda U\Lambda U^\dagger)\right].
  \label{eq:w2-block-norm}
\end{equation}
Since the weight in Eq.~\eqref{eq:w2-source-insertion-saddle} is invariant under right multiplication of $U$ by
\(\operatorname{diag}(V_1,V_2)\), with
\(V_1,V_2\in\mathrm U(n)\),
Eqs.~\eqref{eq:w2-fiber-average-1}--\eqref{eq:w2-block-norm} yield
\begin{align}
  \int_{\mathrm U(2n)}\!\mathrm dU\,
    e^{\frac{|\omega|^2}{4}\tr\,(\Lambda U\Lambda U^\dagger)}
    \tr\,(U_{11})\tr\,(U_{11}^\dagger) &= \int_{\mathrm U(2n)}\!\mathrm dU\,
    e^{\frac{|\omega|^2}{4}\tr\,(\Lambda U\Lambda U^\dagger)}
    \tr\,(U_{22})\tr\,(U_{22}^\dagger) \notag \\
  &=\left(\frac12+\frac1n\frac{\mathrm d}{\mathrm d|\omega|^2}\right)Y_n^{\mathrm A}(|\omega|^2),\\
  \int_{\mathrm U(2n)}\!\mathrm dU\,
    e^{\frac{|\omega|^2}{4}\tr\,(\Lambda U\Lambda U^\dagger)}
    \tr\,(U_{11})\tr\,(U_{11}^\dagger)
    \tr\,(U_{22})\tr\,(U_{22}^\dagger)
  &=\left(\frac14+\frac1n\frac{\mathrm d}{\mathrm d|\omega|^2}
    +\frac1{n^2}\frac{\mathrm d^2}{\mathrm d(|\omega|^2)^2}\right)Y_n^{\mathrm A}(|\omega|^2).
  \label{eq:w2-angular-identities}
\end{align}
Substitution in Eq.~\eqref{eq:w2-source-insertion-saddle} gives
\begin{align}
  &\frac14\left.
  \partial_{\eta_1}^2\partial_{\eta_2}^2
  Z_{2,n}^{\mathrm A{,\mathcal W}}(\eta_1,\eta_2)
  \right|_{\eta_1=\eta_2=0}
  =\left[
    \frac{q^4}{n^2}\frac{\mathrm d^2}{\mathrm d(|\omega|^2)^2}
    +\left(\frac{q^4}{n}-2q^2\right)\frac{\mathrm d}{\mathrm d|\omega|^2}
    +\frac{q^4}{4}-nq^2+n^2
  \right]Y_n^{\mathrm A}(|\omega|^2).
  \label{eq:w2-differential-insertion}
\end{align}

Now we take the replica limit $n\to 0$ in Eq.~\eqref{eq:w2-differential-insertion}.
Using the expansion
$Y_n^{\mathrm A}(|\omega|^2)=1+{n|\omega|^2}/{2} + \bigO(n^2)$
established in
Sec.~\ref{subsec:pair_correlation_replica_limit},
together with the known replica limit in Eq.~\eqref{eq:a-o2-poisson-replica},
we obtain
\begin{equation}
  \frac1{4}\lim_{n\to0}
  \left.
  \partial_{\eta_1}^2\partial_{\eta_2}^2
  Z_{2,n}^{\mathrm A{,\mathcal W}}(\eta_1,\eta_2)
  \right|_{\eta_1=\eta_2=0}
  =\frac12 q^4 \left[1+\mathcal P_2(|\omega|^2)\right].
  \label{eq:w2-source-replica-limit}
\end{equation}
Finally, \(\pi\mathcal O_1(z)=q^2/N\) in class A.  Combining
Eqs.~\eqref{eq:w2-replica-extraction} and
\eqref{eq:w2-source-replica-limit}, we obtain
\begin{equation}
  \frac{\mathcal W_2(z_1,z_2)}{\mathcal O_1(z)^2}
  =\frac12\left[1+\mathcal P_2(|\omega|^2)\right],
  \qquad |\omega|>0.
  \label{eq:w2-replica-final}
\end{equation}
This produces Eq.~\eqref{eq:w_by_p} for class A by the replica method.
The same derivation works for \(\mathrm{AI}^\dagger\) and
\(\mathrm{AII}^\dagger\), and produces Eq.~\eqref{eq:w_by_p} for these classes.

\twocolumngrid

\end{document}